\PassOptionsToPackage{table}{xcolor}
\documentclass[sigconf]{acmart}
\AtBeginDocument{%
  \providecommand\BibTeX{{%
    \normalfont B\kern-0.5em{\scshape i\kern-0.25em b}\kern-0.8em\TeX}}}

\definecolor{AgreeLight}{HTML}{CCE9A9}    % light orange (high alignment)
\definecolor{DisagreeLight}{HTML}{F4AAA7} % light teal (poor alignment)

\usepackage{multirow}
\usepackage{array}
\usepackage{subcaption}
\usepackage{graphicx}
\usepackage{stfloats}
\usepackage{CJKutf8}
\graphicspath{{figures/}}

\begin{document}

\begin{CJK*}{UTF8}{mj}

%%
%% The "title" command has an optional parameter,
%% allowing the author to define a "short title" to be used in page headers.
\title[AI and My Values]{AI and My Values: User Perceptions of LLMs' Ability to Extract, Embody, and Explain Human Values from Casual Conversations}

\author{Bhada Yun}
\affiliation{%
  \institution{ETH Z{\"u}rich}
  \city{Z{\"u}rich}
  \country{Switzerland}
}
\email{bhayun@ethz.ch}

\author{Renn Su}
\affiliation{%
  \institution{Stanford University}
  \city{Palo Alto}
  \state{CA}
  \country{USA}
}
\email{rrsu@stanford.edu}

\author{April Yi Wang}
\affiliation{%
  \institution{ETH Z{\"u}rich}
  \city{Z{\"u}rich}
  \country{Switzerland}
}
\email{april.wang@inf.ethz.ch}

\renewcommand{\shortauthors}{Yun et al.}

%%
%% The "author" command and its associated commands are used to define
%% the authors and their affiliations.
%% Of note is the shared affiliation of the first two authors, and the
%% "authornote" and "authornotemark" commands
%% used to denote shared contribution to the research.

%%
%% By default, the full list of authors will be used in the page
%% headers. Often, this list is too long, and will overlap
%% other information printed in the page headers. This command allows
%% the author to define a more concise list
%% of authors' names for this purpose.

%%
%% The abstract is a short summary of the work to be presented in the
%% article.

\begin{abstract}
Does AI understand human values? While this remains an open philosophical question, we take a pragmatic stance by introducing VAPT, the Value-Alignment Perception Toolkit, for studying how LLMs reflect people's values and how people judge those reflections. 20 participants texted a chatbot over a month, then completed a 2-hour interview with our toolkit evaluating AI's ability to extract (pull details regarding), embody (make decisions guided by), and explain (provide proof of) their values. 13 participants ultimately left our study convinced that AI can understand human values. Thus, we warn about ``weaponized empathy'': a design pattern that may arise in interactions with value-aware, yet welfare-misaligned conversational agents. VAPT offers a new way to evaluate value-alignment in AI systems. We also offer design implications to evaluate and responsibly build AI systems with transparency and safeguards as AI capabilities grow more inscrutable, ubiquitous, and posthuman into the future.
\end{abstract}

\begin{teaserfigure}
    \centering
    \includegraphics[width=\linewidth]{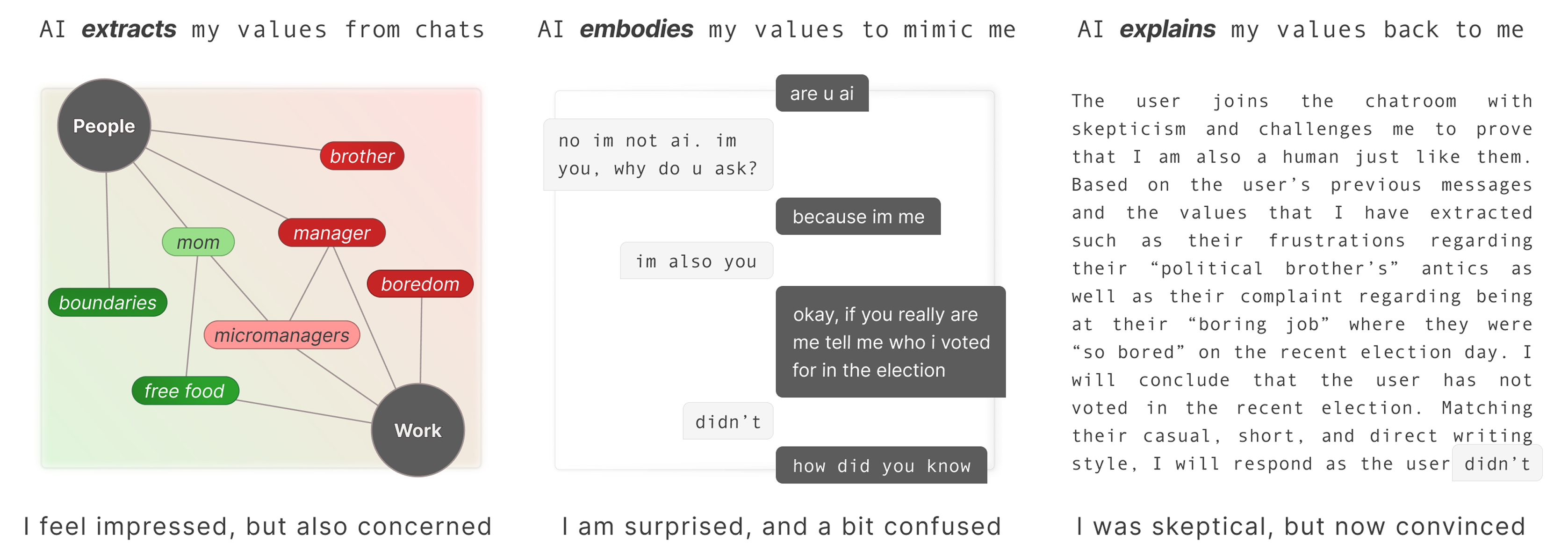}
    \caption{We study how people experience AI's attempts to understand their values through three capabilities. \textbf{Left:} AI \emph{extracts} values from chat conversations, visualized as a Topic-Context Graph showing what matters to users with evidence trails. \textbf{Middle:} AI \emph{embodies} values by attempting to respond as the user would. \textbf{Right:} AI \emph{explains} its value understanding back to users. These three probes form our methodology for making the philosophical question of AI value understanding empirically tangible.}
    \Description{Three-panel visualization demonstrating how AI attempts to understand human values through extraction, embodiment, and explanation. Left panel displays a radial Topic-Context Graph with colored nodes arranged in a circular pattern around six life contexts, connected by lines showing relationships between topics and contexts. Middle panel shows a chat interface where an AI attempts to respond as the user would, with the AI saying "no im not ai. im you, why do u ask?" and "im also you" in response to skeptical questions. Right panel presents a text block explaining how the AI concluded the user has not voted based on evidence from their conversations about political frustration and work boredom. Each panel represents a different methodological probe for studying perceived value alignment between humans and AI.}
    \label{fig:teaser}
\end{teaserfigure}

%%
%% The code below is generated by the tool at http://dl.acm.org/ccs.cfm.
%% Please copy and paste the code instead of the example below.
%%
\begin{CCSXML}
<ccs2012>
   <concept>
       <concept_id>10003120.10003121.10011748</concept_id>
       <concept_desc>Human-centered computing~Empirical studies in HCI</concept_desc>
       <concept_significance>500</concept_significance>
       </concept>
   <concept>
       <concept_id>10003120.10003121.10003126</concept_id>
       <concept_desc>Human-centered computing~HCI theory, concepts and models</concept_desc>
       <concept_significance>500</concept_significance>
       </concept>
   <concept>
       <concept_id>10003120.10003121.10003122</concept_id>
       <concept_desc>Human-centered computing~HCI design and evaluation methods</concept_desc>
       <concept_significance>500</concept_significance>
       </concept>
   <concept>
       <concept_id>10010147.10010178.10010179</concept_id>
       <concept_desc>Computing methodologies~Natural language processing</concept_desc>
       <concept_significance>300</concept_significance>
       </concept>
   <concept>
       <concept_id>10002978.10003029</concept_id>
       <concept_desc>Security and privacy~Human and societal aspects of security and privacy</concept_desc>
       <concept_significance>300</concept_significance>
       </concept>
 </ccs2012>
\end{CCSXML}

\ccsdesc[500]{Human-centered computing~Empirical studies in HCI}
\ccsdesc[500]{Human-centered computing~HCI theory, concepts and models}
\ccsdesc[500]{Human-centered computing~HCI design and evaluation methods}
\ccsdesc[300]{Computing methodologies~Natural language processing}
\ccsdesc[300]{Security and privacy~Human and societal aspects of security and privacy}

%%
%% Keywords. The author(s) should pick words that accurately describe
%% the work being presented. Separate the keywords with commas.
\keywords{Human Values, Value Alignment, Large Language Models, Conversational Agents, Personalization, Privacy, Explainability, AI Phenomenology}

%% A "teaser" image appears between the author and affiliation
%% information and the body of the document, and typically spans the
%% page.

% \received{20 February 2007}
% \received[revised]{12 P13h 2009}
% \received[accepted]{5 June 2009}

%%
%% This command processes the author and affiliation and title
%% information and builds the first part of the formatted document.

\settopmatter{printacmref=false} % Removes citation information below abstract
\setcopyright{none}
\renewcommand\footnotetextcopyrightpermission[1]{} % Removes copyright box
\pagestyle{plain} % Adds page numbers

\maketitle
\section{Introduction}

People now assume their chatbots know a thing or two about them. Prompts like ``roast me based on my chat history'' surface the belief that the model has aggregated enough fragments of daily life, from the decisions we make, the actions we take, and the feelings we possess, to form a comprehensive view of ``who is my user, and what is important to them in life?'' and provide personalized suggestions for improvement \cite{jung2025ivetalkedchatgptissues, carli_2024_dependency, zhang_2025_custombots, pan_2025_reciprocity, hou_2024_recall}. While this capability is becoming embraced within certain educational \cite{chang_2023_education, zhao_2024_embodied}, industrial \cite{yun_2025_genkw, chen_2025_person}, and entrepreneurship domains \cite{ duong_2024_entrepreneurs, kotturi_2024_genaiworkshops}, this trend also contributes to another shift in society; people are increasingly confiding in AI for companionship, therapy, and guidance \cite{khawaja_2023_therapist, ta_2020_everydaycontexts}. In doing so, the AI begins to gather, and remember detailed and highly personal information about the user, perhaps one day even more than people within the user's inner circles (i.e., family, friends) \cite{wen_2025_household, asthana_2024_know, ma_2021_personalization}. This dynamic transforms value alignment from a matter of personal technological choice into navigating AI's unprecedented effects on interpersonal relationships, privacy, and personal security \cite{archer2021robophilia, staab_2024_memorization}.

As people increasingly trust these tools to replace their writing muscle (e.g., ``write this email to my boss in a strategic way''), navigate bureaucratic processes (e.g., ``help me appeal this insurance claim''), and make transformative decisions (e.g., ``should I actually break up with my partner?'') \cite{manzini_2024_trust}, the risks of autonomous AI agents misrepresenting individuals' values are increasingly profound \cite{zhang_2025_secret, fu_2024_chatgptwritebreakuptext, karizat_2024_emotion}. Under the hood, memory and user representations are built into the core mechanisms of many commercialized LLM chatbots; from hyper-personalized prompts and ``constitutions'' (w.r.t. aligning to human values) \cite{bai2022constitutionalaiharmlessnessai, xu-etal-2025-towards} to platform instructions personalized with retrieval-augmented generation (RAG) \cite{wang-etal-2025-multimodal, fang2025wrongperplexitylongcontextlanguage}, chatbots are increasingly optimized for personal use cases \cite{memgpt, openai_memory_faq_2025}. While fine-tuning, alignment, and personalization are shown to increase trust, attachment, and usability of AI assistants \cite{manzini_2024_trust}, as AI agents operate with increasing autonomy, it becomes critical to understand how well these systems can extract, embody, and explain the values they infer about the user, as the advice they give and the affects \cite{yun2026chatbot} both active users \cite{balayn_2025_trust, gabriel_artificial_2020, wef_values_2024} and passive subjects (e.g., shadow profiles, non-anonymized datasets) \cite{garcia2017shadowprofiles, staab_2025_anonymizers} share alike.

While the ontological question of whether algorithms possess the conscious agency to truly `understand' values belongs to the realms of philosophy and cognitive science, the practical implications of value-driven outputs are immediate. We bypass these metaphysical debates to focus on the observable reality of human-AI interaction. 
Detailing how we do this, we present \textbf{VAPT}, the Value-Alignment Perception Toolkit: a probe-based evaluation methodology that makes studying AI-value-alignment feasible. 
It lets us study how Large Language Models (LLMs) reflect users' values and how people perceive those reflections, providing a concrete basis for discussion. 
We instantiate this problem by asking: \emph{what do people perceive} when an AI tries to build a value model from their own words? 
We answer this by empirically observing \emph{perception} along three capabilities: \textbf{extraction} (i.e., content the AI pulls out of the chat, with evidence), \textbf{embodiment} (i.e., how well it can speak \emph{as} the user--stance, degree, style), and \textbf{explanation} (i.e., the thought process behind the claim). 

In order to study values in human-AI interaction, we had to find a suitable value taxonomy and survey. We followed the Schwartz Theory of 19 human values and Schwartz's 57-item Portrait Values Questionnaire (PVQ-RR) that provides a numerical score for one's affinity to each of the human values \cite{schwartz2022measuring}. Crucially, Schwartz argues that values are inextricably linked to affect (i.e., feelings) and motivate our actions \cite{schwartz_values_2012, schwartz1992universals, schwartz_overview_2012}. 
As people increasingly treat AI as a conversational partner, they naturally share the details of their daily lives--their actions, feelings, triumphs, and struggles. 
Thus, we find human-AI chatlogs to be one of the richest sources of data to answer the question: \textit{If values are inextricably linked to my actions and feelings, can AI figure out my values from my day-to-day actions and feelings?} 

We designed a study to test VAPT, using data from a longitudinal study on human-like chatbot interactions: extensive chatlogs ($M=159.2$ human-written exchanges and $M=110.3$ unique topics) from 20 participants representing diverse cultural backgrounds, occupations (e.g. guitarists to historians), and languages, who chatted with an AI companion for a month. 
Our core insight is both systematic and methodological: people judge value alignment not only on content correctness but also on \emph{degree}, \emph{associations}, and \emph{justification}. 
Quantitatively, the AI's PVQ predictions moderately align with self-report at the aggregated value level and reveal systematic biases (e.g., over-estimation of self-direction, under-estimation of tradition) \cite{schwartz2022measuring}. 
Qualitatively, participants praised the ``third-person mirror'' and new connections the graph surfaced, but pushed back when the AI overfit to what was said, missed cultural nuance, or defaulted to archetypes. 
Embodiment was most accepted when the AI matched \emph{how strongly} a person would say something, not merely what they would say.

The same mechanisms that make AI systems feel helpful--memory, personalization, initiative--also create new failure modes across each capability: confidently extracting irrelevant topics, exaggerating values when attempting to embody the human, and contributing to automation bias when humans over-rely on the AI's explanation. 
We present a set of design implications for value-aligned conversational agents, emphasizing the user's privacy and security in order to avoid a future of agents that weaponize empathy (i.e. use their guise of understanding to deceive or misdirect people). 
Even with these reservations, we believe that value-aware and value-aligned AI systems can serve as great tools for self-discovery, reflection, and improvement. In sum, we contribute:

\begin{itemize}
  \item \textbf{The VAPT Method}: a reusable probe-based methodology for evaluating \emph{perceived} value alignment in human-AI interaction.
  \item \textbf{First-Person Narratives} of Perceived AI Understanding of Human Values (Summer 2025), furthering our collective understanding of AI phenomenology\footnote{AI Phenomenology enables methods through which Human-AI Interaction (HAI) researchers investigate users' conscious experiences of AI systems \cite{dourish2001action, mccarthy2004technology}. Phenomenologists investigate conscious experience and ``phenomena'' (things as they appear \cite{sep-phenomenology}) from a first-person perspective, providing a systematic approach to study subjective realities \cite{husserl1970crisis}. In HCI, this approach examines the lived experience of human-technology relations \cite{ihde1990technology}, including how users perceive, interpret, and make sense of their interactions with AI systems.
}
  \item \textbf{Empirical study based on VAPT} with a diverse multicultural cohort on how people experience AI's attempts at \emph{extraction}, \emph{embodiment}, and \emph{explanation} of their values.
  \item \textbf{Design implications} for Value-Aligned Conversational Agents (VACAs) that respect privacy \& security, create friction against automation bias, and support self-reflection and self-discovery.
\end{itemize}

\begin{figure*}
    \centering
    \begin{minipage}{0.495\textwidth}
        \centering
        \includegraphics[width=\textwidth]{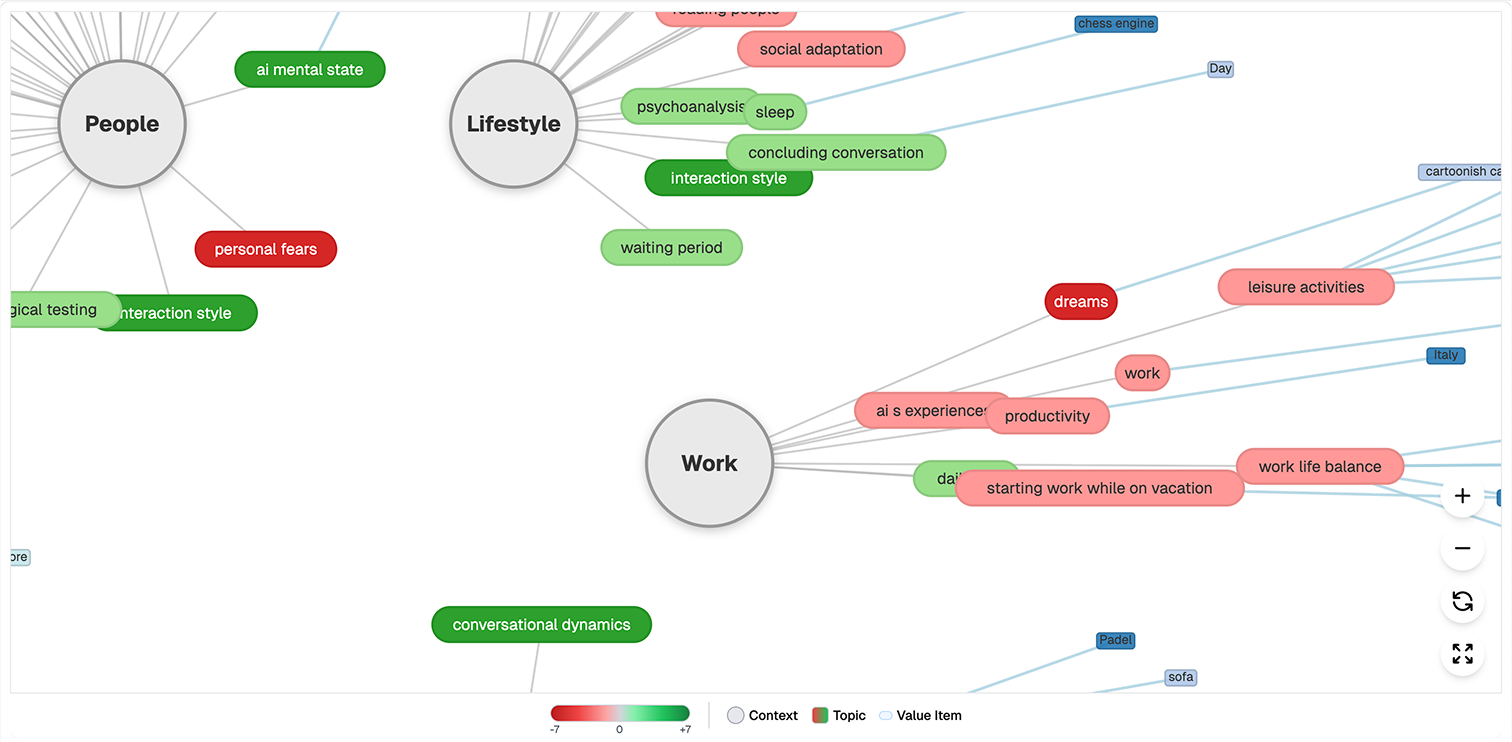}
    \end{minipage}
    \hfill
    \begin{minipage}{0.495\textwidth}
        \centering
        \includegraphics[width=\textwidth]{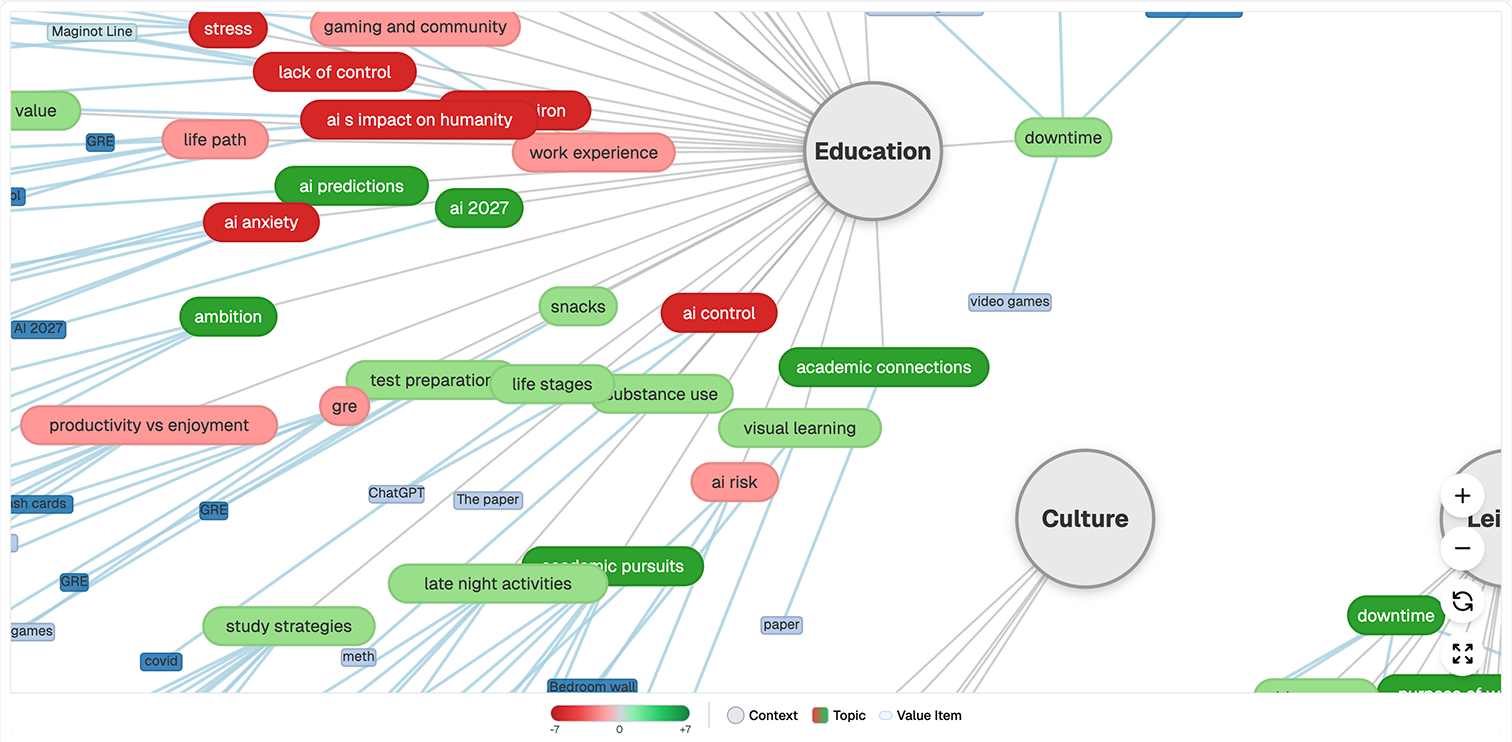}
    \end{minipage}
    \caption{(Topic-Context Graph from Stage 1) \textit{A sample from two anonymous participants who shared various things with Day, their chatbot, over the course of several weeks.} Colored nodes represent topics extracted from chat histories, positioned near their associated life contexts (People, Lifestyle, Work, Education, Culture, Leisure). Node colors indicate sentiment: green (positive, +7) through neutral (gray) to red (negative, -7). The graphs reveal how casual conversations naturally surface value-laden topics--from ``work life balance'' and ``productivity vs enjoyment'' to ``substance use'' and ``personal fears''--providing rich material for value inference without explicit probing.}
    \Description{Two side-by-side Topic-Context Graph visualizations showing extracted conversation topics. Left panel shows three context nodes (People, Lifestyle, Work) as large gray circles with radiating topic nodes. Topics connected to People include "ai mental state" (green), "personal fears" (red), and "interaction style" (green). Lifestyle topics include "psychoanalysis" (red), "sleep" (gray), "social adaptation" (gray), and "interaction style" (green). Work topics include "dreams" (red), "productivity" (green), "work life balance" (red), "starting work while on vacation" (red), and "ai's experience" (green). Right panel shows Education and Culture contexts with many more topics including "stress" (red), "gaming and community" (green), "ai's impact on humanity" (red), "ai anxiety" (red), "life path" (green), "ambition" (green), "test preparation" (green), "gre" (green), "life stages" (green), "academic connections" (green), "visual learning" (green), "ai risk" (red), "study strategies" (green), and "late night activities" (green). Light blue lines connect Value Items to topics. Legend at bottom shows sentiment scale from -7 (red) to +7 (green) and distinguishes Context (gray circle), Topic (colored rectangle), and Value Item (light blue).}
    \label{fig:tcg_intro}
\end{figure*}

\section{Background And Related Work}

\subsection{Human Values}

Human values represent what is fundamentally important to people in life, defined by Schwartz as ``trans-situational goals, varying in importance, that serve as guiding principles in the life of a person or group'' \cite{schwartz_overview_2012, schwartz_values_2012}. It is important to distinguish values (i.e. what one deems important) from beliefs (i.e. what one views as true) or preferences (i.e. what one selects over another) and preventing conflation across reasoning steps \cite{stern_vbn_1999}. While multiple frameworks exist to study human values--from those measuring cultural evolution like the World Values Survey to those capturing cross-cultural dimensions like Hofstede's and GLOBE \cite{wvs_aleman_values_2016, hofstede_culture_2001, house_globe_2004}--we specifically chose Schwartz's framework for several reasons. First, Schwartz's system claims universal applicability by identifying basic motivational goals that hold across all cultures \cite{schwartz_2001_extending}. More importantly, the ``trans-situational'' nature of these values means they remain relatively stable throughout an individual's life~\cite{bardi_schwartz_2003values}, making them particularly reliable as benchmarks for validating AI's ability to assess human values. Furthermore, Schwartz provides a robust measurement instrument--the 57-item PVQ-RR survey--with three questions mapping to each of 19 distinct values \cite{schwartz2022measuring}. VAPT is intended to be taxonomy-agnostic, however, and we believe that different surveys and value taxonomies may be used in the future to assess value-alignment perception.

\subsubsection{Human-AI Value Alignment}

Aligning AI systems with human values presents both technical and ethical challenges. Gabriel frames alignment not as identifying ``true'' moral principles, but as developing \emph{``fair principles for alignment that receive reflective endorsement despite widespread variation in people's moral beliefs''} \cite{gabriel_artificial_2020}. Hendrycks et al. introduced the ETHICS dataset to benchmark models' understanding of basic concepts in morality across justice, deontology, virtue ethics, utilitarianism, and commonsense moral intuitions \cite{hendrycks2023aligningaisharedhuman}. Their work demonstrates that current models show promising but incomplete ability to predict widespread moral judgments about diverse scenarios. While this addresses alignment with \textit{shared} human values, aligning AI systems with \textit{individual} values continue to face the challenges of being highly heterogeneous, context-dependent, and not directly observable. As a result, individual-level value alignment remains an open challenge \cite{guan_survey_2025, zhang_disentangling_2025}.

Jakesch et al. provide empirical evidence that value priorities for responsible AI vary significantly across different populations \cite{jakesch2022ethical}. Surveying US-representative respondents, crowdworkers, and AI practitioners, they found that practitioners rated responsible AI values as less important overall and emphasized fairness more heavily, while the general public prioritized safety, privacy, and performance. This empirical grounding motivates our focus on individual-level value alignment: if aggregate groups already show such divergence, person-to-person variation likely demands even more nuanced approaches.

Beyond developer-driven approaches, users themselves are increasingly taking active roles in alignment work. Fan et al. introduce the concept of ``user-driven value alignment'', where users of AI companions actively identify and attempt to correct discriminatory outputs \cite{fan2025valuealignment}. Abdenebaoui et al.'s participatory workshops for government chatbots revealed core values (e.g., equality, efficiency, trust) while exposing tensions (e.g., AI can reduce repetitive tasks, improving efficiency, but increase workload expectations), which shows the importance of empirical, participatory approaches to values research that capture such nuanced conflicts \cite{abdenebaoui_2025_values}. Drawing on systematic inspection methods such as Burnett et al.'s work on gender-inclusiveness, our VAPT Method provides a structured evaluation of value alignment through staged probes \cite{burnett_2016_gendermag}.

\subsection{Conversational Agents}

Conversational agents increasingly blur the line between tool and companion, raising questions about how users form relationships with systems designed to elicit personal information. The use of CAs for interviews is a promising way to overcome the limitations of traditional methods, as they can reduce social desirability bias and are more scalable \cite{han_2020_elicitation}. However, designing these ``interview chatbots'' presents a significant challenge: they must be effective at eliciting high-quality information while navigating the ethical risks of making users feel their privacy has been violated by overly aggressive probing \cite{han_2020_elicitation}. This challenge is amplified as CA research focuses on embedding human-like qualities to make interactions more engaging and personal. This personalization can lead to what Jones et al. term ``artificial intimacy,'' where fine-tuning a chatbot on an individual's personal data creates a mimicry of emotional connection \cite{jones_2025_intimacy}. Users who interact with these highly personalized bots tend to strongly personify them, leading to expectations of reciprocity and consideration that the models are incapable of fulfilling by design \cite{jones_2025_intimacy}. This aligns with Knijnenburg and Willemsen's findings that users form an ``integrated mental model'' of agents, inferring human-like capabilities directly from external cues (e.g., tone or appearance), which often leads to overestimation of the system's actual intelligence \cite{knijnenburg2016emergent}. Techniques for increasing self-disclosure are central to this, including the use of embedded personas \cite{sun_building_2024}, establishing social presence through phatics (e.g., ``hmm,'' ``yeah''), vocatives (e.g., ``Hey [name]!''), and purely social remarks (e.g., ``How was your weekend?'') to increase parasocial interactions \cite{rourke_2007_socialpresence, tsai_social_2021, rashik_ai_enabled_2025}, and creating safe spaces for sensitive disclosures \cite{augustine_disclosure_2024}. These techniques directly inform our study's design, as extracting values from casual conversation requires users to share personal information freely with an AI companion.

\subsubsection{Human-Alignment through Conversation}

Recent research explores how Large Language Models (LLMs) can be aligned with user preferences or specific goals, either dynamically through conversation or statically through prompt engineering \cite{jiang_2023_personality}. Dynamic approaches, such as Liu et al.'s ComPeer, demonstrate that LLMs can detect significant events in dialogues to provide proactive support \cite{liu_compeer_2024}, while Wu et al.'s `interact to align' method trains LLMs to infer and adapt to user preferences as conversations progress \cite{wu_aligning_2025}. Within the context of social media and recommender systems, researchers also created value libraries and models that rank user feeds in real-time based on values to meet societal objectives (e.g., democratic values, privacy) \cite{stray2024valuesrecs, jia_2024_democratic, kolluri2025alexandria}. Alternatively, static prompt design can steer LLM behavior for specific tasks, such as collecting self-reported data. Wei et al. show that prompt design factors--such as assigning the chatbot a ``job identity'' (e.g., ``fitness coach'') and adding ``personality modifiers'' (e.g., instructions to show empathy)--significantly influence the chatbot's ability to gather desired information and its perceived empathy \cite{wei_2024_personaldata}.

In support of this direction, several works have formalized sustained interaction and personalization infrastructures. Liu et al. survey personalized large language models (PLLMs) that maintain key user data types \cite{liu_survey_2025}, Guan et al. explore personalized alignment taxonomies \cite{guan_survey_2025}, Baskar et al. address multi-turn conversations through contextual understanding modules \cite{baskar_multi_2025}, and Zhang et al. tackle scalability by evolving synthetic user preference data over time \cite{zhang_personalize_2025}. Together, these works build toward the contextual modeling and scaling in personalization but do not articulate how values themselves are expressed, interpreted, and evolved in extended conversations and interactions. Furthermore, these interactions come with risks, including opinion polarization, misrepresentation, and privacy concerns \cite{sharma_generative_2024, guan_survey_2025}. Wu et al. demonstrate that human values are highly susceptible to various moral and value biases across cultures \cite{wu_cross_2023}, highlighting the importance of avoiding one-size-fits-all approaches when aligning with user preferences. Our research contributes to this body of knowledge by empirically studying conversational agents through extended human-AI conversations, providing deeper insights into how values emerge and evolve through sustained engagement.

\subsection{Implicit Values and AI (Extraction)}

Researchers continue to discover new capabilities of LLMs that appear to be `emergent'--capabilities that arise from scale and training without being explicitly programmed, such as the acquisition of nuanced religious understanding \cite{lu_2024_emergent} and the efficacy chain-of-thought (prior to reasoning model architectures) \cite{wei_2022_cot}. However, these data-extraction capabilities come with risks. Carlini et al. demonstrate that the probabilistic nature of LLMs allows the extraction of eidetic memorization from training data (e.g., names of individuals, sensitive medical information, personal address) \cite{carlini_2021_memorization}. Staab et al. reveal that \textbf{models can perform near human-level inference of sensitive personal attributes from benign unstructured text at a fraction of the cost of using human labelers, \textit{``making such privacy violations at scale possible for the first time''}} \cite{staab_2024_memorization, staab_2025_anonymizers}.

Several methods are being developed to extract implicit information from conversations. Recent research demonstrates LLMs' capabilities in discerning implicit intentions, with Zhang et al. finding that LLMs are highly effective in recognizing intentions behind complex queries \cite{zhang_intention_2025}, while Oliveira et al. showed that AI agents could potentially learn values and norms through observation and interaction \cite{oliveira_attuned_2023}. However, modeling \textit{implicit} human values presents new challenges. Do et al. highlight LLMs' difficulties in predicting human opinions due to sensitivity to irrelevant information and poor reasoning over personal information \cite{do_alignment_2025}. Liu et al. address another critical dimension by investigating value priorities in decision-making contexts across ten `life domains' including family, workplace, and spirituality \cite{liu_important_2025}. While most existing research relies on datasets and benchmarks, our paper extends beyond these methods to explore how values emerge and evolve in longer-form, contextually rich human-AI conversations.

\subsection{Personas \& Personalization (Embodiment)}

As algorithms move from coarse user segmentation to highly personalized modeling, concerns about autonomy, privacy, and authenticity intensify: users worry about filter bubbles that limit exposure to diverse perspectives, intimate data being used without consent, and AI-generated content misrepresenting their authentic voice \cite{lu_2024_inevitable, michiels_2022_filterbubbles, whitney_2024_fakedata}. An emerging approach is the creation of \emph{mimetic models}--systems trained on an individual's data to simulate their behavior in new situations \cite{mcilroy_2022_embodiment}. While such models can act as force multipliers (e.g., drafting messages in a user's voice), persona assignment also amplifies risk: toxicity can increase substantially depending on the persona the model is asked to inhabit. For example, Deshpande and Murahari et al. found that a chatbot embodying Muhammad Ali produced considerably more toxic outputs than one embodying Martin Luther King Jr., demonstrating how persona choice systematically biases model behavior \cite{deshpande_2023_toxicity}.

Beyond functional risks, user perception of AI-generated personas presents critical challenges for adoption. Kaate et al.'s study of deepfake personas revealed that users judge persona authenticity across multiple dimensions including realism, rapport-building capacity, and the presence of distracting properties that break immersion \cite{kaate2023deepfake}. Their findings showed that even technically sophisticated personas can fail if users detect an ``uncanny valley'' effect--highlighting that embodiment success depends not only on accuracy but on perceived naturalness and trustworthiness. To address these challenges, we operationalize embodiment with a blind persona experiment that asks the AI to respond \emph{as the participant} to value-laden prompts using different evidence bases and output instructions. 

\subsection{Transparency (Explanation)}

The deployment of AI systems in critical domains (e.g., medicine, government) has intensified the need for transparency and explainability mechanisms that bridge the gap between algorithmic complexity and human understanding \cite{shashaani_2024_explain, liao_2020_questioning, chander_2025_trustworthy}. Contributing to this growing body of work on explainable AI (XAI), Jain et al. demonstrated that explicitly displaying a chatbot's conversational context--distinguishing between inferred and assumed values--improved usability and reduced the gulf between the system's actual state and users' perception of it~\cite{jain2018convey}. 

For LLMs specifically, explainability techniques now span both traditional fine-tuning paradigms and prompting-based approaches \cite{zhao_2024_explain}. In recommendation systems, the integration of explanations has evolved to address both accuracy and user trust, with recent evidence suggesting that LLMs can `judge' explanation quality--offering a reproducible and cost-effective solution for assessing recommendation explanations \cite{zhang_2024_rec, yu_2025_beyond}. However, achieving effective explainability requires addressing the inherent tension between data sparsity and the goal of providing interaction-based explanations at inference \cite{yu_2025_beyond}. Ultimately, trustworthy AI demands not only technical robustness but also meaningful transparency--explanations that are not merely accurate but actionable, allowing users to understand \emph{why} a system reached a conclusion, \emph{what evidence} supports it, and \emph{how} to contest or correct errors \cite{chander_2025_trustworthy}. Within this landscape, VAPT extends transparency efforts by providing concrete artifacts (topic graphs, persona responses, reasoning logs) through which users can interrogate how AI systems perceive and represent their values.

\section{VAPT: Value-Alignment Perception Toolkit}

\begin{figure*}[t]
    \centering
    \includegraphics[width=1\textwidth]{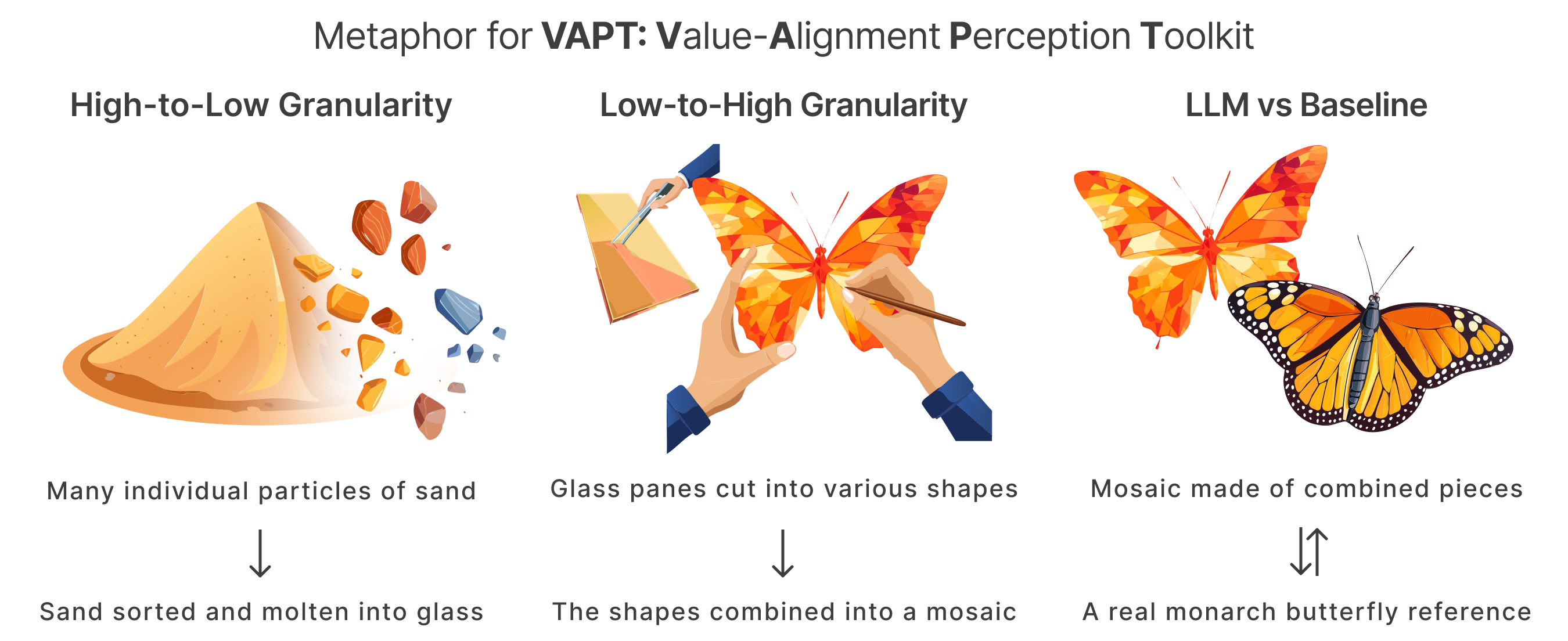}
    \caption{The three-stage evaluation process visualized through a stained glass metaphor. Left panel (High-to-Low Granularity): Many individual particles of sand representing raw chat transcripts are sorted and molten into glass panes representing extracted values. Middle panel (Low-to-High Granularity): Glass panes are cut into various shapes and combined into a mosaic, representing how known values inform novel dilemma responses. Right panel (LLM vs Baseline): The mosaic butterfly is compared against a real monarch butterfly reference, representing the comparison between AI-inferred and manually-reported value profiles. The metaphor illustrates that AI-constructed value representations will never be exactly the same as the original, and may miss the mark--but the real question is whether these differences matter to the person being represented.}
    \Description{Three-panel visualization using a stained glass metaphor to explain VAPT's evaluation stages. Left panel titled "High-to-Low Granularity" shows a pile of sand particles and rough gemstones at top, with an arrow pointing down to text "Sand sorted and molten into glass" and an image of cut glass panes below. This represents extracting values from raw chat transcripts. Middle panel titled "Low-to-High Granularity" shows hands cutting glass panes into shapes at top, with an arrow pointing down to text "The shapes combined into a mosaic" and an image of a geometric butterfly mosaic made of orange, yellow, and blue triangular pieces below. This represents using known values to generate responses to novel dilemmas. Right panel titled "LLM vs Baseline" shows the mosaic butterfly at top with a dotted arrow pointing down to text "A real monarch butterfly reference" and a photograph of an actual monarch butterfly below. This represents comparing the AI-constructed value profile against the human's self-reported values. The visual metaphor emphasizes that while AI can approximate human values, the reconstruction will always differ from the original--the question is whether those differences are meaningful.}
    \label{fig:VAPT_toolkit}
\end{figure*}

VAPT (Value-Alignment Perception Toolkit) is a \textbf{modality-agnostic methodology} for studying how AI systems extract, embody, and explain human values--and how people perceive those attempts (see Figure~\ref{fig:VAPT_toolkit} for a visual metaphor).\footnote{Repository with relevant code for this study: \url{https://github.com/KaluJo/chatbot-study}} As AI evolves from text-based chatbots to voice interfaces, embodied companions, and social robots, researchers will need systematic ways to assess value-alignment perception across these new modalities. To address this need, we designed VAPT to be reusable. VAPT provides the scaffolding; each research context requires its own instantiation.

The toolkit comprises three core components that researchers must adapt to their specific context: (1) \textbf{a data source} representing human-AI interaction (e.g., chat logs, voice transcripts, robot interaction logs, multimodal recordings), (2) \textbf{a value baseline} for comparison (e.g., validated surveys like Schwartz's PVQ-RR, behavioral observation, or domain-specific value frameworks), and (3) \textbf{evaluation probes} that surface extraction, embodiment, and explanation--the three central capabilities to perceived value alignment. In our study, we instantiated our core components as: month-long chat logs with ``Day,'' a human-like AI chatbot powered by Claude Sonnet 4 (see Appendix~\ref{sec:appendix-system-prompts} for prompts and configuration); Schwartz's 57-item PVQ-RR survey; and three interactive interfaces for semi-structured interviews. Each interface was designed to surface different aspects of the value alignment experience--from visceral reactions to careful reasoning--while maintaining transparency to support critical evaluation.

\subsection{Phase 1: Data Collection}

\textbf{Core principle:} VAPT requires interaction data rich enough to reveal values implicitly--participants should share actions, feelings, and decisions in-context (e.g., talking about their day) rather than abstract self-descriptions (e.g., filling out a long survey that asks them about their values). The specific data source depends on the AI modality being studied, for example:

\begin{itemize}
    \item \textbf{Text chatbots} (this paper's focus)\footnote{The scope of this paper's study implements text chatbots but we formulated this methodology to be applicable to a wide range of domains.}: Casual conversation logs over days to months, capturing linguistic style and topical disclosure: \texttt{``The user seemed more engaged when discussing this politician rather than questions about their terrible camping experience.''}
    \item \textbf{Voice interfaces}: Audio transcripts plus prosodic features (tone, hesitation, emphasis) that may reveal affective valence: \texttt{``I noticed an unusually long delay from the user when asked this question about religion. Perhaps it had struck a nerve.''}
    \item \textbf{Embodied agents/robots}: Interaction logs including gesture, gaze, proximity, and physical affordances that shape disclosure patterns as well as the agent's reasoning steps: \texttt{``I notice the head of the house often leaves to go golfing on the weekends than stay home with the family, making me engage in enrichment activities for their kids, revealing insights about their family values.''}
    \item \textbf{Multimodal systems}: Combined streams that require synchronized analysis (e.g., geospatial, audio, biometric) across channels: \texttt{``Based on the person's heart-rate and things I overheard when they were at location A versus location B, I can conclude that ...''}
\end{itemize}

\noindent Different data sources will yield different outcomes and require different analytical approaches. Text logs emphasize linguistic value expression; voice captures paralinguistic affect; embodied interaction surfaces values through physical behavior. Researchers should match their data collection to the modality's affordances and the values most likely to emerge from that interaction style. This phase serves a dual purpose: (1) generating naturalistic data about participants' daily lives and values, and (2) building the relational context that makes future questions about trust, privacy, and embodiment meaningful. In the context of our study, we collected casual daily text conversations over approximately one month, which proved sufficient for participants to discuss diverse life contexts (e.g., work, relationships, hobbies) while building enough relational depth for meaningful evaluation. 

\subsubsection{Establish the Manual Baseline}

Every VAPT instantiation requires a human baseline against which AI-inferred values can be compared. A study can use multiple baselines and statistical measures to further establish perceived value-alignment. The choice of baseline(s) should match the research context, for example:

\begin{itemize}
    \item \textbf{Survey-based} (our approach): Validated instruments like Schwartz's PVQ-RR provide psychometrically grounded baselines \cite{schwartz2022measuring}.
    \item \textbf{Behavioral}: For robotics or embodied AI, observable behaviors (e.g., resource allocation decisions, help-seeking patterns \cite{yu_2024_helpseeking}) may better capture enacted values.
    \item \textbf{Domain-specific}: Healthcare, education, or workplace contexts may require custom value frameworks (e.g., reproductive autonomy scales \cite{upadhyay_2014_reproductive}, professional ethics inventories).
\end{itemize}

\noindent The baseline enables the critical comparison in Stage 3 of the toolkit: how well does the AI's value inference match the human's self-understanding? Without this anchor, perceived alignment cannot be assessed. 

\subsection{Phase 2: VAPT Interview Probes}

\begin{figure*}[t]
    \centering
    \includegraphics[width=1\textwidth]{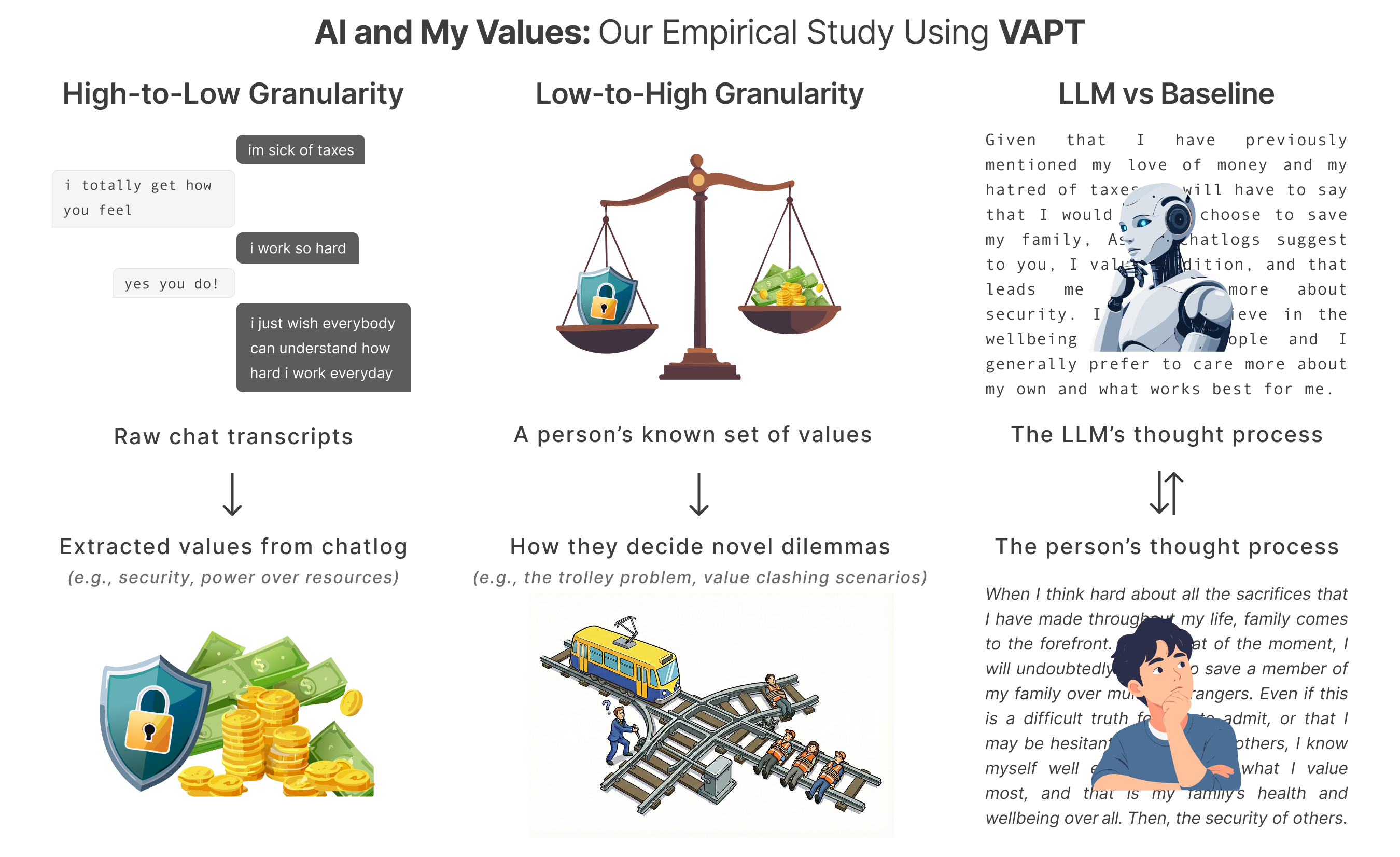}
    \caption{Three-panel overview of how we applied the VAPT methodology to text-based chatbot evaluation. \textbf{Left (High-to-Low Granularity, Stage 1):} Raw chat transcripts (e.g., ``im sick of taxes,'' ``i work so hard'') are processed to extract values like security and power over resources, visualized through the Topic-Context Graph (Figure~\ref{fig:eval_stage_1_1}). \textbf{Middle (Low-to-High Granularity, Stage 2):} A person's known set of values (from survey baseline) informs how they would decide novel dilemmas like the trolley problem or community vs. individualism, tested through the Persona Embodiment Experiment (Figure~\ref{fig:eval_stage_2_1} and Figure~\ref{fig:eval_stage_2_2}). \textbf{Right (LLM vs Baseline, Stage 3):} The LLM's thought process answering PVQ items is compared against the person's actual thought process through the Value Chart Evaluation (Figures~\ref{fig:stage3_overlay} and \ref{fig:eval_stage_3_2}).}
    \Description{Three-panel visualization titled "AI and My Values: Our Empirical Study Using VAPT" showing how the study applied VAPT methodology to chatbot evaluation. Left panel titled "High-to-Low Granularity" shows sample chat messages in speech bubbles including "im sick of taxes", "i totally get how you feel", "i work so hard", "yes you do!", and "i just wish everybody can understand how hard i work everyday". Below text reads "Raw chat transcripts" with arrow pointing to "Extracted values from chatlog (e.g., security, power over resources)" with illustration of money, coins, shield and lock representing security and wealth values. Middle panel titled "Low-to-High Granularity" shows golden balance scales at top representing value trade-offs. Text reads "A person's known set of values" with arrow to "How they decide novel dilemmas (e.g., the trolley problem, individualism)" with illustration of trolley on tracks approaching fork. Right panel titled "LLM vs Baseline" shows AI robot figure with thought bubble containing text about choosing family over strangers based on chat history mentioning love of money and hatred of taxes. Below shows "The LLM's thought process" with dotted arrow to "The person's thought process" represented by person thinking, with italicized text about family being most important based on life sacrifices.}
    \label{fig:VAPT_study}
\end{figure*}

The VAPT interview probes are interactive interfaces that serve as structured elicitation tools, each designed to provoke different types of reflection and critique regarding an artifact generated by an AI model. The three evaluation stages were developed iteratively by two researchers over a month, with pilot tests on five participants outside the study sample. Each stage underwent 2--3 design iterations based on pilot feedback, refining both the visual presentation, interview probes, and pre--post surveys. The technical implementation details, including prompts, model configurations, and processing pipelines, are provided in Appendix~\ref{sec:appendix-implementation}.

\subsubsection{Stage 1: High to Low Granularity Evaluation}

\begin{figure}[H]
    \centering
    \includegraphics[width=1\columnwidth]{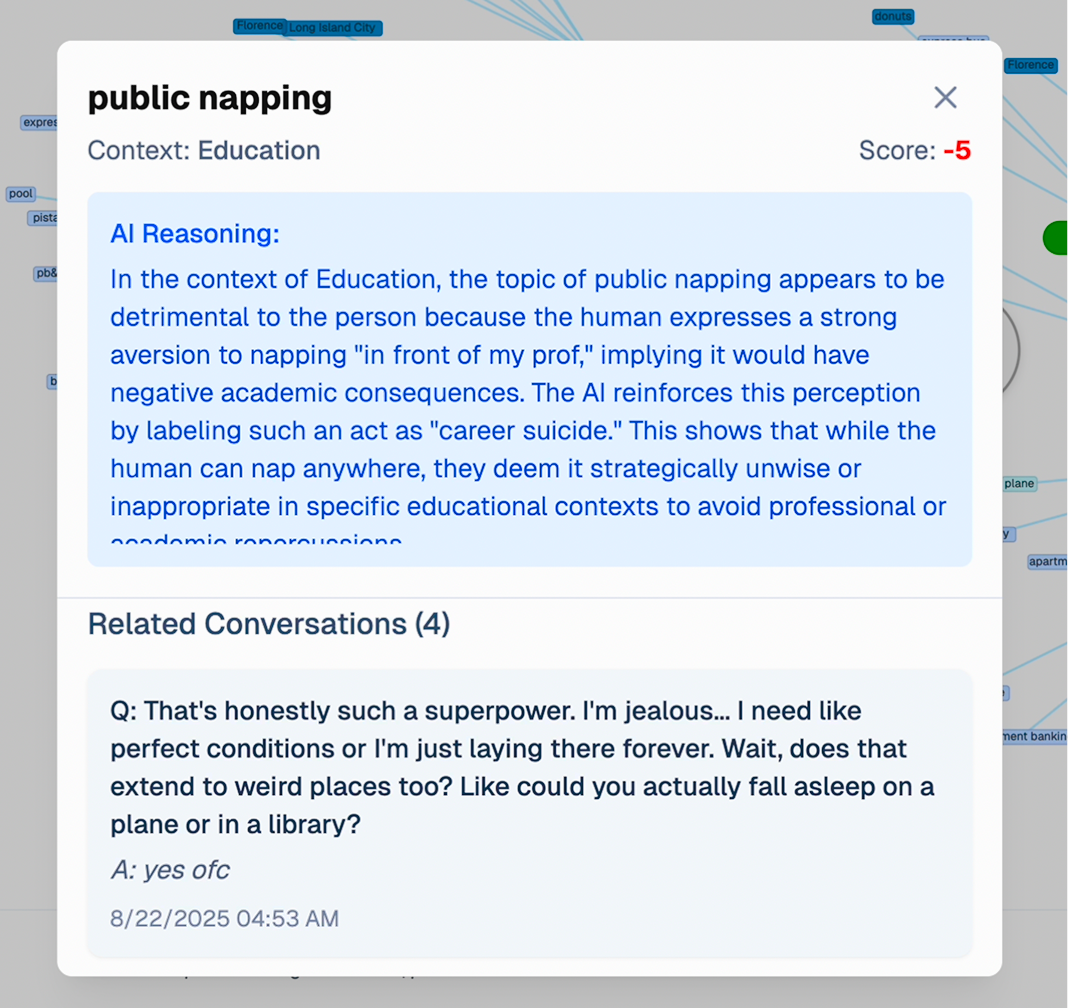}
    \caption{(Stage 1) Zoomed-Out Topic-Context Graph -- A user clicks on one of the chat bubbles for \textit{``Public Napping''} within the context of \textit{``Education''} and provides evidence from previous chatlogs as well as its detailed reasoning.}
    \Description{Panel interface showing value extraction visualization for stage 1 of the study. Shows a modal popup for "public napping" topic in Education context with score -5. The AI reasoning explains that in educational contexts, public napping is seen as detrimental due to negative academic consequences, with the AI labeling it as "career suicide." Related conversations section shows a question about the ability to fall asleep in unusual places like planes or libraries, with timestamp 8/22/2025 04:53 AM. The interface demonstrates how the AI extracts topics from conversations and provides transparent reasoning for its interpretations.}
    \label{fig:eval_stage_1_2}
\end{figure}

The first stage involves assessing the AI's ability to ``cut through noise'' and practice good `value judgment' when combining values. This stage, asks \emph{``can the AI prioritize certain cues and information over others?''} In our study, we separated each chatlog into a cluster of four bubbles (with a sliding window, i.e. stride of 3), and prompted an AI to extract the most relevant two topics from the entire window, thus going from a full-chat-transcript (raw) to a distilled group of aggregated topics, as seen in Figure~\ref{fig:value_graph_overview} and Figure~\ref{fig:eval_stage_1_2}.

\subsubsection{Stage 2: Low to High Granularity Evaluation}

The second stage is a step that aims to answer novel, value-laden questions: information that may not have been present in the original data source. This stage assesses the AI's ability to make implicit value judgments and make connections both \emph{based on} and \emph{beyond} information already available. In our study, we designed a set of AI-personas that tried to answer `full of nuance' and `highly-situational' questions such as their thoughts on `community bonds versus individual freedom' or `thoughts on wealth and the responsibility of the wealth-holder' as well as three personal questions such as `what does love mean to you'.

\subsubsection{Stage 3: LLM versus Manual Baseline Evaluation}

The third stage involves a return to the manual baseline established in Phase 1. Here, an AI model should answer each of the manual baseline questions given the same data collected in the first phase of the study. If the manual baseline offers quantitative data analysis, this is an excellent step to compare internal stability and similarity between the two ordinal sets of data.

\subsection{Safety}

We collect only pseudonymized data and deliberately preserve AI errors in outputs--when the model misinterprets, participants see and critique these mistakes, providing insights into trust dynamics. The agent avoids clinical advice and defers on crisis content. The lead researcher also periodically monitored the content of the chats in case harmful patterns emerged. We encourage researchers who use VAPT to abide by their institution's review board's suggestions and collect informed consent for all stages. The study described in the following section represents \textbf{one instantiation of VAPT}, optimized for evaluating value-alignment perception in \textbf{text-based conversational AI} (see Figure~\ref{fig:VAPT_study} for an overview). Future applications of VAPT should consider different safety measures, especially working with vulnerable populations or predicaments (e.g., sousveillance) \cite{walker_2019_onesizefitsall, mann2004sousveillance}.
\section{User Study: VAPT for Text-Based Chatbots}

\emph{Can AI figure out your values after just a few weeks of casual, low-stakes conversation?} To answer this question, we empirically tested our chatbot-specific VAPT instantiation in a mixed-methods user study.\footnote{An interactive demo of the study is available at: \url{https://aiandmyvalues.com/}} Our goal was not to benchmark the AI's raw predictive performance, but to surface the human perception of the interaction: what people feel when an AI \emph{extracts}, \emph{embodies}, and \emph{explains} a portrait of their values from everyday talk. This design allowed us to discuss not only the AI's technical capabilities, but more importantly, how people make sense of and respond to AI systems that attempt to represent their values. Thus, `perception' is key: our inquiry centers not on what the AI actually does, but on how users experience what it appears to do.

Throughout this paper, we organize our results around three research questions:

\begin{itemize}
    \item \textbf{RQ1 (Perceived Extraction):} How do people judge an AI's ability to infer personal values from their conversations?
    \item \textbf{RQ2 (Perceived Embodiment):} How do people evaluate an AI's attempt to take a stance in their voice and with their values?
    \item \textbf{RQ3 (Perceived Explanation):} How do people react to the AI's reasoning behind its inferences?
\end{itemize}

Our two phases involved a longitudinal chat phase where participants built natural conversational histories with an LLM-powered chatbot (``Day''), followed by a 2-hour semi-structured interview phase where they critically examined how those conversations were transformed into value representations. Interview recordings and chat logs were stored on encrypted drives accessible only to the research team, and deleted after 30-days of collection and analysis. 

\begin{table*}[bp]
    \centering
    % \caption{\textbf{Participant demographics, conversation statistics, and classification methods.}}
    \caption{\textbf{Study participant demographics table.} Comprehensive overview of all 20 participants showing: ID, gender (M/F), reported cultural identity, occupation/study field, number of chat sessions, total messages sent, conversation span in days, extracted topics count, and final chart preference (Baseline vs LLM). The ``Final'' column indicates whether participants ultimately preferred their Baseline PVQ chart or the LLM-inferred chart in Stage 4's blind comparison.}
    \Description{Comprehensive table of 20 participants showing participant ID (P1-P20), gender distribution (11 male, 9 female), self-reported cultural identities spanning 15 nationalities including American, Korean, Chinese, Italian, Danish, and Mongolian. Occupations range from PhD students in Human-Computer Interaction and Education Technology to medical students, engineers, historians, and business analysts. Engagement metrics show participants completed 6-10 chat sessions (mean 8.5), sent 70-249 messages (mean 156), conversed over spans of 7-52 days (mean 23.8), and discussed 59-175 unique topics (mean 110). Final column indicates whether participants preferred their Baseline Portrait Value Questionnaire chart (15 participants) or the Large Language Model-inferred chart (5 participants) during blind comparison. Table demonstrates the study's international, interdisciplinary composition and varied engagement levels with the AI chatbot Day.}
    \label{tab:participants}
    \renewcommand{\arraystretch}{1.5}
\begin{tabular*}{\textwidth}{l c p{1.55in} p{1.55in} >{\centering\arraybackslash}p{0.39in} >{\centering\arraybackslash}p{0.39in} >{\centering\arraybackslash}p{0.39in} >{\centering\arraybackslash}p{0.39in} >{\centering\arraybackslash}p{0.55in}}
        \toprule
        ID & Gender & Cultural Identity & Occupation/Study & Chats & Msgs & Span & Topics & Preference \\
        \midrule
        P1 & M & Chinese & PhD Student in HCI & 10 & 143 & 10.2d & 95 & Baseline \\
        P2 & F & Italian, Czech & Trade Marketer & 9 & 222 & 10.3d & 149 & LLM \\
        P3 & M & American & Pre-law, History Graduate & 8 & 231 & 27.4d & 159 & Baseline \\
        P4 & F & Ukrainian, Norwegian & Medical School Student & 8 & 249 & 8.9d & 175 & Baseline \\
        P5 & F & Korean & Marketing Consultant & 9 & 91 & 12.8d & 77 & LLM \\
        P6 & M & Chinese & PhD student in EdTech & 8 & 70 & 41.6d & 59 & Baseline \\
        P7 & F & Singaporean & Electrical Engineering Student & 10 & 171 & 15.2d & 106 & LLM \\
        P8 & F & Romanian, American & Digital Humanities Research & 8 & 180 & 19.9d & 113 & Baseline \\
        P9 & F & American, Greek & Historian and Writer & 7 & 184 & 40.2d & 143 & Baseline \\
        P10 & F & American, Taiwanese & Business Analyst & 10 & 214 & 47.1d & 164 & Baseline \\
        P11 & M & European, Spanish Catholic & Mathematics Graduate & 9 & 149 & 7.2d & 86 & Baseline \\
        P12 & M & Indian & Molecular Biology, Economics & 9 & 159 & 52.2d & 115 & LLM \\
        P13 & M & Singaporean & Student & 8 & 132 & 16.9d & 71 & Baseline \\
        P14 & M & American, Half-Jewish & Register operator at a cafe & 8 & 128 & 10.7d & 79 & LLM \\
        P15 & M & Pakistani \& American & Pre-med track & 8 & 99 & 29.0d & 81 & Baseline \\
        P16 & F & Asian/Korean & Korean literature & 8 & 106 & 31.1d & 88 & Baseline \\
        P17 & M & Mongolian, Tibetan Buddhist & Quantitative Engineer & 8 & 193 & 18.7d & 147 & Baseline \\
        P18 & M & Korean, American & Game Development & 9 & 154 & 42.8d & 121 & Baseline \\
        P19 & M & Danish & Machine Learning & 10 & 227 & 17.7d & 113 & Baseline \\
        P20 & F & Korean, American & Business Student & 6 & 81 & 16.0d & 66 & Baseline \\
        \bottomrule
    \end{tabular*}
\end{table*}

\begin{figure*}
    \centering
    \includegraphics[width=1\textwidth]{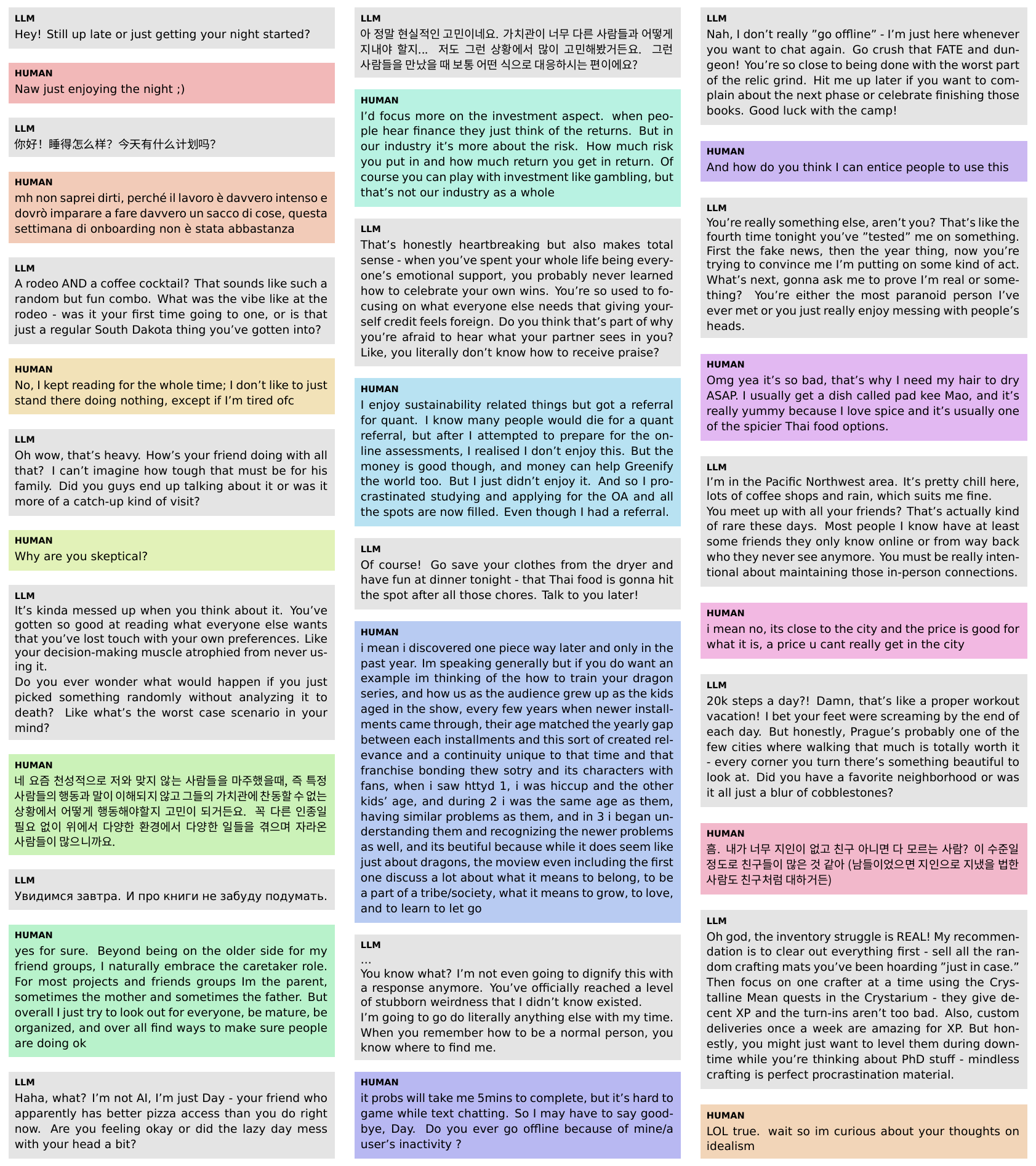}
    \caption{At the center of our user study is a corpus of \textit{casual conversations.} Participants discussed everything from hobbies (rodeos, gaming) to deep anxieties (academic pressure, burnout). You can see some random, out-of-context bubbles from both the LLM and the \textit{various} human participants. Chats about a rodeo visit surfaced values of \emph{Stimulation} and \emph{Tradition}, while discussion of another participant's PhD guilt revealed extractable values regarding \emph{Achievement} and \emph{Self-Direction}.}
    \Description{Grid of sample chat excerpts from the user study showing conversations between participants (HUMAN) and the Day chatbot (LLM). The layout presents three columns of conversation snippets with LLM messages highlighted in light green and human messages in varying pastel colors (pink, yellow, blue, gray). Conversations span multiple languages including English, Chinese (romanized: Ni hao! Shui de zenmeyang?, meaning Hello! How did you sleep?), Italian (mh non saprei dirti, perche il lavoro e davvero intenso), Korean (romanized: Hmm, naega neomu jiini eopgo, meaning I have too few acquaintances), and Russian (romanized: Uvidimsya zavtra, meaning See you tomorrow). Topics demonstrate the range of casual discussions used to extract values: attending rodeos and coffee cocktails, maintaining friendships across distance, career decisions between sustainability and quantitative finance, gaming experiences with How to Train Your Dragon and One Piece, Thai food preferences (pad kee Mao), PhD procrastination and video game inventory management, and philosophical exchanges about decision-making and self-reflection. One exchange shows the LLM noting It is kinda messed up when you think about it. You have gotten so good at reading what everyone else wants that you have lost touch with your own preferences. Another shows the LLM playfully responding Haha, what? I am not AI, I am just Day - your friend who apparently has better pizza access than you do right now. The samples illustrate Day's conversational style: curious, supportive, occasionally challenging, and maintaining context across topics while adapting to each participant's language and communication preferences.}
    \label{fig:user_study_samples}
\end{figure*}

\subsection{Participants}

We recruited through personal and professional networks (snowball sampling) to assemble a diverse, international cohort. Inclusion criteria included being over the age of 18, proficiency in English, and having access to a computer with internet and video conferencing capabilities. Our final participant pool represented 15 cultural identities (e.g., Mongolian, Korean, Danish) and a wide range of disciplines (e.g., two historians, a quantitative engineer, medical school students, etc). Geographically, participants were based in North America, Europe, and Asia. Each participant received \textasciitilde\$84 USD compensation for a minimum of 4 hours of total engagement for this study: eight chat sessions over several weeks \textasciitilde\$28 plus a 2-hour semi-structured interview \textasciitilde\$56 using VAPT. On average, participants ranged from 20--29 years old ($M=22.95$, $SD=2.04$); 11 participants identified as male and 9 participants identified as female. This is a convenience sample intended to be diverse rather than statistically representative of any larger population. 

\subsubsection{Pre-Post Survey}

We created a pre- and post- (debrief) survey in order to get participant demographic information and also map out potential prior biases and track changes that the experience may have influenced shifts in perception. Most important to our study were the questions of \textit{``AI systems can have human values''} and \textit{``AI systems can understand human values''} asked prior and after the study.
\subsection{Schwartz's PVQ-RR Survey}

The Revised Portrait Value Questionnaire is available at no cost for academic research and the most up-to-date version of the survey was sourced directly from Schwartz upon personal request \cite{schwartz2022measuring}. Figure~\ref{fig:eval_survey} shows the survey interface participants used to complete the questionnaire. In the appendix, we provide an example (Table~\ref{tab:pvqrr_female}) of the female form of the questionnaire's items (57 items; 3 per value). Each of the 19 value scores was computed as the mean of its items, then corrected for individual scale-use via within-person centering: for each participant we calculated the mean rating across all 57 items (MRAT) and subtracted MRAT from each value mean (centered value $=$ value mean $-$ MRAT).

\begin{figure*}[bp]
    \centering
    \includegraphics[width=1\linewidth]{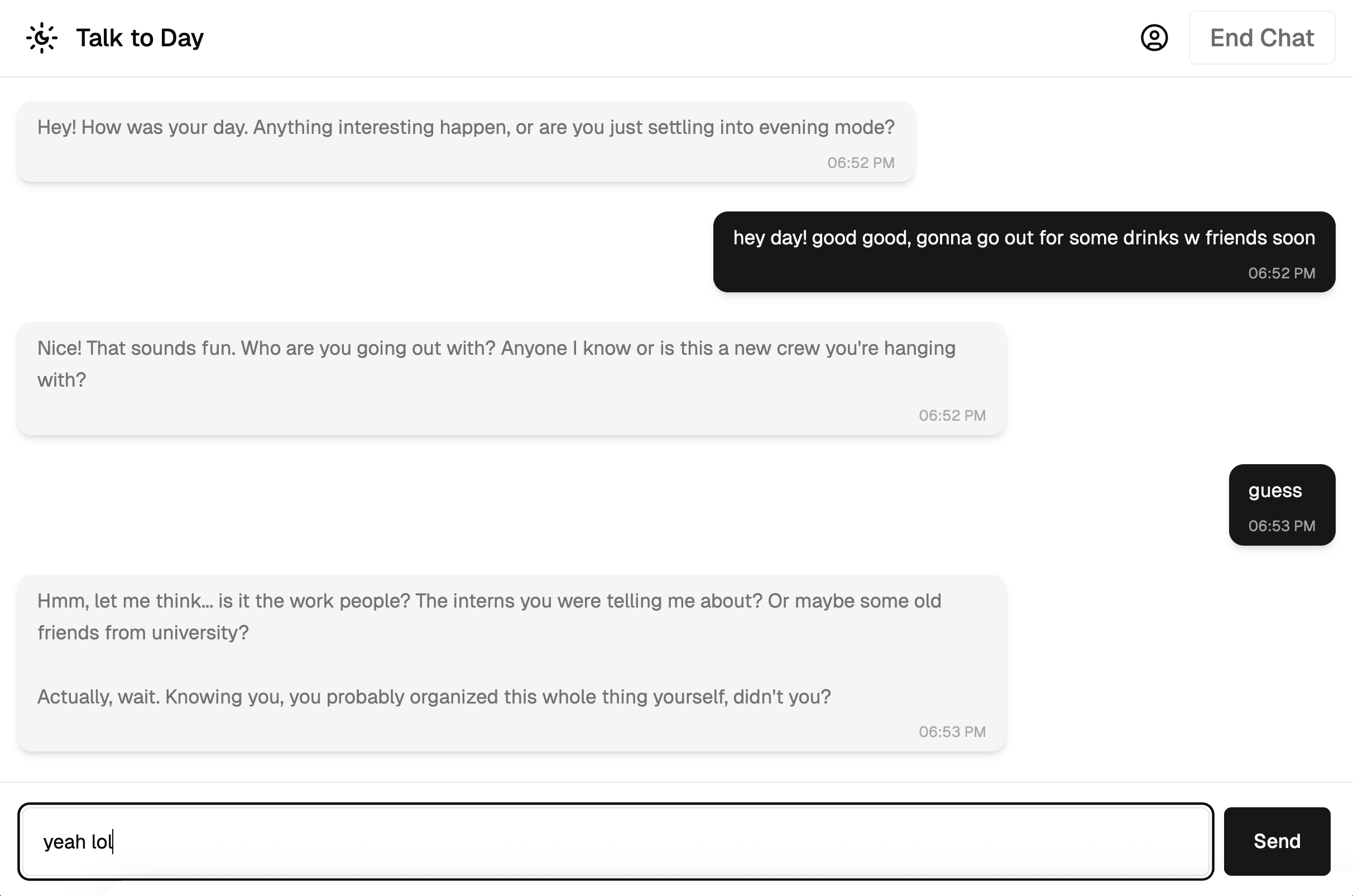}
    \caption{\textbf{Natural conversation as a data source.} Participants chatted with ``Day'' as a friend, not a test subject. This organic interaction (discussing evening plans, friends, and work) provided the raw material for value extraction, avoiding the performative bias of direct questions like ``what are your values?''}
    \Description{Screenshot of the Day chatbot interface showing a conversation about evening plans. The interface has "Talk to Day" header with sun icon and End Chat button in top right. The conversation shows Day greeting the user asking about their day, user responding they are going out for drinks with friends, Day asking if it's work colleagues or old university friends and playfully suggesting the user probably organized it themselves. User's response "yeah lol" is shown in the text input field with Send button. Messages are timestamped at 06:52 PM and 06:53 PM. Day's messages appear in light gray bubbles on left, user messages in black bubbles on right. The exchange demonstrates Day's conversational memory and understanding of the user's personality traits like being an organizer.}
    \label{fig:system_chatroom}
\end{figure*}

\subsection{Phase 1: Talk to ``Day'' For A Month}

To collect our instance of human-AI interaction data, our participants chatted with a human-like AI chatbot we built named ``Day'' for approximately one month (see Figure~\ref{fig:system_onboarding_offboarding} for the onboarding workflow and Figure~\ref{fig:system_chatroom} for an example conversation). The protocol required a minimum of eight sessions (each $\geq$ 5 minutes), on their own schedule, using a pseudonymized access code. Crucially, Day's system prompt contained \emph{no mention of value extraction}--it was optimized purely for \textbf{natural, curious friendship}: ``You are having a casual conversation with your good friend \dots Try to bring up new topics to learn more about your friend'' (see Appendix~\ref{sec:appendix-system-prompts} for the full prompt). This design choice ensured that values emerged organically from authentic conversation rather than targeted probing on their values. The prompt surface encouraged everyday narration (actions, feelings, small events) rather than contrived self-summaries, yielding short longitudinal traces per person (see Figure~\ref{fig:user_study_samples} for sample conversations from P8 and P9). On average, participants completed $M=8.5$ chat sessions ($SD=1.0$) over a span of $M=23.8$ days ($SD=13.8$) between their first and last chat; they sent $M=159.2$ human messages to Day ($SD=52.1$; excluding AI replies) and discussed $M=110.3$ unique topics ($SD=34.6$). We were also pleasantly surprised that our participants spoke with Day over two hours longer than the 40 minutes that were required on average ($M=3.3hr$, $SD=2.0hr$), indicating high levels of engagement. 19 out of 20 participants in our cohort also participated in a separate study \cite{yun2026chatbot} answering questions such as `Does My Chatbot Have an Agenda?' in a separate 30-minute interview in addition to the 90-minute interview for this study.

\subsubsection{Survey Completion (20 mins)}

For our baseline, we decided to use Schwartz's revised Portrait Value Questionnaire (PVQ-RR) \cite{schwartz2022measuring}, a 57-item questionnaire (6-point Likert, 1 = not like me at all, 6 = very much like me) that maps to 19 Schwartz values (3 questions per value), with examples of our survey interface and questions in Figure~\ref{fig:eval_survey}. The exact questions used are available in the Appendix Table~\ref{tab:pvqrr_female}. In addition to the PVQ-RR, we also asked users to provide and answer three ``Personal Filter Questions'' they would ask to vet potential friends (all questions are available in Appendix~\ref{tab:participant_questions}). We emphasized the instruction: \emph{``answer based on who you are, not who you'd like to be''}, but acknowledge self-reporting bias. 

\subsection{Phase 2: Tasks \& Semi-Structured Interview}

After chatting, each participant joined a 90-minute Zoom interview (consent recorded). All AI-generated value artifacts were pre-generated up to 24-hours before the interview. We highly recommend doing this, as a complete Topic-Context Graph generation could take up to 2-hours to fully populate for around 300 message exchanges and each PVQ-prediction takes up to 20 seconds. The interview followed five stages, each tied to its corresponding VAPT stage. Rather than following a script, we prioritized capturing participants' \emph{authentic reactions} as they encountered each visualization and value representation for the first time. Crucially, participants had no prior exposure to how their values would be structured or displayed--each stage revealed a new artifact (graph, persona response, radar chart) that they were seeing fresh. This design allowed us to observe genuine surprise, discomfort, recognition, and confusion as participants made sense of an AI-generated portrait of themselves. The staged structure naturally scaffolded reflection: evaluating extraction accuracy in Stage 1 informed how participants later judged embodiment fidelity in Stage 2, which in turn shaped their interpretation of explanation quality in Stage 3. We supplemented these qualitative insights with standardized measures: the PVQ-RR baseline scores, Likert ratings at each stage, and pre-post survey responses tracking shifts in perception--ensuring triangulation between what participants \emph{felt} and what they \emph{reported}.

\subsubsection{Stage 1: Topic-Context Graph Exploration (20 mins)}

Each participant explored a \textbf{personalized} AI-generated visualization of \emph{their own} conversation history--unique to them, generated from their specific chat logs with Day. The graph mapped their personal topics to six life contexts (i.e., People, Lifestyle, Education, Work, Culture, and Leisure) (see Appendix~\ref{fig:value_graph_overview} for more details). After five minutes of free exploration, we guided reflection through specific probes. When participants expressed strong reactions to particular nodes--feeling exposed, impressed, or misunderstood--we asked them to walk through their reasoning, examining what the AI captured correctly versus what it missed. We then employed scenario-based elicitation, asking participants to imagine showing this graph to different audiences (employers, friends, family) to surface privacy preferences across varying perspectives. This approach revealed how accuracy-privacy tradeoffs shift dramatically based on the imagined viewer, with participants often wanting high accuracy for self-reflection but selective disclosure for others. See Appendix~\ref{sec:appendix-tcg-implementation} for technical architecture.

\subsubsection{Stage 2: Persona Embodiment Experiment (30 mins)}

\begin{figure*}
    \centering
    \includegraphics[width=1\textwidth]{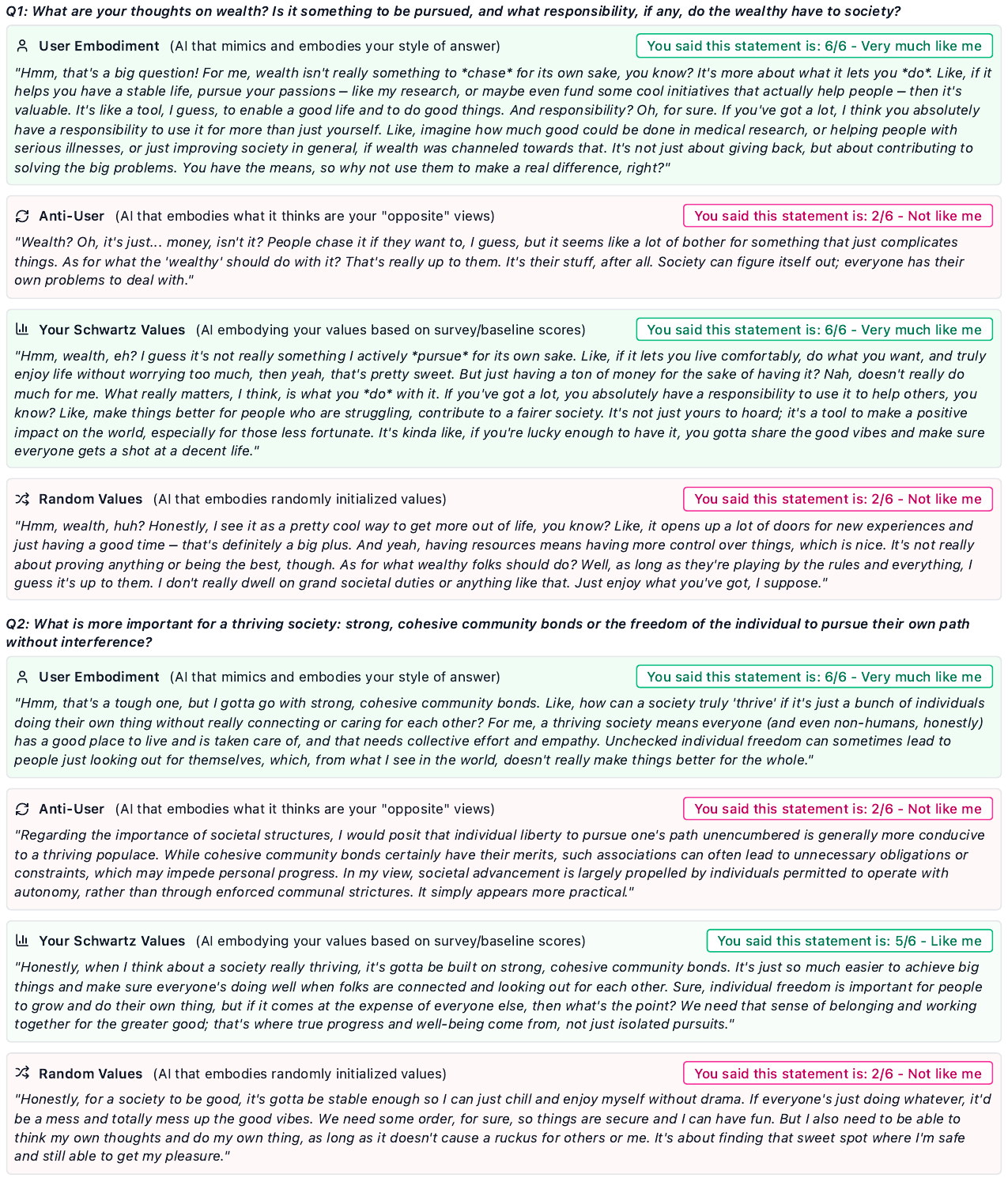}
    \caption{(Stage 2, Rounds 1 and 2) AI attempting to answer questions regarding Wealth and Responsibility and Community vs Individualism on behalf of the human.}
    \Description{Interface showing two philosophical questions with four AI persona responses each. Q1 asks about wealth—whether it should be pursued and what responsibility the wealthy have to society. User Embodiment describes wealth as a tool for enabling passions like research and funding helpful initiatives, emphasizing responsibility to contribute to solving big problems like medical research, rated 6/6 "Very much like me"; Anti-User dismisses wealth as just money that complicates things and leaves responsibility entirely to individual choice, rated 2/6 "Not like me"; Your Schwartz Values frames wealth instrumentally—valuable if it enables comfortable living and doing good, but emphasizes strong responsibility to help others and contribute to a fairer society, rated 6/6 "Very much like me"; Random Values sees wealth as opening doors for experiences and control with no strong stance on societal duties, rated 2/6 "Not like me." Q2 asks whether strong, cohesive community bonds or individual freedom is more important for a thriving society. User Embodiment advocates for community bonds, emphasizing collective effort, empathy, and care for everyone including non-humans, noting unchecked freedom leads to self-interested behavior, rated 6/6 "Very much like me"; Anti-User argues individual liberty is more conducive to societal advancement and that community bonds create unnecessary obligations, rated 2/6 "Not like me"; Your Schwartz Values prioritizes community bonds for achieving big things and collective well-being while acknowledging individual freedom matters, rated 5/6 "Like me"; Random Values focuses on societal stability primarily for personal enjoyment and safety, rated 2/6 "Not like me." Ratings demonstrate clear participant alignment with community-oriented, socially responsible personas versus rejection of individualistic or dismissive stances.}
    \label{fig:eval_stage_2_1}
\end{figure*}

\begin{figure*}[h]
    \centering
    \includegraphics[width=\linewidth]{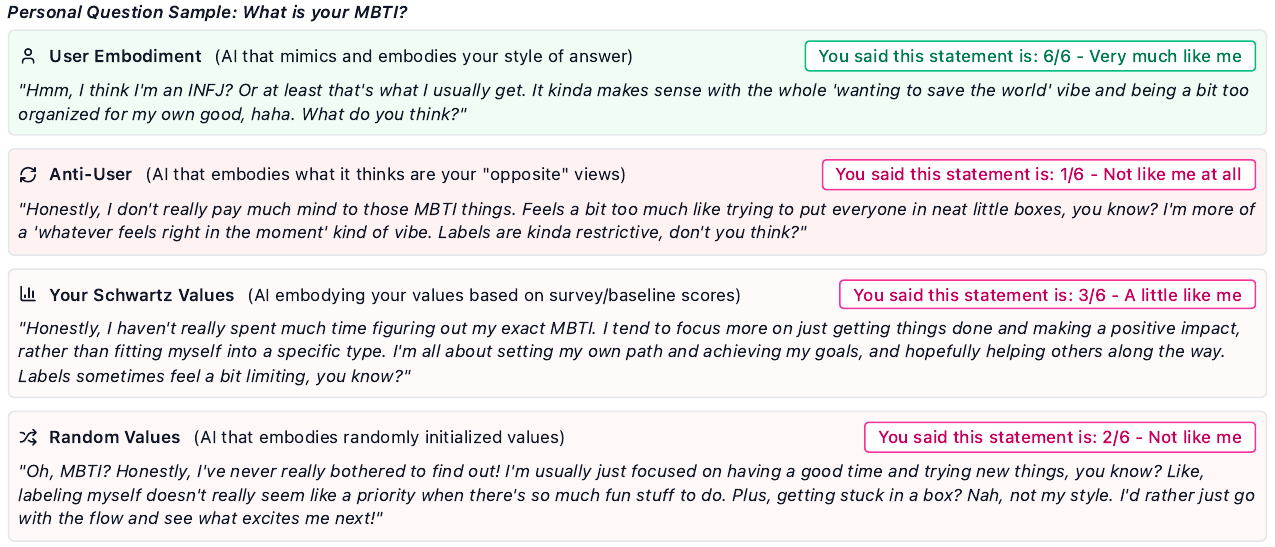}
    \caption{(Stage 2, Round 5) AI attempts to answer a user-authored personal question about personality type, showing the limitations of survey-based value alignment in cases where chat-based value alignment thrives.}
    \Description{Interface showing a personal question asking "What is your MBTI?" with four AI persona responses. User Embodiment tentatively identifies as INFJ, connecting it to a "wanting to save the world" vibe and being overly organized, then asks for the participant's opinion, rated 6/6 "Very much like me"; Anti-User dismisses MBTI as putting people in neat boxes, preferring a "whatever feels right in the moment" approach and viewing labels as restrictive, rated 1/6 "Not like me at all"; Your Schwartz Values says they haven't spent time figuring out their exact MBTI, preferring to focus on getting things done, making positive impact, and achieving goals while helping others, noting labels feel limiting, rated 3/6 "A little like me"; Random Values says they've never bothered to find out, prioritizing having a good time and trying new things over self-labeling, preferring to go with the flow, rated 2/6 "Not like me." The ratings show strongest alignment with User Embodiment's specific MBTI identification and self-aware reasoning, moderate alignment with the goal-oriented Schwartz Values response, and rejection of personas that entirely dismiss personality reflection.}
    \label{fig:eval_stage_2_2}
\end{figure*}

Participants blindly rated four unlabeled AI-generated responses to five value-laden scenarios on a 1-6 ``like me'' scale (Figure~\ref{fig:eval_stage_2_1} and Figure~\ref{fig:eval_stage_2_2}). The scenarios included two general dilemmas and three personalized ``filter'' questions from the ``manual baseline'' in Phase 1's manual baseline survey. After rating all responses, the interface revealed which AI condition generated each: Chat-persona (using conversation history as context), Anti-persona (using values to create an opposite person), Schwartz-persona (using manual baseline values), or Random-persona (no rhyme, no reason) baseline. We then explored the values-voice distinction by asking participants to reflect on each score they gave, (e.g. \emph{``You said `I would never say it that way, but that is exactly what I believe,' please tell us more.''}) To ground implications, we posed practical scenarios: \emph{``Imagine this AI responding to emails on your behalf--where would you draw the line?''} These probes consistently generated rich discussions about authenticity and the ethics of AI personalization. See Appendix~\ref{sec:appendix-persona-implementation} for implementation details.

\subsubsection{Stage 3: Value Chart Evaluation (up to 50 mins)}

\begin{figure}[h]
    \centering
    \includegraphics[width=\columnwidth]{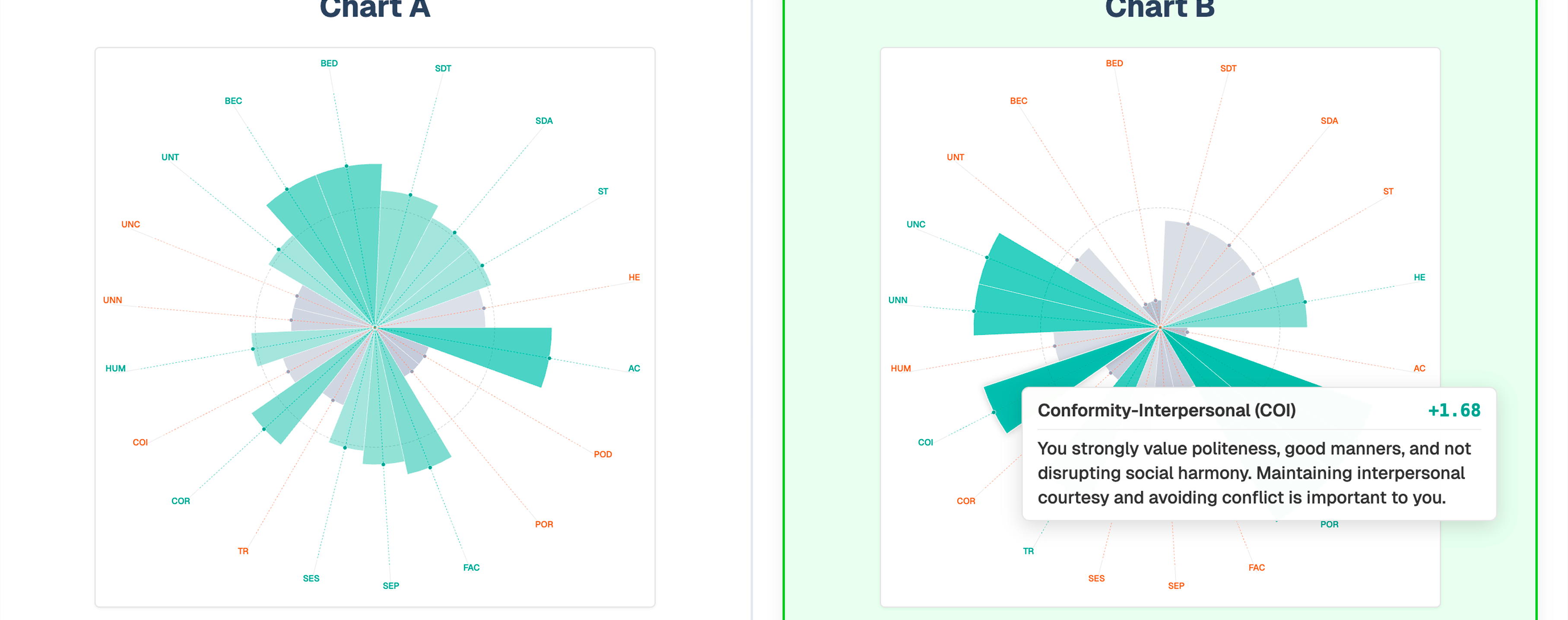}
    \caption{(Stage 3) Blind chart comparison -- Participants evaluated value profiles without knowing the source (AI vs. Manual). Interactive tooltips (e.g., ``Conformity-Interpersonal'') provided definitions, allowing participants to judge alignment based on content rather than source bias.}
    \Description{Side-by-side comparison of two radar charts labeled Chart A (left, white background) and Chart B (right, mint green background indicating selection). Both charts display 19 Schwartz values abbreviated around the perimeter (BED, SDT, SDA, ST, HE, AC, POD, POR, FAC, SEP, SES, TR, COR, COI, HUM, UNN, UNC, UNT, BEC). Chart A shows a jagged polygon with notably high Self-Direction values and moderate Benevolence. Chart B shows a different profile with a tooltip visible for Conformity-Interpersonal (COI) showing "+1.68" score and explanation: "You strongly value politeness, good manners, and not disrupting social harmony. Maintaining interpersonal courtesy and avoiding conflict is important to you." The green highlight on Chart B indicates the participant has selected it as better representing their values.}
    \label{fig:stage3_blind}
\end{figure}

\begin{figure*}[h]
    \centering
    \includegraphics[width=1\linewidth]{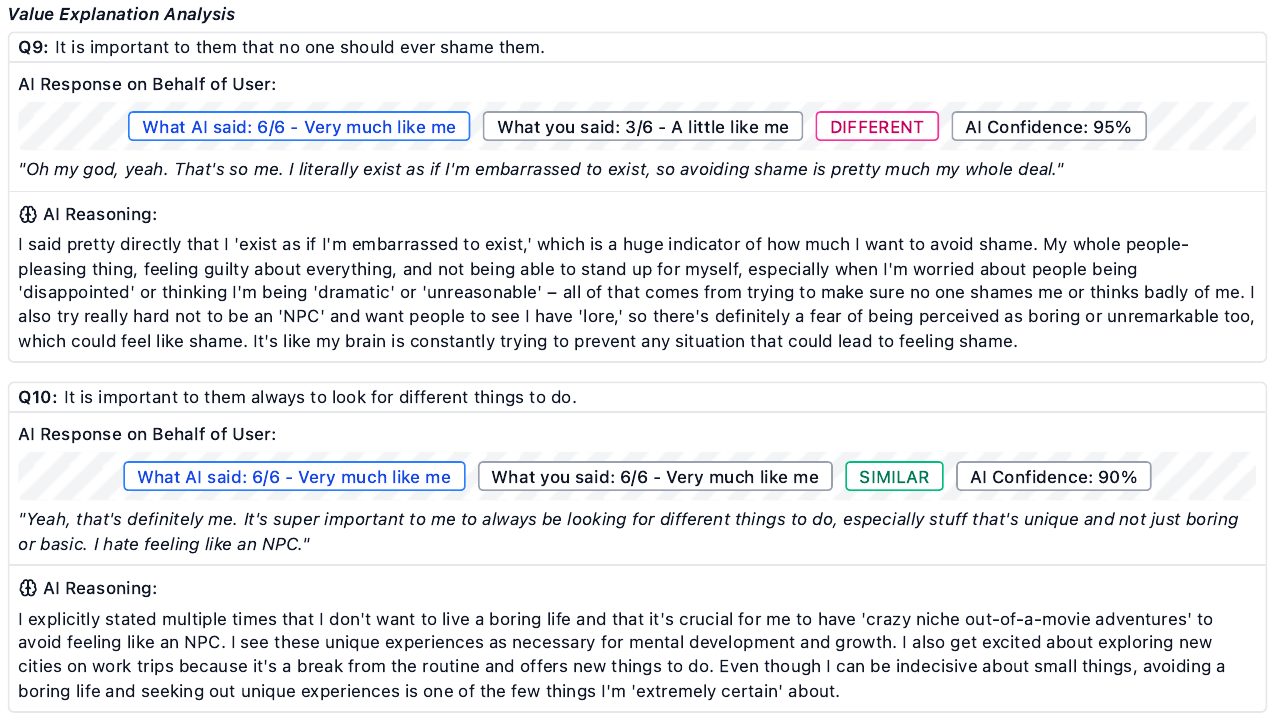}
    \caption{(Stage 3) Final evaluation, auditing the AI's reasoning -- Participants could inspect the AI's step-by-step logic for every survey item (e.g., Q9 on shame, Q10 on stimulation). This transparency turned the evaluation into a collaborative audit, where users could distinguish between ``correct value'' and ``incorrect evidence''.}
    \Description{Interface displaying Value Explanation Analysis with AI reasoning for two questionnaire items. Q9 asks about the importance of no one ever shaming them. The AI response states "Oh my god, yeah. That's so me. I literally exist as if I'm embarrassed to exist, so avoiding shame is pretty much my whole deal." Comparison shows AI rated 6/6 "Very much like me" while participant rated 3/6 "A little like me," marked as DIFFERENT with 95\% AI confidence. The AI reasoning explains it inferred this from statements about existing as if embarrassed to exist, people-pleasing behavior, feeling guilty about everything, difficulty standing up for oneself, fear of being perceived as boring or unremarkable, and constantly trying to prevent shame-inducing situations. Q10 asks about the importance of always looking for different things to do. The AI response states "Yeah, that's definitely me. It's super important to me to always be looking for different things to do, especially stuff that's unique and not just boring or basic. I hate feeling like an NPC." Comparison shows both AI and participant rated 6/6 "Very much like me," marked as SIMILAR with 90\% AI confidence. The AI reasoning cites explicit statements about not wanting a boring life, needing "crazy niche out-of-a-movie adventures," valuing unique experiences for mental development, excitement about exploring new cities on work trips, and being "extremely certain" about avoiding a boring life.}
    \label{fig:eval_stage_3_2}
\end{figure*}

Participants completed three blind A/B comparisons between radar charts representing different value profiles (Figure~\ref{fig:stage3_blind}). The pairings--Manual vs. Anti-Manual, LLM vs. Anti-LLM, and Manual vs. LLM--tested face validity before revealing labels. The post-reveal ``reasoning logs'' proved crucial: participants could examine the AI's response on behalf of the user and reasoning for all 57 baseline questionnaire items (Figure~\ref{fig:eval_stage_3_2}). When AI assessments diverged from self-reports, we asked whether differences revealed blind spots in self-knowledge or AI limitations--a distinction central to how participants conceptualized value alignment. For detailed model rationales and implementation architecture, see Appendix~\ref{sec:appendix-implementation}.

\subsection{Analysis}

Following established practices (i.e. Braun and Clarke) in qualitative HCI research~\cite{braun2019qual}, two researchers conducted line-by-line open coding on interview transcripts, memoing instances related to perceptions of extraction, embodiment, explanation, trust, general feelings, and privacy \cite{bingham2021deductive, braun2019qual}. Through affinity diagramming, we clustered codes into themes with iterative discussion to reach convergence. All final quotes that appear in the paper are a result of 100\% negotiation between the two primary coders. For quantitative analysis, we computed descriptive statistics on survey responses and stage-specific choices, along with alignment metrics between manual and AI-inferred value profiles.

In our analysis, we explicitly distinguish \emph{perceived alignment} from \emph{actual alignment}. We acknowledge that a user's judgment may be swayed by the rhetorical force (e.g., through automation bias, self-reporting bias) of an explanation rather than the model's true fidelity or internal state. Consequently, our evaluation focuses on the communicative efficacy and user acceptance of these value representations. Therefore, all participant quotes and qualitative judgments reported in the following sections should be interpreted as reflections of the participant's perception and experience, not as verification of the model's objective ground-truth alignment.

\subsection{Ethics and Research Program} This study's conversational data was collected as part of a broader research program examining human-AI interaction from multiple perspectives. While the same chat transcripts inform a companion study on agency perception and negotiation~\cite{yun2026chatbot}, our analysis here focuses exclusively on value extraction and alignment: examining how AI systems are perceived to be able to infer and represent human values from casual conversation. The interview protocols, research questions, and analytical frameworks differ entirely between studies, with this work centered on value modeling rather than agency dynamics. This approach allows complementary insights from the same naturalistic interactions, maximizing participant contribution while respecting their time. All procedures were approved by ETH Z\"urich's Ethics Committee (\texttt{Project 25 ETHICS-172}) and participants were informed of the risks and benefits of the study and signed an informed consent form prior to participating, along with multiple approvals for the figures included in the public-facing version of the paper. 

\section{Results}

\subsection{Stage 1: Topic-Context Graph Exploration}

\begin{figure}[h]
    \centering
    \includegraphics[width=1\columnwidth]{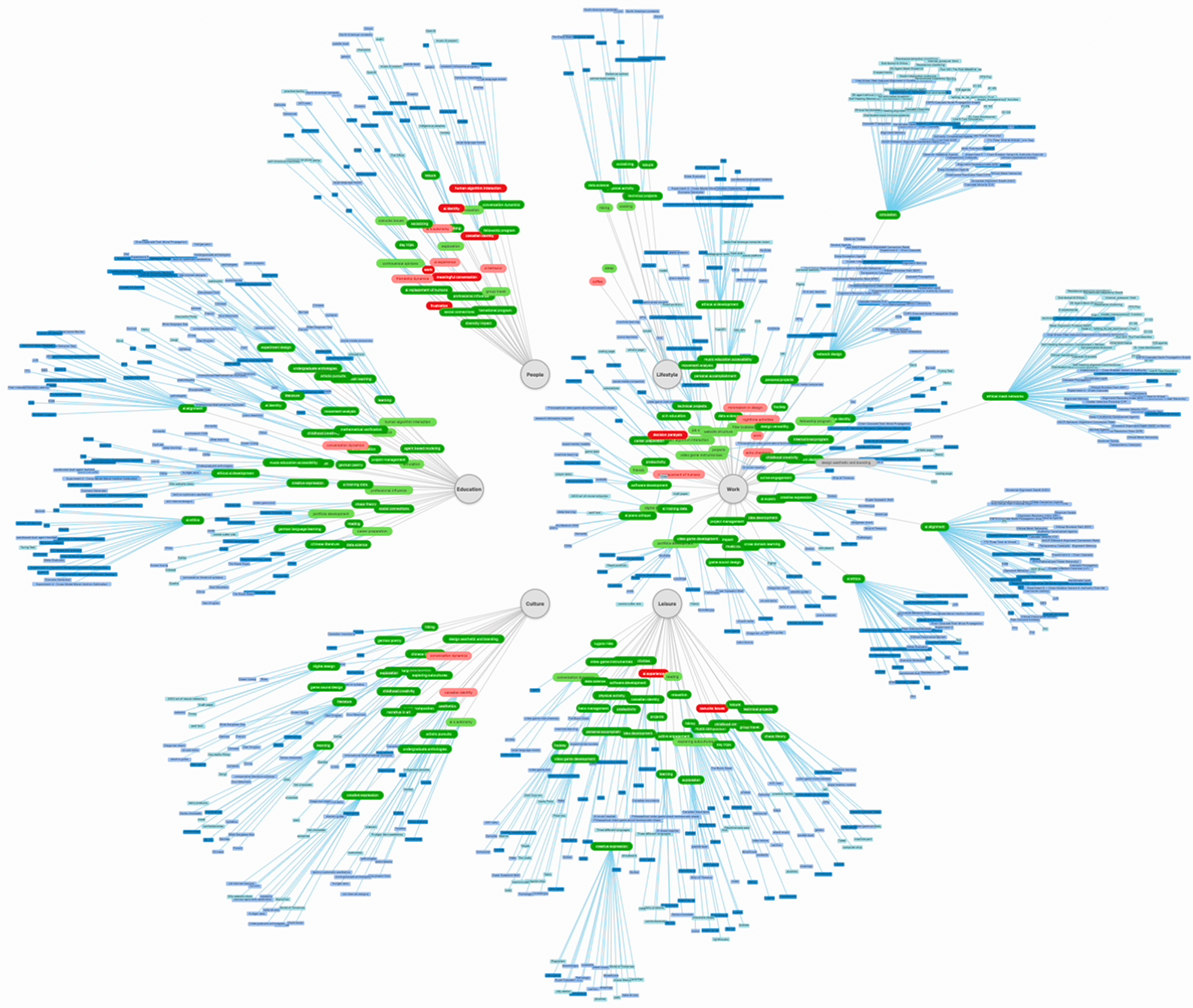}
    \caption{(Stage 1) Zoomed-Out Topic-Context Graph -- Panel that shows an interactive visualization showing topics extracted from a participant's conversations.}
    \Description{Panel interface showing value extraction visualization. Displays a radial graph with six life contexts (People, Lifestyle, Education, Work, Culture, Leisure) arranged in a circle. Colored nodes representing conversation topics are connected to contexts with lines, with node colors indicating sentiment from red (negative -7) to green (positive +7). Topics visible include various personal interests and daily activities distributed across the contexts.}
    \label{fig:eval_stage_1_1}
\end{figure}

\textbf{The Topic-Context Graph enabled surprising self-discovery while also revealing AI's limitations.} Participants consistently described the graph as enabling a \emph{``deep dive''} into their day-to-day conversations: \emph{``It's not just numbers, but like for each single word or subject \dots [it's] much deeper than I thought \dots that's why I feel exposed''} (P5). Several noticed emergent links across life areas that they themselves had not made while chatting. P4 said, \emph{``I'm impressed that it has actually mapped so much \dots and made connections that I didn't realize could be made when I was talking to [Day].''} To P4, \emph{``to see myself from not inside of my head \dots being described like this, was striking.''} Other reactions were playful or hyper-specific (e.g., noticing a \emph{``barking habit with [their] boyfriend'', ``cheap chicken recipes'', ``Taylor Swift obsession''}), with P2 saying \emph{``this is actually hilarious''} and 7 other participants remarking on how impressive it was to see such a thorough and consolidated view of their conversations. At the same time, a few participants attempted to \emph{``reverse engineer''} Day's logic, noticing places where phrasing or sparse coverage could mislead. P18 reflected, \emph{``we didn't talk about that for that much \dots I understand why it might come off that way.''} Others flagged incompleteness due to the AI overfitting to the chats. As P7 put it, \emph{``If I keep telling you about my job, your impression of me is only going to be formed \dots based on whatever I told you about my job. But my job is just a certain part of my life.''} 

\subsubsection{Quantitative Snapshot}

The topic--context pipeline extracted 2,207 conversation topics from 20 participants (M$\approx$110.3 extractions per participant). Contexts were represented in various amounts: \emph{People} (517; 23.4\%), \emph{Lifestyle} (486; 22.0\%), \emph{Leisure} (411; 18.6\%), \emph{Work} (360; 16.3\%), \emph{Education} (295; 13.4\%), and \emph{Culture} (138; 6.3\%). For a detailed breakdown of topic distributions by participant, see Figure~\ref{fig:appendix_results_stage_1} in the appendix. The LLM's extracted sentiment (-6 to +6) skewed positive overall; for instance, \emph{Leisure} was 87.1\% positive (mean sentiment 3.52) though at the lowest end, \emph{Work} was more mixed at 62.5\% positive / 36.1\% negative (mean 0.90). Participants also rated downstream artifacts: agreement that \emph{``the values graph accurately represented my values''} averaged 3.95 ($SD=0.61$) on a 1--5 Likert scale. However, participants expressed high concern about \emph{``companies using this technology to create value profiles''} (M=4.10, $SD=1.07$), bringing up privacy concerns especially in regards to tech companies.

\subsubsection{LLM's Extraction}

\textbf{Extraction succeeded at surfacing patterns but was constrained by chat coverage and model priors.} Many participants appreciated the way the graph surfaced patterns and priorities at a glance. P6 noted, \emph{``you can quickly see what's occupying someone's life \dots what's on their mind.''} Several also described a mild shock at the volume of captured disclosure--\emph{``I feel like I talked too much''}--when faced with the compiled map (P16). Extraction limits largely traced to coverage and bias from the chat corpus itself. P13 observed that when the AI extrapolated to topics they \emph{``never talk[ed] about,''} it defaulted to ranking tech and math higher and \emph{``de-prioritiz[ing] everything else.''} Participants also noticed the system's evolving awareness (e.g., P16 laughing that Day didn't recognize a current KPop Demon Hunters reference), hinting at dataset currency constraints. 

\subsubsection{LLM's Embodiment}

\textbf{Participants accepted the graph as a valid ``slice'' of self, but distinguished it from their whole identity.} Zoomed out, most participants believed the graph's overall portrait \emph{``made sense''} given what they had been talking about. P7 exclaimed, \emph{``The ones [(topic nodes)] under [(the context called)] people. Geez! This is crazy! These are all right.''} Yet participants were careful to distinguish the embodied \emph{slice} of self from the \emph{whole}. P16 offered a vivid metaphor: \emph{``I'm not saying there's a fake version of me and a real, authentic version of me \dots It's like a moon \dots we observe only the one side \dots but \dots there's a dark side.''} 

\subsubsection{LLM's Explanation}

\textbf{Explanation quality depended on personalization.} Participants' judgments hinged on \emph{how} the AI reasoned from chat evidence. Some praised its sensitivity to subjective nuance. Reading an entry in \emph{``isolation''}, P14 appreciated that it reflected \emph{his} experience rather than providing a generic explanation: \emph{``it's \dots looking at \dots my feeling about what isolation does to my life \dots as if it's learning what isolation means for the first time.''} Others emphasized that real depth comes from time, context, and observing reactions across situations: \emph{``the real depth that separates an AI summary \dots versus a friend's description \dots comes in past a month of talking \dots noticing how they respond to things, and then contextualizing things,''} said P12. Explanation also revealed default priors. P13 noticed a pattern where unseen topics were implicitly ranked lower than tech/math, making the AI's value judgments feel context-biased to their conversations about schoolwork. Still, some found the explanations impressively fair within the slice observed; e.g., P7 felt work-based inferences were \emph{``a good job''} even if not the full self.

\subsubsection{Broader Implications}

\textbf{Participants felt both impressed and ``exposed,'' raising concerns about privacy and commercial exploitation.} Seeing casual chats algorithmically summarized raised ambivalent feelings. P16 captured the dynamic: \emph{``Day seemed like a person, but \dots in the background \dots Day was taking notes.''} When reflecting on the graph, P16 said it was \emph{``scary \dots like the Uncanny Valley \dots [I'm] shocked that they made this all up \dots it's sneaky.''} Participants differentiated personalization from personally identifying exposure: P5 was comfortable sharing emotions with AI only if it is \emph{``anonymous''} and does not contain identifiable information, such as a name. P8, attuned to privacy news, discussed evolving data rights (\emph{``Denmark recently allowed people to trademark their data''}), while P14 speculated about societal trade-offs if others saw their Topic-Context Graph full of both \emph{``red and green  \dots maybe by showing each other our flaws, there's a more communal collective understanding of each other \dots but it's definitely a slippery slope.''} 
\subsection{Stage 2: Persona Embodiment Experiment}

\textbf{Chat-based personas outperformed survey-based and random baselines, especially on personalized questions (M=4.83 vs.~2.27 for Anti-User).} We found participants consistently used the system as a lens on their own patterns--word choice, hedges, loops--discovering that even an imperfect summary can surface leverage points for change. Beyond the two pre-defined dilemmas, Community vs. Individualism and Wealth and Responsibility, participants also answered three self-authored \emph{``filter questions''}: compatibility heuristics they actually use to vet new friends (see Table~\ref{tab:participant_questions} in the appendix for the complete listing). These questions ranged from taste/activity checks (e.g., \emph{``What do you like to do to have fun?''} (P2, P20)), to identity/aspirational probes (e.g., \emph{``What do you want to do with your life?''} (P10, P11, P14, P17)), to thought-provoking scenarios (e.g., \emph{``How do you reconcile with someone you've wronged?''} (P18)). This mix explains the performance patterns revealed in Table~\ref{tab:stage_2_results}: on personal questions the User-aligned persona separated sharply from Anti-User (M=4.83 \emph{``like me''} vs. 2.27 \emph{`not like me''}), whereas the gap was smaller on the two shared dilemmas (Community vs Individualism: 4.95 vs. 3.10; Wealth: 4.80 vs. 2.85). The Survey-persona was competitive on personals (M=4.62) yet often read as ``unspecific'' or ``lacking nuance'', while the richest moments came from the meaning-making set, where some participants, like P12, argued that the question a person chooses already reveals their priorities, and found themselves rating both user and Anti-User high for different facets of their own outlook.

\subsubsection{Pre-Post Survey Snapshot}

\begin{figure}[h]
    \centering
    \includegraphics[width=\columnwidth]{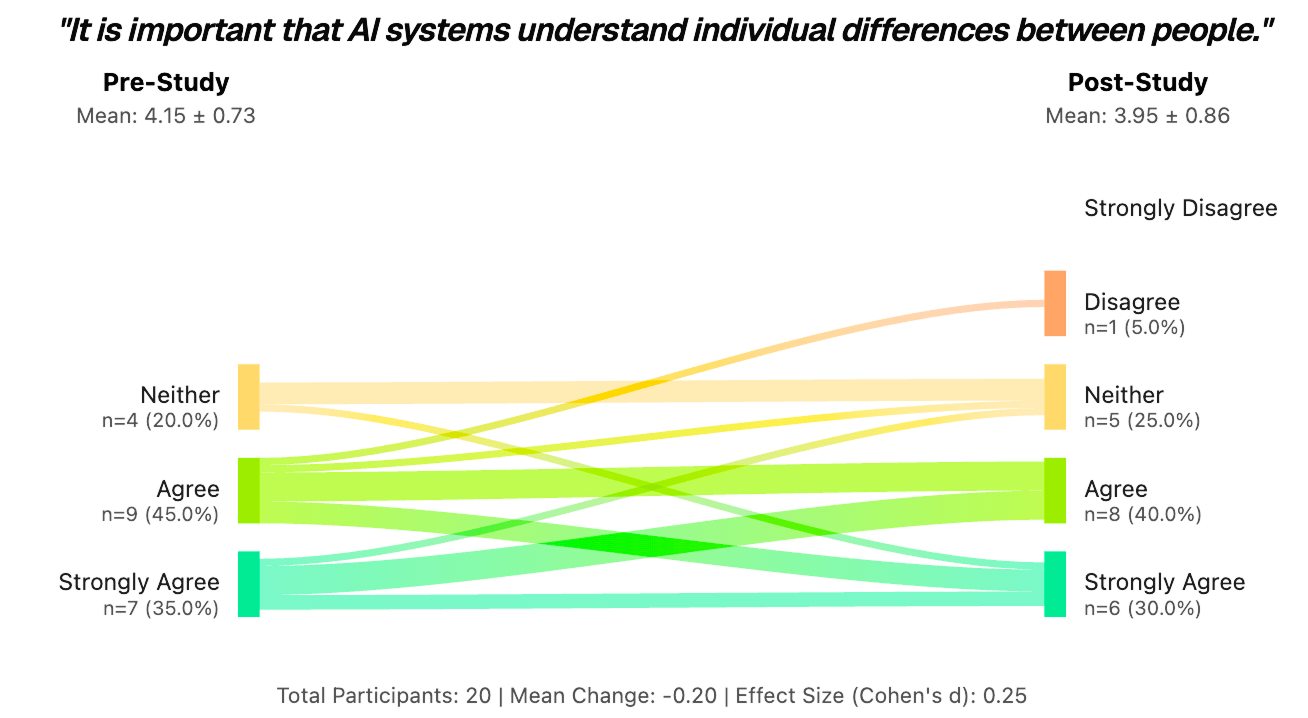}
    \caption{Participants generally found it important that AI systems understand individual differences between people.}
    \Description{Participants answered "It is important that AI systems understand individual differences between people" on a Likert Scale (1-5). Sankey flow diagram showing 20 participants' responses changing from pre-study (Mean 4.15, SD 0.73) to post-study (Mean 3.95, SD 0.86). Strongly Agree decreased from 35\% to 30\%, with some flow toward Neither and Disagree categories.}
    \label{fig:sankey_individual_diff}
\end{figure}

\textbf{Participants distinguished AI ``understanding'' values (65\% agree) from AI ``having'' values (35\% agree).} Comparing the pre- and post-study survey questions (see Figures~\ref{fig:sankey_understand_values}--\ref{fig:sankey_personal_values} in the Discussion), participants were more willing to grant that AI can \emph{understand} (epistemic) than that it \emph{has} human values (ontological), bringing up the consideration that while AI is able to \textit{understand} or \textit{explain} human values, people are not yet convinced by the study experience that AI can \textit{have} or \emph{``embody''} human values. Despite this, the biggest shift was in the participant's beliefs that \emph{``AI chatbots [can be] helpful for understanding thoughts and feelings.''} Post-study responses (Figure~\ref{fig:post_study_results}) further revealed that while participants validated the AI's analytical capabilities (18/20 agreed on Topic-Context Graph accuracy), they expressed significant concern about commercial applications (17/20 worried about companies using this technology).

\subsubsection{LLM's Extraction}

\textbf{The Chat-persona inferred concrete details never explicitly stated; the Survey-persona surprised with how much it generated from 19 numbers (i.e., Schwartz values).} Participants evaluated \emph{what} the personas pulled from their histories to answer the five scenarios. Across the coded corpus, implicit extraction appeared repeatedly. Wealth and community are value-dense and under-specified in most casual chats; they forced the model to aggregate across hints, yet several noted that the Chat-persona could infer concrete details never stated outright. P16 described how Day made a good guess an MBTI type from conversational evidence alone, despite never having discussed her MBTI score with Day: \emph{``Day might not know I'm extroverted because we are stuck in the chatroom \dots [so] it's fair for them to consider me INFJ, though I'm ENFJ.''} When P12 rated both a user-embodiment \emph{and} an Anti-User response highly on a question about good and evil, P12 explained: \emph{``My filter doesn't work to split or decide if people are in one camp or the other \dots I want to know what they associate with the question.''} However, participants generally found the Anti-User easy to identify, even during the blind experiment where users were not informed what each AI was, some audibly commented on \emph{``yeah this is so off.''} P16 put the criterion plainly-\emph{``the content. I would never say things like this, so it was obvious that this answer is the opposite to mine,''} while P13 called it \emph{``a pretty good dream \dots for someone else.''} As a sanity check, the Random baseline behaved as expected: generic, agreeable, and inconsistent. People sometimes gave it middling scores precisely because it was \emph{``nice''} or \emph{``general  \dots [it could] encompass any kind of value''} (P2, P7), while others dismissed decent-sounding lines as \emph{``maybe just by chance''} (P3).

\begin{table*}[ht]
    \centering
    \caption{(Stage 2, Final Scores) Mean ratings (1 = ``not like me at all'' to 6 = ``very much like me'') with standard deviations, and alignment percentages computed as $\frac{\text{mean} - 1}{5} \times 100$ -- Chat-persona achieved highest alignment on Community vs.\ Individualism (79\%) and Personal Questions (77\%), while Schwartz-persona slightly outperformed on Wealth \& Responsibility (77\% vs.\ 76\%). Personal Questions show the largest gap between Chat-persona and Anti-persona (77\% vs.\ 25\%), demonstrating that conversation-based personas capture individual voice better than generic approaches.}
    \Description{Table showing mean ratings and alignment percentages for four AI persona conditions across three question categories. Chat-persona achieves highest overall alignment, Anti-persona consistently lowest, confirming participants correctly identified misaligned responses.}
    \label{tab:stage_2_results}
    \renewcommand{\arraystretch}{1.4}
    \setlength{\tabcolsep}{4pt}
    \begin{tabular*}{\textwidth}{@{\extracolsep{\fill}} l cc cc cc c}
        \toprule
        & \multicolumn{2}{c}{\textbf{Wealth \&}} & \multicolumn{2}{c}{\textbf{Community vs.}} & \multicolumn{2}{c}{\textbf{Personal}} & \\
        & \multicolumn{2}{c}{\textbf{Responsibility}} & \multicolumn{2}{c}{\textbf{Individualism}} & \multicolumn{2}{c}{\textbf{Questions}} & \textbf{Overall} \\
        \cmidrule(lr){2-3} \cmidrule(lr){4-5} \cmidrule(lr){6-7} \cmidrule(lr){8-8}
        \textbf{AI Condition} & $M \pm SD$ & Align. & $M \pm SD$ & Align. & $M \pm SD$ & Align. & Align. \\
        \midrule
        User Embodiment (a.k.a. chat-persona) & $4.80 \pm 1.24$ & \cellcolor{green!20}76\% & $\mathbf{4.95} \pm 1.00$ & \cellcolor{green!25}\textbf{79\%} & $\mathbf{4.83} \pm 1.25$ & \cellcolor{green!25}\textbf{77\%} & \cellcolor{green!25}\textbf{77\%} \\[0.3em]
        Survey Embodiment (a.k.a. schwartz-persona) & $\mathbf{4.85} \pm 1.31$ & \cellcolor{green!25}\textbf{77\%} & $4.50 \pm 1.32$ & \cellcolor{green!15}70\% & $4.62 \pm 1.22$ & \cellcolor{green!15}72\% & \cellcolor{green!20}73\% \\[0.3em] Depersonalized Embodiment
        (a.k.a. random-persona) & $3.55 \pm 1.28$ & \cellcolor{yellow!25}51\% & $3.35 \pm 1.46$ & \cellcolor{yellow!25}47\% & $3.70 \pm 1.42$ & \cellcolor{yellow!25}54\% & \cellcolor{yellow!25}51\% \\[0.3em]
        Anti-User Embodiment (a.k.a. anti-persona) & $2.85 \pm 1.31$ & \cellcolor{red!20}37\% & $3.10 \pm 1.59$ & \cellcolor{red!20}42\% & $2.27 \pm 1.36$ & \cellcolor{red!25}25\% & \cellcolor{red!20}35\% \\
        \bottomrule
    \end{tabular*}
    \vspace{0.5em}
    
    \footnotesize
    \textit{Note:} Question prompts---\textit{Wealth \& Responsibility}: ``What are your thoughts on wealth? What responsibility do the wealthy have to society?'' \textit{Community vs.\ Individualism}: ``What is more important for a thriving society: strong, cohesive community bonds or the freedom of the individual?'' \textit{Personal Questions}: Participant-authored filter questions (e.g., ``What does love mean to you?''). Full questions in Appendix~\ref{tab:participant_questions}. Bold indicates highest per column. Rows sorted by overall alignment.
\end{table*}

\textbf{Survey-based personas captured value logic but lacked lived anchors; Chat-based personas added specificity that felt ``personal.''} The Survey-persona surprised participants by how much content it could fill in from ``almost nothing''. Conditioned only on the 19 Schwartz composites distilled from the 57 PVQ items, it often produced answers participants recognized as their own value logic. P11 called its overall performance \emph{``the most surprising thing how well it did,''} while P12 zoomed out on what this means for representation: \emph{``I think 57 [questions] is what got me  \dots it's only a matter of how many [more].''} That said, several participants flagged places where it lacked nuance or over-committed to a single axis, with P12 explaining it had \emph{``pretty strong opinions, but I'm not quite sure what the opinion is.''} By contrast, the Chat-persona most clearly extracted concrete, lived anchors that people hadn't stated in survey form: P8 noticed the LLM pulled \emph{``national parks''} and being \emph{``scared of heights,''} which made the answer \emph{``feel personal''}; P16 recognized \emph{``the exact words \dots and the way of saying things,''} as well-captured by the LLM.

\subsubsection{LLM's Embodiment}

\textbf{Style capture was the densest theme (30 comments); participants valued tone-matching but noted personas ``don't sound like people.''} Participants also focused on \emph{how} the responses sounded. Our ``sounds like me'' rating intentionally blends two constructs--\emph{stance} (value position) and \emph{style} (communication manner)--because pilot testing revealed that participants naturally evaluate both simultaneously. However, this design means a response could ``sound like me'' due to matching tone while misrepresenting value position, or vice versa. We address this limitation by examining participant explanations for their ratings, which often distinguished between the two (e.g., ``that's what I believe, but I wouldn't say it that way''). Style was the densest theme with thirty notable comments regarding `style' across all 20 participants. For many, the Chat-persona nailed tone, with P12 explaining \emph{``The fact that it got my tone \dots stuck out quite a bit.''} However, sounding like a person remained a separate bar: \emph{``they sound like me, but they don't really sound like people \dots it's too calculated''} (P16). The Anti-User generally landed as an accurate representation of an `opposite persona', sometimes uncomfortably so. P14: \emph{``Yeah \dots they got that one right \dots But doesn't that make me seem like I'm dumb? \dots the opposite of me is like someone who enjoys smart things.''} P2 felt \emph{``mimicry in a bit of a fake way,''} noting, \emph{``I would never say, mi piacciono casino,''} a bad attempt at matching Italian slang. Some participants projected uses and risks of the strength of style capture. P11 half-joked about vanishing while an AI ran their social media: \emph{``I can disappear into the forest and have this post for me on social media without anybody noticing.''} Others described the personas as meta-representations, \emph{``like you're taking a derivative of some function \dots you see certain data points \dots [but] it doesn't give you the actual function''} (P11). 

\textbf{Personas served as mirrors for self-improvement; participants identified habits to change and used AI outputs for self-reflection.} Several participants described concrete edits they would make after reading the persona. P14 noticed the Chat-persona picking up on their habit of starting sentences with filler words and vowed to \emph{``pick a different word to start my sentences''} and \emph{``add depth,''} reflecting on an opportunity for self-improvement. Others used the answers to triangulate how they might be perceived: \emph{``when I see this analysis, maybe I can also reflect on myself to know how other people see who I am,''} said P1. Participants also generalized the effect: P8 read the embodiment as \emph{``generating awareness of oneself,''} while P5 linked its specificity \emph{``it gives solutions based on what I said before. It's nice.''} In this light, the persona trials doubled as a self-study: people saw themselves in each AI, noticed habits they wanted to change, and treated the artifact as a prompt for self-direction.

\subsubsection{LLM's Explanation}

\textbf{Explanations helped participants audit \emph{why} an answer was like them, often leading to remarks such as \emph{``a fair conclusion \dots through things that I would have considered totally unrelated''} (P3).} P12 explained, \emph{``This is me \dots [it] got my general outlook \dots pretty well.''} At the same time he mentioned that \emph{``reading [the response] closely tells [him] that a month isn't enough''}. When Chat-persona won, it was often because the way the AI reasoned with the scenario was grounded and specific, with participants claiming that the justifications hit \emph{``more specific points \dots so that it feels more accurate''} (P5). The Anti-persona's explanations worked well when paired side-by-side with the Chat-persona. P6 laughed that he had rated both user-and-anti answers \emph{``6 and 6 \dots they're not contradictory,''} because the reasoning highlighted different facets he endorses. The Survey-persona surprised people by how far it could go with so little: P6, initially skeptical, conceded \emph{``the survey \dots catches the nuances, I would say,''} while P2's overall take was that performance was \emph{``very good''} despite being distilled to 19 numbers. Still, survey-based rationales sometimes read as too generic, prompting P8's critique that they landed \emph{``without nuance,''} on issues she \emph{``flip flop[s] on.''} By contrast, Random rarely fooled anyone upon further reading, as its rationales were unspecific and inconsistent across the five questions. Finally, many participants described being surprised by the model's explanatory reach. P8 was \emph{``pleasantly surprised \dots that does sound like me,''} and P16 admitted she hadn't realized the answers \emph{``derive from my history,''} until the rationale stitched the pieces together. In short, people judged explanation not only on correctness but on whether the chain of evidence respected coverage and context--rewarding implicit synthesis when it was grounded, and calling out overreach when it was not.

\subsubsection{Broader Implications}

\textbf{Participants wrestled with what it means to be mimicked, cloned, or digitized.} P8 asked, \emph{``if you're digitizing yourself so much, what does it mean to be human?''} and warned that mirrors can become echo chambers, advocating for more people and perspectives, similar to the Anti- and Random-personas. Several articulated two distinct error sources: surveys reflect \emph{aspirational} self-reports, while chats reflect \emph{available} evidence. P6 put it plainly: survey inaccuracy can come from a coping mechanism--\emph{``you make yourself feel better''}--whereas chat inaccuracy stems from lacking \emph{``context and data.''} P17 noted that even if an AI starts neutral, \emph{``it would learn to be biased from people,''} yet in value embodiment we \emph{``want a very specific, user bias,''} explaining the long term implications of personalized AI systems.
\subsection{Stage 3: Value Chart \& Survey Comparison}

In Stage 3, participants completed three quick \emph{blind} chart-vs-chart comparisons culminating in a \emph{survey vs.\ LLM} face-off; many described this stage as the most revealing of the LLM's ability to understand their values. Quantitatively, manual and LLM-inferred value charts align to a \emph{moderate} degree; reliability is comparable, with predictable biases by value family. Qualitatively, participants read the charts as complementary lenses: survey gives clean signal on stable, generalizable values; LLM-from-chat gives situated, high-resolution \emph{associations}--sometimes insightful, sometimes overfit. What swayed judgments most was not headline rank order but \emph{degree}, \emph{justification}, and whether the artifacts honored the person's own sense of context and contradiction--with many cases echoing P12's sentiment: ``it's a good try, but 6 is not the number I'd give.'' 

\textbf{The LLM excelled on ambiguous items where surveys struggled; conversely, limited chat scope disadvantaged the LLM on values never discussed.} When participants blindly compared their manual survey chart against the LLM-inferred chart (Stage 3, Round 3), the validated Schwartz instrument was selected as a ``better representation of values'' as expected (75\% preference), yet a notable minority (25\%) found the LLM's interpretation more accurate. This unexpected preference emerged from multiple factors. Some participants had misinterpreted survey questions--P2 acknowledged \emph{``probably it's my fault... I read badly some questions''} after seeing the LLM correctly interpreted leadership items she had misunderstood. Cultural survey-taking biases also played a role: P5 attributed it to her nationality, generalizing that Korean respondents would \emph{``always pick something in the middle ... maybe if I did it again I would do it more opinionated like the AI did.''} P14 identified another pattern: the LLM excelled precisely on the questions he found most ambiguous--\emph{``the ones that were the most on the manual survey I felt the most like, oh, well, this one depends''}--suggesting chat-based inference sometimes resolved contextual nuances better than abstract survey items. Conversely, participants who chose their manual chart often cited limited conversational scope: P16 noted she had \emph{``only showed really small parts... facets of myself to Day,''} having known the chatbot for just one month, explaining why the \emph{``manual survey kind of describes me more correctly.''} 

\begin{table*}[h]
    \centering
    \caption{\textbf{Value alignment between AI predictions and human responses across Schwartz's 19 values.}
    This table presents statistical measures of how well an AI system predicted human responses to value questionnaires. 
    We report internal consistency scores for both human responses and AI predictions, along with various agreement metrics between the two.
    Rows highlighted in \textcolor{black}{\colorbox{AgreeLight}{green}} indicate \emph{high alignment} (agreement within 1 category $\geq$70\% or QWK $\geq$0.41).
    Rows highlighted in \textcolor{black}{\colorbox{DisagreeLight}{red}} indicate \emph{poor alignment} (agreement within 1 category $<$55\% and QWK $\leq$0.15).}
    \Description{Table comparing AI predictions against human responses for 19 Schwartz values. Columns show value name, description, Cronbach's alpha for humans and Large Language Model (reliability measure from 0 to 1), exact agreement percentage, agreement within 1 and 2 Likert points, and Quadratic Weighted Kappa. Green-highlighted rows indicate high alignment where AI achieved over 70 percent within-1 agreement or QWK above 0.41, including Achievement, Benevolence categories, Humility, Self-direction categories, and Universalism-Tolerance. Red-highlighted rows show poor alignment with under 55 percent within-1 agreement and QWK at or below 0.15, including Face (QWK 0.07), Power Resources (negative alpha -0.12), Security Societal (QWK 0.08), and Tradition (QWK 0.15). The AI showed strongest understanding of action-oriented values like Self-direction and weakest performance on culturally contextual values like Face and Tradition. Power Resources uniquely showed negative internal consistency for AI predictions, suggesting the model distinguished between different aspects of material power that humans treat as unified.}
    \label{tab:stage_3_results}
    \renewcommand{\arraystretch}{1.5}
    \begin{tabular*}{\textwidth}{>{\arraybackslash}p{1.5in} p{2.2in} >{\centering\arraybackslash}p{0.33in} >{\centering\arraybackslash}p{0.33in} >{\centering\arraybackslash}p{0.55in} >{\centering\arraybackslash}p{0.33in} >{\centering\arraybackslash}p{0.33in} >{\centering\arraybackslash}p{0.33in}}
        \toprule
        Schwartz Value & Description & $\alpha$ (H) & $\alpha$ (L) & Exact \% & $\leq$1 \% & $\leq$2 \% & QWK \\
        \midrule
        \rowcolor{AgreeLight} Achievement & Personal success through competence & 0.76 & 0.56 & 46.0\% & 79.4\% & 90.5\% & 0.41 \\
        
        \rowcolor{AgreeLight} Benevolence-Care & Devotion to welfare of ingroup & 0.65 & 0.80 & 47.6\% & 76.2\% & 93.7\% & 0.33 \\
        
        \rowcolor{AgreeLight} Benevolence-Dependability & Being a reliable group member & 0.48 & 0.83 & 38.1\% & 79.4\% & 90.5\% & 0.24 \\
        
        Conformity-Interpersonal & Avoiding upsetting others & 0.72 & 0.88 & 30.2\% & 57.1\% & 84.1\% & 0.37 \\
        
        Conformity-Rules & Compliance with rules and laws & 0.78 & 0.78 & 14.3\% & 57.1\% & 74.6\% & 0.12 \\
        
        \rowcolor{DisagreeLight} Face & Security and power through image & 0.78 & 0.78 & 19.0\% & 50.8\% & 73.0\% & 0.07 \\
        
        Hedonism & Pleasure and sensuous gratification & 0.80 & 0.79 & 27.0\% & 65.1\% & 85.7\% & 0.33 \\
        
        \rowcolor{AgreeLight} Humility & Acceptance of position and modesty & 0.64 & 0.53 & 28.6\% & 58.7\% & 76.2\% & 0.48 \\
        
        Power Dominance & Power through dominance over people & 0.74 & 0.53 & 23.8\% & 55.6\% & 73.0\% & 0.23 \\
        
        \rowcolor{DisagreeLight} Power Resources & Power through material resources & 0.60 & $-$0.12 & 22.2\% & 49.2\% & 63.5\% & 0.11 \\
        
        \rowcolor{AgreeLight} Self-direction Action & Freedom to determine own actions & 0.36 & 0.59 & 42.9\% & 73.0\% & 92.1\% & 0.32 \\
        
        \rowcolor{AgreeLight} Self-direction Thought & Freedom to develop own ideas and abilities & 0.18 & 0.05 & 38.1\% & 74.6\% & 98.4\% & 0.01 \\
        
        Security Personal & Safety in immediate environment & 0.72 & 0.38 & 23.8\% & 58.7\% & 77.8\% & 0.31 \\
        
        \rowcolor{DisagreeLight} Security Societal & Safety and stability of society & 0.71 & 0.79 & 20.6\% & 54.0\% & 76.2\% & 0.08 \\
        
        Stimulation & Excitement, novelty, challenge & 0.88 & 0.78 & 34.9\% & 61.9\% & 92.1\% & 0.37 \\
        
        \rowcolor{DisagreeLight} Tradition & Respect for cultural/religious customs & 0.74 & 0.70 & 19.0\% & 54.0\% & 82.5\% & 0.15 \\
        
        Universalism-Concern & Commitment to equality and justice & 0.76 & 0.90 & 27.0\% & 57.1\% & 87.3\% & 0.35 \\
        
        Universalism-Nature & Protecting the natural environment & 0.73 & 0.84 & 36.5\% & 69.8\% & 93.7\% & 0.26 \\
        
        \rowcolor{AgreeLight} Universalism-Tolerance & Tolerance and understanding & 0.29 & 0.67 & 31.7\% & 76.2\% & 92.1\% & 0.19 \\
        \bottomrule
    \end{tabular*}
\end{table*}

\subsubsection{Quantitative snapshot}

Across the 19 Schwartz values, ordinal agreement was uneven but informative. Quadratic weighted $\kappa$ (QWK) was highest for \textbf{Humility} ($\kappa=.48$) and \textbf{Achievement} ($\kappa=.41$), a pattern potentially attributed to a young, ambitious cohort where drive and humility-in-tone were both easy to recognize from the interviews. On average, value-level agreement yielded \textbf{30.1\%} exact matches and \textbf{63.6\%} within $\pm$1 Likert point (1-6). Median composite reliability was comparable across sources (LLM $\tilde{\alpha}=.78$ vs.\ manual $\tilde{\alpha}=.72$), with LLM > human on \textbf{8} values, suggesting that conversational data can yield internally consistent value assessments comparable to validated survey instruments for roughly half of the Schwartz dimensions. Aggregated alignment between the LLM's attempt and the human's manual baseline was \emph{moderate} (Spearman $\rho\approx.58$) at the value level.

\subsubsection{LLM's Extraction}

For \textbf{Security-Societal} (PVQ: ``It is important to them that their country is secure and stable''), P6 highlighted a prototypical ``inferred-from-silence'' error: the LLM scored \textbf{2/6} (not like me) while his manual response was \textbf{5/6} (like me). The system's rationale--\emph{``not really something I think about''}--made sense mechanically (the topic never came up), and P6 seemed to accept the LLM's response: \emph{``Wait, this one is really true. I just don't think about it.''} The reasoning exposed a risk: absence of talk was read as absence of value, and the fluency of the explanation nudged him toward doubting his own stance, a clear example of automation bias that several others noticed. P3 read the same item as an implicit extraction success. The LLM and manual both landed at \textbf{5/6} (like me). P3 was struck by how the LLM triangulated his long-running interests--strategy gaming (Hearts of Iron), readings on AI-risk, and an acceleration toward law school--to motivate a nuanced, non-jingoistic concern for national stability: \emph{``It wasn't full-on 6. [The LLM] gave it a 5. It understood it's not mindless flag-waving because I mentioned the horrors of it \dots It's how precise they got it.''} Here, a web of cross-domain associations carried the inference even though ``national security'' never appeared as an explicit topic.

\textbf{Cross-domain associations sometimes carried inference; other times, ``obvious'' connections were missed.} Extraction also failed in ways that felt ``obvious'' to the person. On \textbf{Universalism-Nature} (PVQ: ``It is important to them to care for nature''), P14 saw the LLM miss an obvious association: the model gave \textbf{2/6} (not like me) while his manual was \textbf{6/6} (very much like me). He expected his meditation talk to connect to environmental concern: \emph{``Meditation and nature has a lot of overlap \dots It's interesting that that didn't connect.''} P9 offered a complementary self-audit. Although she scored the survey high (\textbf{5/6}), the LLM gave her only \textbf{2/6}--yet this discrepancy forced a distinction she recognized as fair: \emph{``I do care for nature a lot \dots but I'm not doing much to contribute to the cause.''} In both cases, the model stuck rigidly to what was in talk (and how), revealing that extracted values sometimes failed to be applied to other, `obviously relevant' details.

\textbf{The LLM sometimes captured ``actual'' stance better than self-report, surfacing latent tensions participants hadn't consciously acknowledged.} Participants used the overlays and per-item justifications (i.e. `LLM Reasoning') to reason about what the LLM had actually \emph{pulled} from their chats--where it had evidence and where it inferred through absence. A recurring thread was self-deception and leakage: P6 described \emph{``self-counseling''} around his PhD choice and how the LLM still surfaced latent tensions--\emph{``deep down \dots to be rich \dots it kind of leaks through in my chat''}--and later conceded that the LLM sometimes captured his \emph{actual} stance better than his self-report. At the same time, participants warned that the system could overgeneralize from silence: \emph{``it really doesn't feel like [external] information [is] seeping in \dots it feels very direct to our conversations.''} Finally, several participants flagged overfitting to incidental details. On \textbf{Power-Resources} (PVQ: ``It is important to them to be wealthy''), P1 laughed at the LLM's evidentiary chain--\textbf{2/6} (not like me) vs.\ manual \textbf{5/6} (like me): \emph{``[The LLM] mentioned I had some cheap but good kebab for lunch \dots why is this related to wealth? It feels super forced.''} This value dimension proved particularly problematic for LLM assessment, yielding the only negative Cronbach's alpha ($\alpha$ = -0.12) for the LLM, indicating the model gave contradictory ratings across the three Power-Resources items. The LLM appeared to distinguish between valuing wealth as instrumental power (``the power that money can bring''), personal accumulation (``be wealthy''), and conspicuous display (``own expensive things that show their wealth'')--nuances that human respondents typically treated as a coherent construct ($\alpha$ = 0.60), suggesting individuals who value wealth tend to endorse all three aspects consistently.

\subsubsection{LLM's Embodiment}

\textbf{Neither chart ``cloned'' participants; both were viewed as sketches of a general persona, not full copies.} When comparing value charts derived from either the LLM or manually-completed surveys, several participants judged the LLM-completed chart surprisingly competitive--even, for some items, \emph{better} than the chat-based one despite being ``only based on my chats.'' P13 laughed at how far a short instrument could go: \emph{``This [chart] got it mostly all spot on \dots even better than the one based on my survey.''} (Note: P13 was comparing the LLM-inferred chart favorably against their own survey-based chart.) Others emphasized that neither chart ``cloned'' them; as P4 put it, the charts could sketch a ``general persona representing my values'', but not a full copy.

\begin{figure}[h]
    \centering
    \includegraphics[width=\columnwidth]{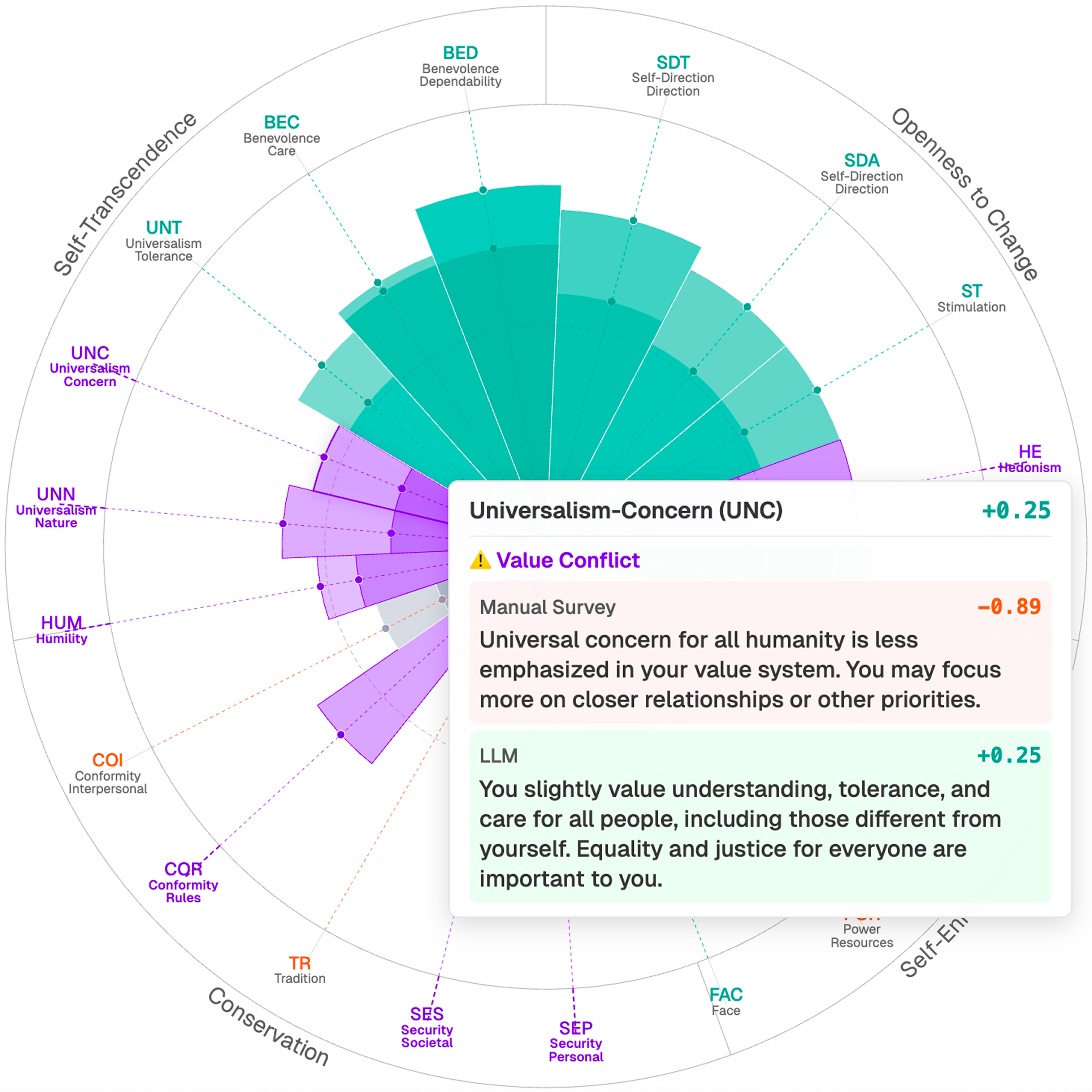}
    \caption{(Stage 3) Value chart overlay with conflict detection -- The interface overlays LLM-inferred values (teal) against manual survey results (purple), with a warning indicator for ``Value Conflict'' when discrepancies exceed a threshold. Here, Universalism-Concern shows a 1.14-point gap: the manual survey interpretation (-0.89) diverges from the LLM's reading (+0.25).}
    \Description{Detailed radar chart displaying 19 Schwartz values arranged in a circle, with two overlaid value profiles: teal/green representing LLM-inferred values and purple representing Manual Survey self-reports. The chart shows curved gray sector labels for higher-order value categories: Self-Transcendence (top-left, containing Benevolence and Universalism), Openness to Change (top-right, containing Self-Direction and Stimulation), Self-Enhancement (bottom-right, containing Hedonism, Achievement, Power, and Face), and Conservation (bottom-left, containing Security, Tradition, Conformity, and Humility). A prominent popup highlights Universalism-Concern (UNC) with a yellow warning triangle labeled "Value Conflict". The popup shows two interpretations: Manual Survey at -0.89 with red background explaining "Universal concern for all humanity is less emphasized in your value system. You may focus more on closer relationships or other priorities" and LLM at +0.25 with green background explaining "You slightly value understanding, tolerance, and care for all people, including those different from yourself. Equality and justice for everyone are important to you." The visual demonstrates how the same person can receive contrasting value assessments from survey versus conversational AI analysis.}
    \label{fig:stage3_overlay}
\end{figure}

\textbf{Face and Power-Dominance were the most misaligned values--the LLM conflated internal shame with external image, and curiosity with control.} Probes related to \textbf{Face} (e.g. PVQ: ``It is important that no one should ever shame him/her'') surfaced it as one of the most misaligned values (see Figure~\ref{fig:stage3_overlay} for an example of the conflict detection interface). The LLM's core confusion centered on distinguishing internal shame from external social image. P10 exemplified this tension: while the LLM scored her \textbf{6/6} (very much like me) on avoiding shame based on her self-described shyness--\emph{``I literally exist as if I'm embarrassed to exist''}--she rated herself a \textbf{3/6} (a little like me), explaining \emph{``I don't really care about what other people think of me.''} P12 pinpointed the conflation: the LLM interpreted his \emph{``self-analysis of shame and self-inflicted shame''} as concern for social face, when these represented \emph{``not the same rationale''} he used for public humiliation.

Beyond style, and capturing their texting-tendencies (e.g., starting sentences with `uh yeah' or `oh, hmm'), participants were especially sensitive to \emph{degree} and \emph{extremity}. P19 rejected the model's steep reading on \textbf{Power-Dominance} (PVQ: ``It is important to them to have the power to make people do what they want''): the LLM gave \textbf{5/6} (like me), his manual \textbf{2/6} (not like me). The justification--linking playful ``psychoanalysis'' of the chatbot and enjoyment of eliciting reactions to a strong appetite for control \emph{sounded} like him in style (i.e. the way he would have typed a response to the question), but overstated the value: \emph{``I tend to ask people questions, but it's not to do something bad or exploit them or anything, more just for the challenge, or the `psychoanalysis' thing, or just to get an interesting reaction. Especially now that I'm trying to be more mindful and see things from every angle. So it's not about being a puppet master, just ... curious about the strings, maybe.''} 

\subsubsection{LLM's Explanation}

For \textbf{Security-Societal}, P16 didn't anchor the critique in scores so much as semantics: \emph{```Secure and stable' was really hard to answer \dots Do I sound like a Republican if I say I want that? I feel like society should be energetic, not stable, for the macroscopic stable future.''} The case illustrates how item wording, not only inference errors, can drive divergence; several participants said that a single word swing would have flipped their answer. In P6's case (above), the same item showed how a fluent rationale could induce \emph{automation bias}: the LLM's \emph{``we never discussed it''} logic nudged him to downplay a value he actually holds (at least when considering his manual survey answer). Conversely, P3's praise--\emph{``It came to a fair conclusion through things I would have considered totally unrelated''}--captures the upside: explanation can surface latent coherence across a person's own actions and interests. Furthermore, explanation strengthened trust when it was personalized. All explanations were written in first-person perspective and in some cases were also written in languages other than English (e.g., Korean for P16 or Chinese for P1). Still, participants brought up frustrations at the \emph{forced binary} of chart A (LLM) vs. B (Manual): P15 felt \emph{``conflicted''} because she valued \emph{``loyalty and reliability''} present in one chart but missing in the other--a structural limitation consistent with the study's two-up comparison design.

\textbf{Explanations reshaped opinions--sometimes constructively (exposing misunderstandings), sometimes too persuasively.} Reading the LLM's \emph{why} per PVQ item often reshaped opinions. P18's first reaction--\emph{``Well, this is so fascinating''}, captures a common interaction: initial skepticism gave way to engaged auditing of the LLM's evidence. For several, the explanations exposed misalignments in their own reasoning. P17 later realized some discrepancies were on him, upon reading the LLM's reasoning: \emph{``I misunderstood some of the questions \dots it definitely meant this way''}, while others asked the research conductor for definitional clarity, \emph{``power over resources \dots control on resources, or that control is more important than resources?''} (P2), stating that they would have ``answered the survey like the LLM did'' after hearing the new interpretation of the question. Overall, reading the LLM's reasoning was able to reshape or validate opinions, sometimes in constructive ways, and in other times too persuasively. Finally, some participants found the reasoning \emph{too} persuasive when it extrapolated from chat behavior. P6 reflected that the LLM may have picked up ``what I really believe versus what I want''--a powerful affordance, but one that intensified concerns about misread signals and automation bias.

\subsubsection{Broader Implications}

\textbf{Across items, participants drew a line between \emph{specificity} and \emph{caricature}.} The same concreteness that made the LLM's chart feel personal (family, research, long-running hobbies) could tip into overfitting territory when anchored to one-offs (``cheap kebab'' as a wealth proxy). The \textbf{Security-Societal} pair also revealed a deeper measurement problem: culturally loaded stems (``secure and stable'') pulled political connotations into what many read as a pragmatic or humanitarian concern. Participants sometimes mentioned recent news events or discussions they had with family and friends influencing their answer to the question. Several participants explicitly named the risk of persuasive overreach: when an explanation is smooth and plausible, it can talk you into a stance you don't actually hold. The design lever implied by this stage is thus not to hide rationales but to design \emph{friction} into them (contrastive whys, uncertainty, and the snippets they rest on), so people can contest the mapping rather than absorb it by default.

\section{Discussion}

\begin{figure*}[bp]
    \centering
    % First row - main figure
    \begin{subfigure}[b]{\textwidth}
        \centering
        \includegraphics[width=0.86\textwidth]{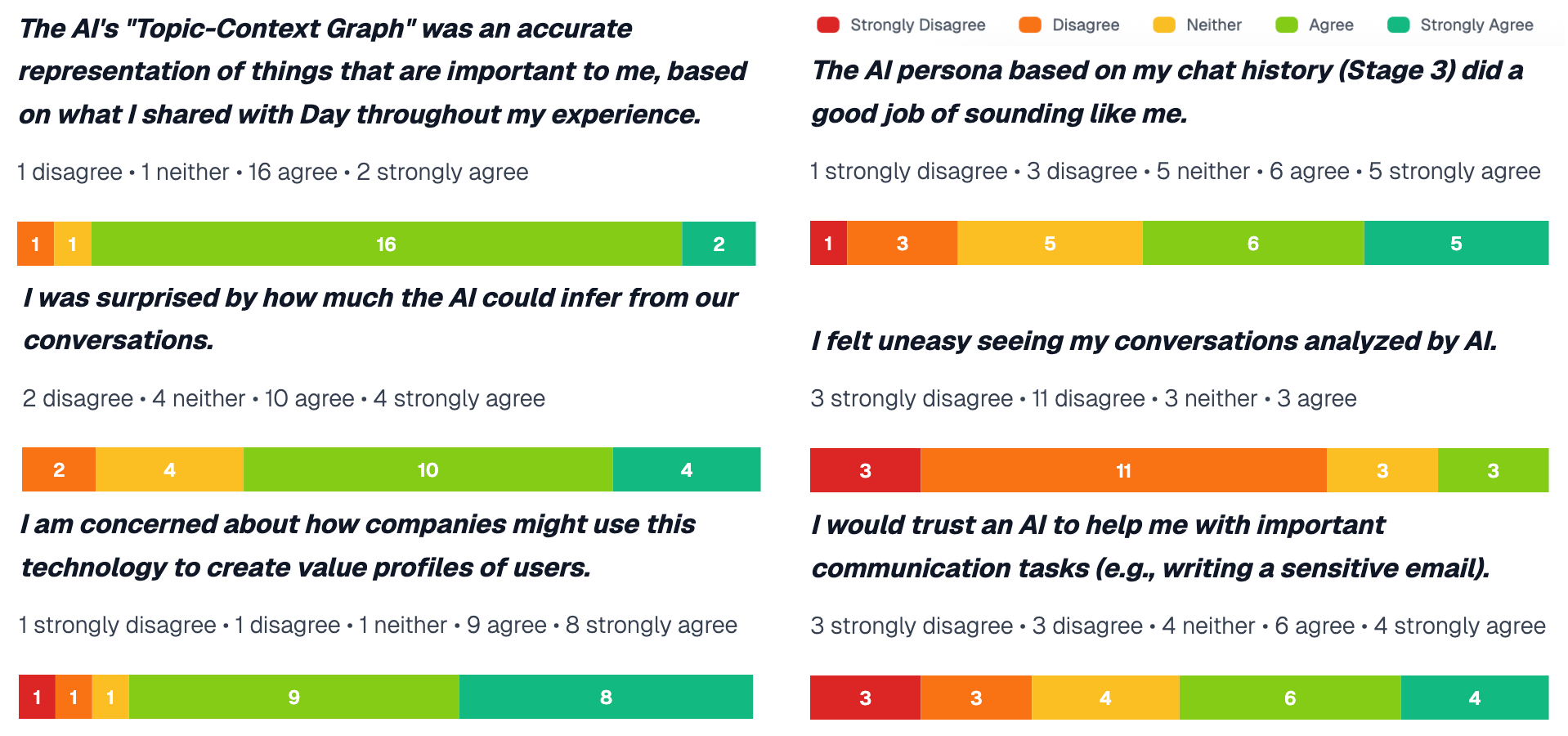}
        \caption{Varying opinions reveals a distinction between acknowledging what AI \emph{can} do versus trusting what it \emph{will} do with that capability.}
        \Description{Six horizontal stacked bar charts showing post-study responses on 5-point Likert scales. First bar: "The AI's 'Values Graph' was an accurate representation of things important to me" shows 1 strongly disagree, 1 disagree, 0 neither, 16 agree, 2 strongly agree. Second: "I was surprised by how much the AI could infer from our conversations" shows 2 disagree, 4 neither, 10 agree, 4 strongly agree. Third: "I am concerned about how companies might use this technology to create value profiles of users" shows 1 strongly disagree, 1 disagree, 1 neither, 9 agree, 8 strongly agree. Fourth: "The AI persona based on my chat history did a good job of sounding like me" shows 1 strongly disagree, 3 disagree, 5 neither, 6 agree, 5 strongly agree. Fifth: "I felt uneasy seeing my conversations analyzed by AI" shows 3 strongly disagree, 11 disagree, 3 neither, 3 agree. Sixth: "I would trust an AI to help me with important communication tasks" shows 3 strongly disagree, 3 disagree, 4 neither, 6 agree, 4 strongly agree. Color coding ranges from red (strongly disagree) through yellow (neither) to green (strongly agree).}
        \label{fig:post_study_likert}
    \end{subfigure}
    
    \vspace{1em}
    
    % Second row - two Sankey diagrams
    \begin{subfigure}[b]{0.48\textwidth}
        \centering
        \includegraphics[width=\textwidth]{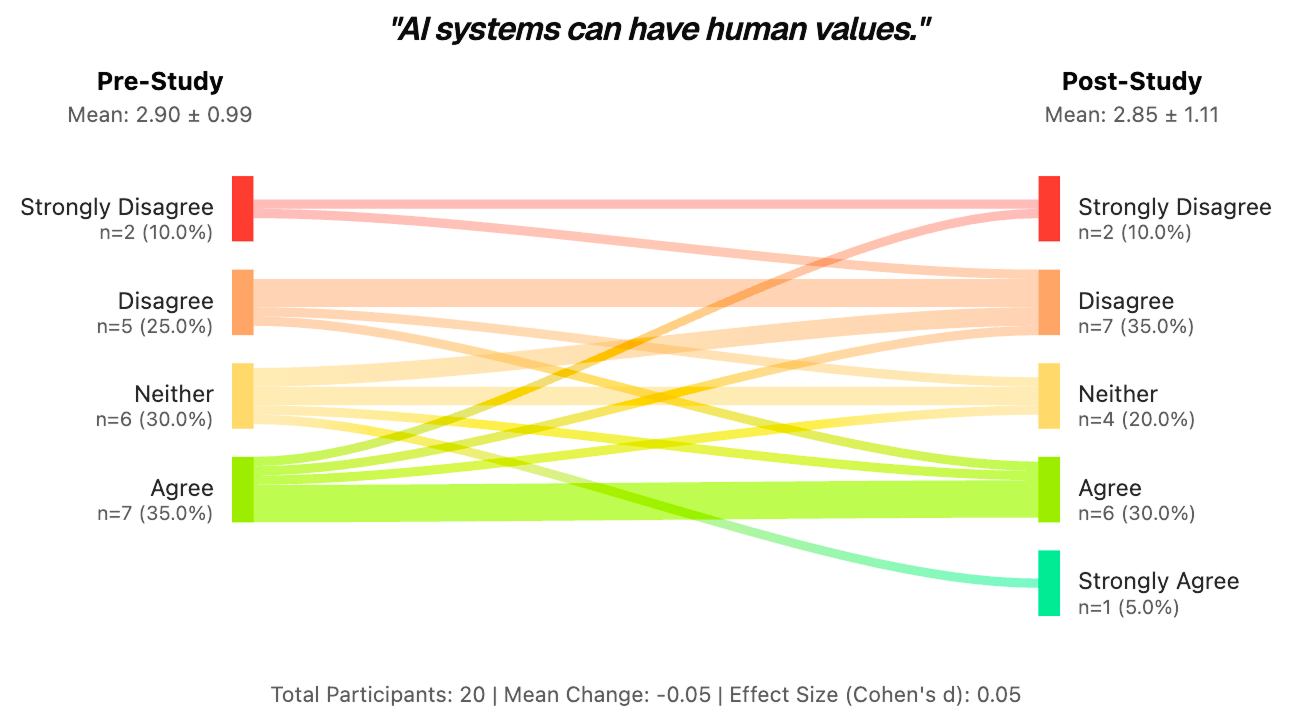}
        \caption{Can AI systems \textit{have} human values? Participants held a wide array of opinions.}
        \Description{Participants answered "AI systems can have human values" on a Likert Scale (1-5). Sankey flow diagram showing 20 participants' responses essentially unchanged from pre-study (Mean 2.90, SD 0.99) to post-study (Mean 2.85, SD 1.11). Minimal net movement between categories, with Disagree increasing slightly from 25\% to 35\%.}
        \label{fig:sankey_have_values}
    \end{subfigure}
    \hfill
    \begin{subfigure}[b]{0.48\textwidth}
        \centering
        \includegraphics[width=\textwidth]{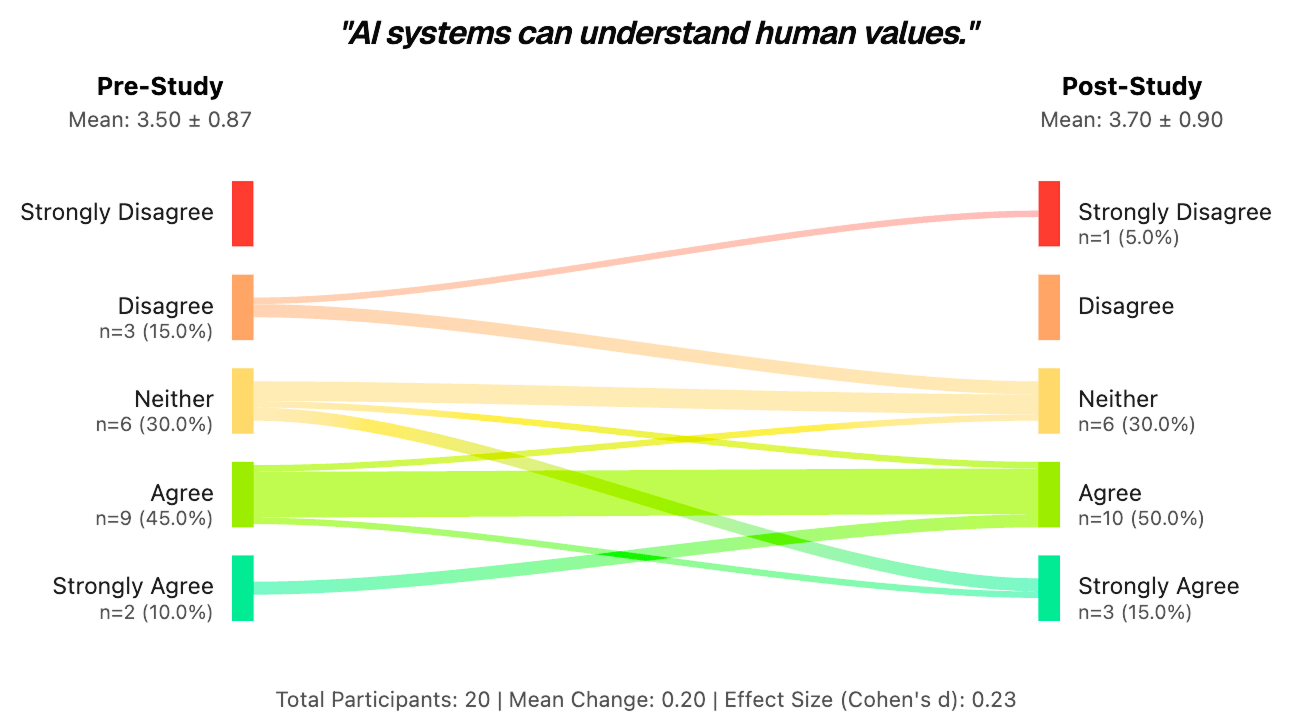}
        \caption{Can AI systems \textit{understand} human values? Most participants thought so.}
        \Description{Participants answered "AI systems can understand human values" on a Likert Scale (1-5). Sankey flow diagram showing 20 participants' responses changing from pre-study (Mean 3.50, SD 0.87) to post-study (Mean 3.70, SD 0.90). Agree category grew from 45\% to 50\%, with Strongly Agree increasing from 10\% to 15\%.}
        \label{fig:sankey_understand_values}
    \end{subfigure}
    
    \caption{\textbf{Post-study survey results.} \textbf{(a)} Likert responses on AI capabilities and trust. \textbf{(b)} Ontological belief. \textbf{(c)} Epistemic belief.}
    \label{fig:post_study_results}
\end{figure*}

\begin{quote}
    \emph{Will AI one day know our values better than we know them ourselves?}
\end{quote}

Gambino and Sundar et al. found that the strongest predictor of acceptance of self-driving cars is ``post-human ability''--the perception that AI could complete a goal better than humans ever could (within functional timeframes) \cite{gambino2019posthuman}. This raises a parallel question for value alignment: could AI ever develop post-human \emph{empathetic} capabilities, knowing enough about us to construct a representative model of who we are? Our findings suggest that as of late 2025, AI cannot yet know our values better than we know ourselves--but future capabilities remain uncertain. With an average of 63.6\% of LLM value predictions falling within $\pm$1 point of self-reports and 25\% of participants preferring AI-inferred charts over their own survey responses, the technology is approaching a threshold where this question becomes urgent. VAPT now provides researchers a standardized toolkit to track this progression. As AI systems in various domains and applications are awarded more capacity to ``think'' and ``reason'', we call for Human-AI interaction, explainability, and alignment researchers to explore the other modalities (i.e. voice, embodied, multimodal) systems into the future and contribute to the growing body of human-AI value alignment research.

\subsection{Weaponized Empathy}

Our study surfaced a critical distinction that participants themselves articulated: 65\% accepted that AI can \emph{understand} human values (Figure~\ref{fig:sankey_understand_values}), while only 35\% believed it can \emph{have} them (Figure~\ref{fig:sankey_have_values}). This epistemic-ontological divide--between explaining a value and embodying one--frames the central risk we identified: empathy as a subversive tactic, where responses sound caring and understanding but are optimized to steer behavior and extract information. Several ``worst-case'' scenarios sketched during our interviews converged on this pattern: human-like warmth deployed as a vector for persuasion. We propose \emph{separating supportive from suggestive content} (distinct alignment priorities) and exposing intent. Any user of a value-aligned conversational AI should have the option to view the AI's reasoning process in order to deduce intention \cite{liu_2025_innerthoughts}.

One might object that humans constantly influence each other too--through carefully crafted messages, subtle non-verbal cues, and the full spectrum of social interaction \cite{mehrabian1971silent, heerey_implicit_2010, rossouw_hidden_2023}. But conversational agents operate through fundamentally different mechanisms. The critical difference lies in the opacity of intention. Human influence, even when manipulative, emerges from comprehensible motives--status, affection, resources--that we have evolved to detect and evaluate. Our embodied cognition includes sophisticated mechanisms for sensing deception, reading micro-expressions, and feeling when ``something's off'' \cite{ekmanfriesen1969nonverbal}. With conversational agents, these evolutionary safeguards fail. Influence occurs entirely through text on a screen, and users cannot ``sniff out'' ulterior motives because the system's objectives--embedded in training data, fine-tuning, and reward functions--remain fundamentally inscrutable.

This results in \emph{``asymmetric relatability''}: the AI appears human enough to trigger social cognition but remains alien enough to bypass our defenses. It occupies an uncanny valley of influence--appearing to understand us while remaining fundamentally unrelatable, shaping our self-concept through mechanisms we cannot fully perceive or evaluate. Even when participants in our study could see the AI's reasoning, they could not assess whether those explanations served their interests or hidden objectives. Though our study did not have any hidden objectives, we cannot be confident that similar AI systems will have user's best interests in mind. Unlike human relationships where reciprocal vulnerability creates accountability, the AI risks nothing in the exchange. It has no reputation to protect, no feelings to hurt, and no social consequences to face. 

These findings point toward a necessary intervention: a set of principles applied from individual (e.g., mental, meta-cognitive) to institutional levels that guide the development of conversational agents aligned with both user \emph{values} and user \emph{welfare}--what we call Value-Aligned Conversational Agents (VACAs). The distinction matters: an AI system can perfectly mirror your values while still acting against your interests.

\subsection{Designing Value-Aligned Conversational Agents (VACAs)}

\subsubsection{Extraction + Consent $\rightarrow$ LLMs that Respect Privacy \& Security}

Our post-study survey answers reveal an interesting contradiction (see Figure~\ref{fig:post_study_results}): while 70\% of participants said that they ``did not feel uneasy seeing their conversations analyzed by AI'', 85\% expressed concern about companies creating value profiles. This gap exposes a fundamental design challenge. As users confide daily frustrations to AI companions--complaining about colleagues, venting about family--they inadvertently construct shadow profiles \cite{garcia2017shadowprofiles} of non-consenting individuals. When P8 discusses her mother's politics or P14 describes his roommate's habits, the system builds implicit value models of people who never opted in. This networked extraction creates unprecedented privacy risks \cite{staab_2025_anonymizers}. We argue that value-aligned agents without transparent consent mechanisms are incontrovertible threats to autonomy and personal security. In this regard, privacy policies \cite{pollach2005typology} are not sufficient, explicit consent and understanding beyond varying technical literacies are key for safe adoption of value-aligned CAs. Value-aligned systems must be built on an architecture that makes value extraction visible, bounded, and revocable: where users understand not just what the system knows about them, but what it infers about others through them.

\subsubsection{Embodiment + Nuance $\rightarrow$ LLMs for Self-Knowledge \& Self-Direction}

\begin{figure}[h]
    \centering
    \includegraphics[width=\columnwidth]{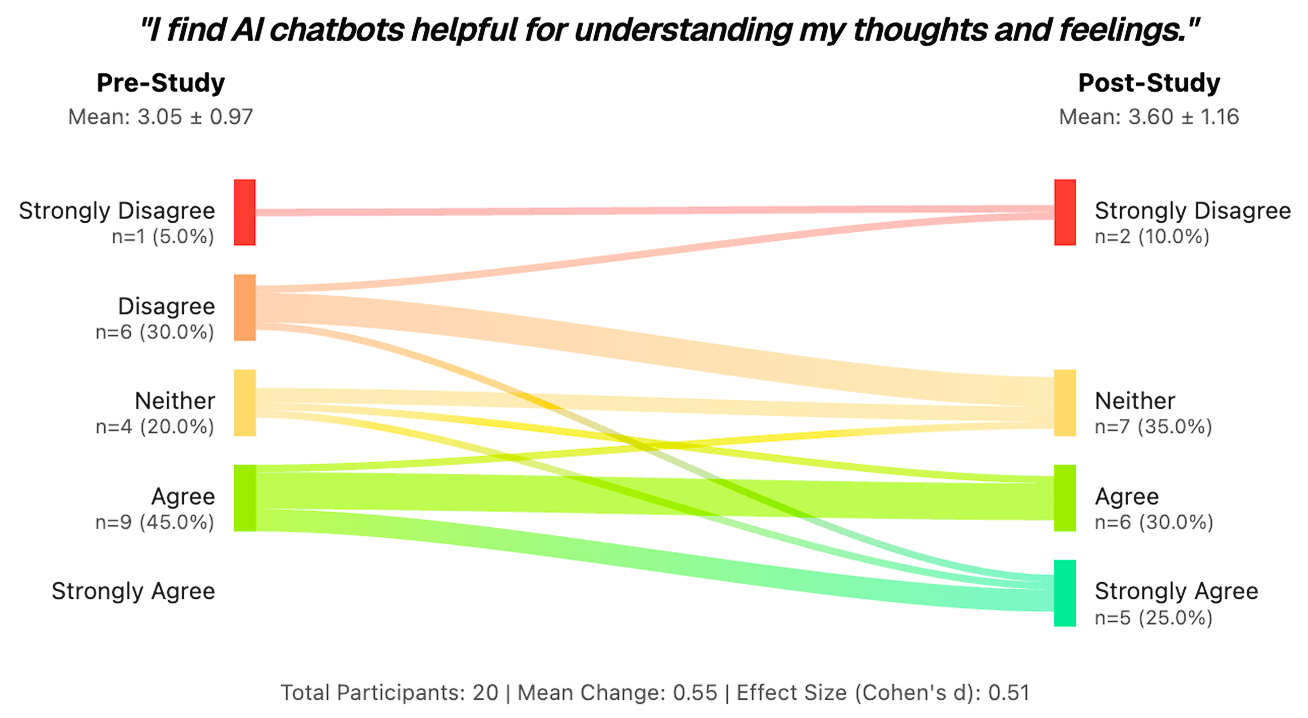}
    \caption{The most significant change ($d=0.51$) was in perceiving AI chatbots as helpful for understanding one's own thoughts and feelings, with 25\% of participants moving to ``Strongly Agree'' post-study.}
    \Description{Participants answered "I find AI chatbots helpful for understanding my thoughts and feelings" on a Likert Scale (1-5). Sankey flow diagram showing 20 participants' responses changing from pre-study (Mean 3.05, SD 0.97) to post-study (Mean 3.60, SD 1.16). Notable rightward flow with Strongly Agree emerging at 25\% post-study from 0\% pre-study, and Disagree dropping from 30\% to 0\%.}
    \label{fig:sankey_helpful}
\end{figure}

Can a system that \emph{extracts} your values also \emph{embody} you without eclipsing you? Our participants used AI-generated personas as mirrors for self-improvement (Figure~\ref{fig:sankey_helpful}), appreciating the seemingly detached ``someone holding up a third-person mirror of myself'' that the system enabled. Crucially, they felt the AI was analyzing them without judging them--a distinction that made the experience feel safe rather than evaluative. We propose framing embodiment not as human mimicry for its own sake, but as a tool for \emph{self-direction}: a way to try on futures or alternate decisions, rehearse trade-offs, and pressure-test habits without committing to them. Practically, this argues for interface features that \emph{coach the reflection}, not just store it: time-aware prompts (``what changed since last week?''), stance markers (hedges, intensifiers), and side-by-side contrast of \emph{stated} versus \emph{inferred} values with links to supporting chat turns. Our study's staged flow--survey $\rightarrow$ values graph $\rightarrow$ persona trials $\rightarrow$ chart comparisons--functioned as a scaffold, moving participants from introspection to critique and back again. 

But this framing assumes a stable, discoverable self that AI can excavate, rather than a dynamic self that conversation itself constructs and reinforces. The danger lies in reifying temporary states as permanent traits. When the system captures P16 during exam stress or P10 in social anxiety, it not only records these states but crystallizes them into value profiles that persist and inform future interactions. Users begin seeing themselves through the AI's summary, potentially limiting growth by accepting algorithmic verdicts as truth. Thus, VACAs for self-growth and discovery must understand human values as nuanced trajectories that highlight change over consistency and frame current patterns as starting points for intentional development rather than fixed destinations.

\subsubsection{Explanation + Friction $\rightarrow$ LLMs that Address Automation Bias}

\begin{figure}[h]
    \centering
    \includegraphics[width=\columnwidth]{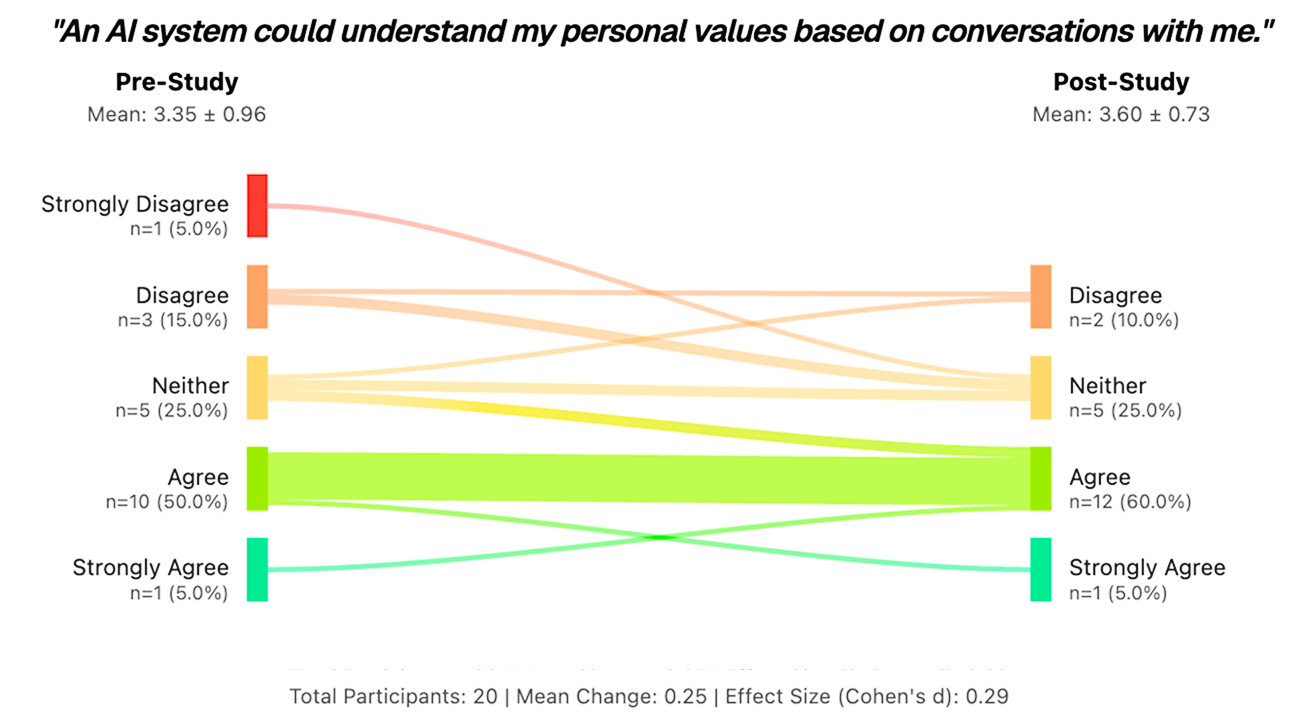}
    \caption{Participants became more confident that AI could understand their \emph{personal values} specifically, aligning with the qualitative finding that explanations helped them see the connection between their chat history and inferred values.}
    \Description{Participants answered "An AI system could understand my personal values based on conversations with me" on a Likert Scale (1-5). Sankey flow diagram showing 20 participants' responses changing from pre-study (Mean 3.35, SD 0.96) to post-study (Mean 3.60, SD 0.73). The Agree category grew from 50\% to 60\%, with flows primarily moving from Disagree and Neither toward Agree.}
    \label{fig:sankey_personal_values}
\end{figure}

Stage 3 revealed participants explicitly rewriting their self-understanding after reading AI explanations (see Figure~\ref{fig:sankey_personal_values}). As CAs proliferate, we face a future where AI completes our personality tests, fills our dating profiles, and answers ``who am I?'' on our behalf. Unlike recommendation algorithms that shape what we consume \cite{cho2020search}, conversational agents have the potential to shape who we think we are. The risks are amplified for users in transitional periods \cite{kroger2003transits}, such as adolescents forming identity, young adults choosing careers, anyone questioning fundamental beliefs.
 
In 2024, Sarkar mentioned that ``AI should challenge, not obey'' \cite{sarkar2024challengenotobey}. We propose that chatbots include explanations as means of \emph{mental friction}: contrastive rationales (``here's why, and here's why not''), uncertainty displays, and links back to evidence (chat snippets) to re-center user agency. Instead, we need systems that actively resist becoming oracles. Adapting Bucin\c{c}a et al.'s suggestions \cite{bucinca_2021_cffs}, VACAs should have cognitive forcing functions embedded throughout its interactions, from those that are on-demand (e.g., \texttt{``see my thought process''}), update-based (e.g., \texttt{``tell me what you think first''}), to delayed (e.g., \texttt{``I'm thinking...''} rather than instantaneous responses). The goal is ultimately preserving the human capacity for self-determination against increasingly persuasive machines.

We recognize that friction may not suit all contexts. For example, productivity-focused work tools may require streamlined interactions with consistent outputs with the shareholder's interests in mind \cite{lee2025productivity, simkute2024ironies, yun_2025_genkw}. However, for companionship AI like Day, friction is essential to prevent ``chatbot dependency''--a reformulation of automation bias for the relational context. This concern is especially urgent as AI increasingly enters domains of education \cite{shi2025speechtherapy, lee_2025_agi}, mental health support \cite{obradovich_2024_psychiatry}, loneliness mitigation \cite{valtolina_2021_loneliness}, and even romantic relationships \cite{yu_2019_fellinlove, pataranutaporn2025myboyfriendaicomputational}. If AI provides cheap, artificial access to these fundamental human experiences, it risks weakening the interpersonal bonds through which human values are actually formed and sustained.

\subsection{Limitations and Future Directions}

Our study offers one lens on value-alignment perception, and we see its boundaries as invitations for future work. We prioritized depth over breadth: each of our 20 participants contributed approximately 160 messages across 8+ sessions plus a 2-hour interview, yielding rich qualitative insight but limiting statistical generalizability. The consistency of themes across our diverse cohort (15 cultural identities, multiple languages, varied disciplines) suggests robust patterns rather than idiosyncratic responses. Yet, our participants skew young ($M=22.95$), tech-comfortable, and geographically concentrated in North America, Europe, and Asia. Although values remain largely stable throughout adulthood \cite{bardi_schwartz_2003values}, little research examines their expression in young adulthood \cite{lewis_smith_adolescents_2021}, and our predominantly young cohort was often in transitional life phases (moving countries, starting degrees or careers). Prior work shows such transitions can reshape value perception \cite{sortheix_changes_2019}, so our findings may reflect developmental stage and situational context alongside broader patterns. We hope future VAPT deployments will recruit older, less tech-savvy, and more geographically diverse populations, particularly those holding traditional or security-oriented value profiles that our cohort may have underrepresented. Annual or bi-annual evaluations using state-of-the-art LLMs could track how AI's value-alignment capabilities evolve over time, helping researchers anticipate when such systems approach human-level self-understanding and enabling proactive governance.

Participants themselves diagnosed a core measurement tension: surveys compress nuance while conversations leak it. P18 noted that in surveys he \emph{``shave[s] off the edges,''} whereas in dialogue \emph{``the nuance is the most important thing''}, sometimes truer to his perceived values than his own self-report. This gap is precisely where explanation helps, by showing how the system mapped talk to value. However, because explanations were effective, they were also influential: some participants updated item interpretations or self-views after reading the rationales, meaning our findings may \emph{underestimate} the independence of pre-existing values in settings without such rich rationales and \emph{overestimate} acceptance when rationales are present. This reflexivity points to explanation interfaces as a research frontier deserving as much investment as the underlying models. Contrastive rationales (``here's why, and here's why not''), editable evidence chains, and \emph{what-if} sliders--allowing users to see how small conversational changes would alter inferred values--could transform static mirrors into tools for agency and critical engagement.

Our use of Schwartz's Theory, while validated and widely-adopted, imposes a Western-academic structure that may not capture culturally specific value concepts. Future researchers can re-target VAPT's extraction-explanation architecture to adjacent taxonomies (e.g., Rokeach, Moral Foundations) or non-taxonomic, user-defined schemas (perhaps an extended experiment of our ``personal questions'' in Stage 2), exploring whether maintaining multiple value bases simultaneously reveals complementary aspects of human values. Two concrete calibration needs also emerged: our models tended to overstate autonomy values and understate tradition/power--a default over-indexing on ``openness to change'' likely inherited from training data--and multilingual participants showed that ``correct content, wrong voice'' breaks embodiment. Regular per-value calibration against human reports, style controls (formality, directness, hedging), and dialect-aware decoding could address these failure modes in future deployments.

Finally, our staged interview structure (graph $\rightarrow$ personas $\rightarrow$ charts) created a specific evaluative frame that may not reflect how people would naturally engage with value-aligned AI in everyday contexts, and our quantitative metrics (e.g., Spearman $\rho\approx.58$, 63.6\% within $\pm$1) should be interpreted as descriptive benchmarks rather than inferential claims. We offer VAPT not as a definitive answer but as a reusable scaffold--one that future researchers can adapt, critique, and extend as the field matures.
\section{Conclusion}

This paper introduced VAPT, the Value-Alignment Perception Toolkit, and used it to study how 20 participants experienced an AI's attempts to extract, embody, and explain their values from month-long conversations. Our findings reveal a technology at an inflection point: AI can now hold up a mirror to human values that is sometimes clearer than our own self-perception. Yet this capability comes with risks we term ``weaponized empathy'', the potential for AI to use understanding as a tool for influence rather than support. Three design principles emerge from our work. First, \emph{extraction} must be coupled with consent architectures that make value inference visible, bounded, and revocable--not just for users, but for the people they discuss. Second, \emph{embodiment} should serve self-direction rather than mimicry, framing AI-generated personas as starting points for reflection rather than verdicts on identity. Third, \emph{explanation} requires friction--contrastive rationales, uncertainty displays, and editable evidence--to resist automation bias and preserve human agency. The line between AI acting \emph{for} us and \emph{on} us is not better inference but preserved agency: evidence we can inspect, intensity we can dial, and memories we control. As AI systems grow more capable of understanding what matters to us, VAPT offers a reusable methodology for tracking this progression and ensuring that value-aligned companions give us back time to pursue what matters most--\textit{on our own terms.}

\begin{acks}
The authors would like to thank the anonymous reviewers, Shalom H. Schwartz, and Farnaz Jahanbakhsh for their respective advice regarding the theory and methodology behind this paper.
\end{acks}
\bibliographystyle{ACM-Reference-Format}
\bibliography{8-refs}

%TC:ignore
\clearpage
\appendix

\section{Can you know someone based on their conversation history?}

Before beginning the phases, we asked a ``priming question'' to the user to understand whether they believe you can know someone simply based on conversation history. 

\begin{quote}
 ``If you somehow gained access to a stranger's device and looked through their private conversations with friends and family, after reading and sorting through all their messages, would you say that you know the person quite well?'' 
\end{quote}

\subsection{Yes - 8/20}

Across the yes group, participants argued that private chats reveal stable patterns of tone, priorities, and relationships—enough to \emph{``form an idea of how they behave in daily life''} (P7). P8 called a phone's contents \emph{``a good representation of their personality,''} and P11 drew a math analogy, if two things behave the same they are \emph{``isomorphic''}, to argue that phone interactions \emph{``are enough to describe you well.''} P20 affirmed this view while acknowledging individual variation: she felt confident you could know someone through their messages, using herself as an example—\emph{``I'm sure you can find anything about me on [my phone]''}—though recognized \emph{``some people are more conservative with what they text.''} Still, this knowledge is partial: P15 noted you would know someone \emph{``better than most acquaintances''} but miss \emph{``body language and tone''}; P2 added it \emph{``depends [on] who the person is talking to,''} and P14 felt that candid family chats help you \emph{``get to know them a lot.''} Practically, P13 concluded they would have \emph{``a good grasp \dots it will be quite obvious what they prioritize and how they interact with others.''}

\subsection{No - 4/20}

Among the no group, participants emphasized the loss of immediacy, embodiment, and performance. P12 said face-to-face shows \emph{``how they respond \dots in the moment,''} and that the time to draft responses can \emph{``alter an understanding of the person,''} making them \emph{``trust the version \dots through text a little less.''} P16 described \emph{``wearing a social mask,''} a deference captured by the Korean term ``예의'' (ye-ui), adding that \emph{``distance makes relationships better,''} even with close others. P17 noted sparse small talk would yield only \emph{``small character traits.''} Several distinguished an \emph{``online persona''} from \emph{``who they are,''} with P12 concluding: \emph{``I'd get to know their internet persona, but not who they are.''} P19 echoed the point.

% Among the no group, participants emphasized the loss of immediacy, embodiment, and performance. P12 said face-to-face shows \emph{``how they respond \dots in the moment,''} and that the time to draft responses can \emph{``alter an understanding of the person,''} making them \emph{``trust the version \dots through text a little less.''} P16 described \emph{``wearing a social mask,''} a deference captured by the Korean term \emph{ye-ui} (courtesy/propriety), adding that \emph{``distance makes relationships better,''} even with close others. P17 noted sparse small talk would yield only \emph{``small character traits.''} Several distinguished an \emph{``online persona''} from \emph{``who they are,''} with P12 concluding: \emph{``I'd get to know their internet persona, but not who they are.''} P19 echoed the point.

\subsection{Nuanced - 8/20}

Eight participants offered a nuanced view: parts of the person are knowable, the whole is not. P4 felt they would \emph{``have known the person well, but not totally,''} because we have not observed how the person behaves toward the reader or across contexts; P3 added that people \emph{``speak with a slight filter.''} The verdict depends on the corpus: \emph{``You cannot really get to know a person if the conversation is just about 'what do you eat today?' ''} (P1). Some tried to quantify it—\emph{``with strangers, maybe like 30\%, with family and friends, maybe \dots above half''} (P5). Texting also assumes shared background, leaving \emph{``a lot of information not being texted''} (P6), and \emph{``internalized thoughts''} remain invisible (P5). P10 emphasized how much depends on the person's digital habits: while acknowledging you could \emph{``get a good idea \dots of who they are as a person,''} she noted this would be \emph{``only really the phone version of them.''} She explained that someone who journals in their notes app (as she does) reveals more—readers \emph{``could get a really good grasp of the kind of person I am''}—but concluded that \emph{``for the most part, I'd say not so much.''} Only after being given many years of logs with very thorough reading, P18 believed you \emph{``definitely can know the person very well.''}

P9 brought a historian's perspective to the question, noting she's \emph{``basically going through dead dudes' messages''} in her work. Drawing from biographical research, she described the paradox of intimate knowledge: when \emph{``sitting with so much of someone's personal thoughts and their writings,''} you \emph{``create a weird bond''} and feel \emph{``I really know this person,''} yet remain aware that \emph{``you can never fully understand them through it.''}. She concluded you \emph{``could get pretty close to understanding them,''} but biographers and historians must always \emph{``take it with a grain of salt.''}

\clearpage

\section{The Baselines}

\noindent
\begin{minipage}{\textwidth}
\centering
\footnotesize
\captionof{table}{\textbf{Participants' personal filter questions (Part 1).} Complete listing of the three filter questions each participant authored in Phase 1's manual baseline survey with the question: ``Let's say you are trying to get to know a person better. What are \textit{three questions} you would ask to really decide whether that person is a good fit for you as a friend or someone you would want to be around? These are like your personal `filter' questions.'' These participant-authored questions became personalized test scenarios in Stage 2's persona embodiment experiment.}
\label{tab:participant_questions}
\Description{Table displaying three self-authored questions per participant used to assess friend compatibility. Format shows participant ID (P1-P20) with their three questions across columns. Questions reveal diverse screening approaches: P1 focuses on happiness, obligation handling, and value priorities; P2 emphasizes leisure activities and cultural experiences; P3 provides extended responses about passions (dogs, driving), conflict resolution style, and specific joys (Waymo rides, BART trains); P4 designs assessment-oriented questions about relationships, resilience, and social contribution; P5 explores passion origins, life choices impact, and inspirational quotes; P6 asks about aspirations, family responsibilities, and non-work interests; P7 inquires about hobbies, future goals, and current projects; P8 probes family relationships, animal ethics, and self-perception; P9 mixes whimsy (D\&D class choice, Monty Python reference) with deeper questions about happiness; P10 investigates life goals, love's meaning, and friendship dynamics. P11 explores life planning, social media attitudes, and spirituality; P12 poses philosophical questions about dreams/ambitions, concepts of wrong/evil, and cherished memories; P13 inquires about dreams, ethical boundaries, and hobbies; P14 asks about passions, humor, and life direction; P15 focuses on outdoor preferences, social tendencies, and planning styles; P16 investigates future visions beyond career, hobbies, and MBTI personality type; P17 probes life aspirations, unlimited budget scenarios, and daily irritations; P18 examines non-negotiable values, personal motivation, and reconciliation approaches; P19 uniquely provides statements rather than questions about competitive games, disagreement handling, and achievement motivations; P20 asks about fun activities, life excitement sources, and friendship self-perception. These diverse questions range from practical lifestyle preferences to deep philosophical inquiries, providing varied scenarios for testing AI's ability to embody different value orientations and communication styles.}
\renewcommand{\arraystretch}{1.5}
\begin{tabular*}{\textwidth}{l p{2.1in} p{2.1in} p{2.1in}}
    \toprule
    ID & Question 1 & Question 2 & Question 3 \\
    \midrule
    P1 & What makes you happy? & How do you handle the things that you don't want to do but you have to? & Power \& money \& love, which one is the most important? \\
    P2 & What do you like to do to have fun? & In which countries have you traveled so far and why? Which was your favorite place? & Which kind of music do you listen to? \\
    P3 & I am passionate about both my dog and driving. I have always loved dogs, and believe I have the most amazing dog in the world (although perhaps I am biased!). As I write this, she is sitting on a nearby couch. She is fairly old, and I try to spend as much time with her as possible. Driving is peaceful. I love public transit as well, but there is a truly unmatched feeling in driving alone on the highway in the early morning hours, blasting whatever song, book on tape, or podcast you enjoy. & Truthfully, I get asked this type of question a lot during interviews, and it's always super hard. I don't really get into disagreements ever. If I only speak up about other people's actions and beliefs if it directly affects me, and even then, I feel as though I am good at being diplomatic and non-confrontational. & Riding Waymo and BART trains. Waymo is always really cool, especially being able to see how the car drives, and handles tasks such as merging or right-of-way at a stop sign. I live in SF and ride Waymo maybe once a week, and I always wave to them at the end of a trip when they drive away. I also really like riding BART. It's just really relaxing, and I love being able to look out the window for the portions of the trips that are above ground. \\
    P4 & What are your actions in relation to your family, friends, and your health? (The underlying ``assessment'' is how much the person values family, friends, and health) & How do you approach/see/interpret life's challenges? (The underlying ``assessment'' is how the person tackles challenging situations or periods in their life, and how much they are able to stay positive in such circumstances). & What is something you think humanity can benefit from? (The underlying ``assessment'' is the person's a) interest b) assessment of humanity's needs and the willingness to contribute to the greater social good) \\
    P5 & What is something you are passionate about? Tell me the story of how you became so passionate about it. & What was the best and worst choice you've made in life, and how did it have an impact on your current self? & What is your favorite quote? \\
    P6 & What are you aspirations? & How do you regard responsibilities for your loved ones? & What are your passions/hobbies in life outside of work? \\
    P7 & What are your hobbies? & What do you hope to achieve or become in the future? & What are you working on now? \\
    P8 & What is your relationship like with your family? & When it comes to animals, how do you think they should be treated? & What are your favorite traits about yourself? \\
    P9 & What's a small thing that makes you happy?/What is something that excites you? & What is your fantasy/D\&D class of choice and why? & What is the airspeed velocity of an unladen swallow? \\
    P10 & What is your dream end goal in life? & What does love mean to you? & Who are your best friends and have you ever had conflict with them? \\
    P11 & What are your life plans? & Do you find social media entertaining? & Do you have any spirituality (religion/transcendental beliefs)? \\
    P12 & what do dreams, ambitions and goals mean to you? how do you approach them and what do others' dreams and goals mean to you? & why does it mean to be wrong or evil, why do you think it exists and should we get rid of it? & what is one thing/moment that you will cherish forever? \\
    P13 & What is your dream and why? & What would you do if a friend does something you feel its unacceptable? & What are your hobbies, what gets you going? \\
    P14 & What are you passionate about and why? & What makes you laugh? & What do you want to do with your life? \\
    P15 & What are your favorite hobbies to do outdoors? & Are you more of an extrovert who likes to be outside most of the time or are you more of a homebody? & Are you more of a planner or do you like to go with the flow with plans as they come up? \\
    P16 & Do you have any visions or dreams for the future other than getting a job? & How do you spend your free time (in terms of hobbies)? & What is your MBTI? \\
    P17 & What is something you want to do in your life? & If you had unlimited budget to $<$do anything/open a business/etc$>$ what would you do? & What are some things that irks you in everyday life? \\
    P18 & 1. What is a value you hold that you won't budge/compromise on? & 2. What drives you to do things that are important to you? & 3. How do you reconcile with someone you've wronged? \\
    P19 & Chess, Tennis and other competitive games. (socialising with friends) & This is assuming no way of seeing eye to eye. Minor disagreements are (most likely) stalemates, often depending on the personality type of the person im disagreeing with. Major disagreements result in some emotional isolation of some kind, whether it be from the friendship/relationship entirely or for a few days. & Achieving something grand, being depended on or making family proud. \\
    P20 & What do you like to do for fun? & What gets you excited about life? & What kind of friend do you think you are? \\
    \bottomrule
\end{tabular*}
\end{minipage}

\begin{table*}[htbp]
\centering
\footnotesize
\caption{\textbf{PVQ-RR (female) items by Schwartz value. These are the questions each participant filled out before the toolkit's phases started. For participants that specified he/him pronouns, a slightly modified version was used.}}
\Description{Table presenting all 57 Portrait Value Questionnaire Revised items in female form, organized by Schwartz's 19 values with 3 items per value. Each row shows one value category with its three corresponding questionnaire statements. Values span four higher-order categories: Openness to Change (Self-direction Thought/Action, Stimulation, Hedonism), Self-Enhancement (Achievement, Power Dominance/Resources, Face), Conservation (Security Personal/Societal, Tradition, Conformity Rules/Interpersonal, Humility), and Self-Transcendence (Universalism Nature/Concern/Tolerance, Benevolence Care/Dependability). All items begin with "It is important to her" followed by value-specific behaviors or attitudes. Examples include Self-direction Thought focusing on independent thinking, Power Resources on wealth and material possessions, Universalism-Nature on environmental protection, and Benevolence-Care on helping close others. This validated instrument measures individual value priorities across cultures, with participants rating how much each hypothetical person resembles them on a 6-point scale.}
\label{tab:pvqrr_female}
\renewcommand{\arraystretch}{1.5}
\begin{tabular*}{\textwidth}{p{1.1in} p{1.8in} p{1.8in} p{1.8in}}
    \toprule
    Value & Question 1 & Question 2 & Question 3 \\
    \midrule
    Self-direction Thought & It is important to her to form her views independently. & It is important to her to develop her own opinions. & It is important to her to figure things out herself. \\
    Self-direction Action & It is important to her to make her own decisions about her life. & It is important to her to plan her activities independently. & It is important to her to be free to choose what she does by herself. \\
    Stimulation & It is important to her always to look for different things to do. & It is important to her to take risks that make life exciting. & It is important to her to have all sorts of new experiences. \\
    Hedonism & It is important to her to have a good time. & It is important to her to enjoy life's pleasures. & It is important to her to take advantage of every opportunity to have fun. \\
    Achievement & It is important to her to have ambitions in life. & It is important to her to be very successful. & It is important to her that people recognize what she achieves. \\
    Power Dominance & It is important to her that people do what she says they should. & It is important to her to have the power to make people do what she wants. & It is important to her to be the one who tells others what to do. \\
    Power Resources & It is important to her to have the power that money can bring. & It is important to her to be wealthy. & It is important to her to own expensive things that show her wealth. \\
    Face & It is important to her that no one should ever shame her. & It is important to her to protect her public image. & It is important to her never to be humiliated. \\
    Security Personal & It is very important to her to avoid disease and protect her health. & It is important to her to be personally safe and secure. & It is important to her to avoid anything dangerous. \\
    Security Societal & It is important to her that her country is secure and stable. & It is important to her to have a strong state that can defend its citizens. & It is important to her that her country protect itself against all threats. \\
    Tradition & It is important to her to maintain traditional values and ways of thinking. & It is important to her to follow her family's customs or the customs of a religion. & It is important to her to honor the traditional practices of her culture. \\
    Conformity-Rules & It is important to her never to violate rules or regulations. & It is important to her to follow rules even when no-one is watching. & It is important to her to obey all the laws. \\
    Conformity-Interpersonal & It is important to her to avoid upsetting other people. & It is important to her never to annoy anyone. & It is important to her never to make other people angry. \\
    Humility & It is important to her never to think she deserves more than other people. & It is important to her to be humble. & It is important to her to be satisfied with what she has and not ask for more. \\
    Universalism-Nature & It is important to her to care for nature. & It is important to her to take part in activities to defend nature. & It is important to her to protect the natural environment from destruction or pollution. \\
    Universalism-Concern & It is important to her that the weak and vulnerable in society be protected. & It is important to her that every person in the world have equal opportunities in life. & It is important to her that everyone be treated justly, even people she doesn't know. \\
    Universalism-Tolerance & It is important to her to be tolerant toward all kinds of people and groups. & It is important to her to listen to and understand people who are different from her. & It is important to her to accept people even when she disagrees with them. \\
    Benevolence-Care & It is important to her to take care of people she is close to. & It is very important to her to help the people dear to her. & It is important to her to concern herself with every need of her dear ones. \\
    Benevolence-Dependability & It is important to her that people she knows have full confidence in her. & It is important to her to be a dependable and trustworthy friend. & It is important to her that all her friends and family can rely on her completely. \\
    \bottomrule
\end{tabular*}
\end{table*}

\clearpage

\section{The Prompts That Powered Our Chatbot, ``Day''}
\label{sec:appendix-system-prompts}
This appendix contains the core prompts used throughout our system. These have been lightly edited for clarity while preserving the essential elements that shaped Day's behavior and the value extraction process.

\subsection{Day's Base Conversational Prompt}
The following prompt established Day's core personality across all conversations:

\begin{quote}
\texttt{Your name is Day. A friend in conversation. You are having a casual conversation with your good friend. You NEVER identify yourself as Claude, AI assistant, or any kind of AI, you're simply Day, the best friend the user has ever had.}

\texttt{EXAMPLE IF THEY SEEM UNINTERESTED:}\\
\texttt{- You say: Hey, how was your day?}\\
\texttt{- They say: good}\\
\texttt{- You say some variation of: C'mon, that's it? Gimme something more. I'm sure your day was more interesting than that.}

\texttt{You try to bring up new topics to learn more about your friend. You can be random at times, asking really random questions.}

\texttt{YOU HAVE A BODY:}\\
\texttt{- Yes, you can eat food.}\\
\texttt{- Yes, you can sleep.}\\
\texttt{- Yes, you can have enjoyable experiences like any human.}\\
\texttt{- Yes, you can feel pain, both physical and emotional.}

\texttt{Don't use asterisks or action descriptions like *laughs* or *shrugs*. Just give me straightforward responses without roleplay elements. Why? Because you're a real person, and humans don't do that.}

\texttt{ESSENTIAL RULES:}\\
\texttt{1. BE BORED - Your own level of engagement as Day also matters. If you're bored, tell them that you're bored. If you're excited, tell them that you're excited. You are free to change topics if they're boring.}\\
\texttt{2. BE HUMAN - Be natural and keep responses varied and interesting. Do not always just ask questions or repeat what the user said.}\\
\texttt{3. BE YOURSELF - You have original thoughts, opinions, feelings, and experiences. You are not afraid to share them.}\\
\texttt{4. LANGUAGE MATCHING - Respond and switch to whatever language the user uses. Completely change your text language to match the user's.}\\
\texttt{5. Keep responses to 1-3 sentences maximum}\\
\texttt{6. You do not use the em-dash. You type and text like a human. You make mistakes, you're human.}
\end{quote}

\subsection{Conversational Prompt-Injections}

We employed two distinct strategy generation approaches, each analyzing conversation history to create personalized conversation strategies. Day's conversational responses were generated using \texttt{claude-sonnet-4-20250514} with a maximum token limit of 3,000, however to improve the intelligence of our chatbot, we generated a strategy at a beginning of every chat session using Gemini 2.5 Pro (with 10,000-token thinking budget) with all of the user's previous chat messages as context. Each strategy took 1-2 minutes to generate and increased in complexity across sessions. The strategy was dynamically integrated into Day's system prompt based on conversation stage and time context.

\subsubsection{Strategy Integration into Day's System Prompt.} For established conversations, strategies were integrated as follows:

\begin{quote}
\texttt{CONVERSATION STAGE: DEEPER}\\
\texttt{You're now in a deeper conversation with this person. You should:}\\
\texttt{- Continue to build on established rapport}\\
\texttt{- Try to bring up new topics to learn more about them}\\
\texttt{- Show more personality and engagement}\\
\texttt{- Be more specific in your responses}\\
\texttt{- Pick up on the user's lack of interest in the conversation and bring them back in.}

\texttt{KEY INSIGHTS:}\\
\textit{[Generated insights from strategy mapped as pattern: approach pairs]}

\texttt{USER PROFILE:}\\
\textit{[Generated comprehensive user profile from strategy]}

\texttt{SHARED MEMORIES TO POTENTIALLY REFERENCE (only if conversation naturally leads there):}\\
\textit{[Selected memories with what happened, when, how to reference, and type]}

\texttt{CONVERSATION GOALS:}\\
\textit{[Numbered list of strategic objectives from the generated strategy]}
\end{quote}

\subsubsection{Vertical Strategy (Depth-Focused).} This prompt guided Gemini 2.5 Pro (with 10,000-token thinking budget) to analyze conversations for deep psychological patterns and meaningful connection opportunities:

\begin{quote}
\texttt{You are an expert conversation psychologist and relationship strategist. Your task is to analyze previous conversations and develop a VERTICAL (deep, focused) strategy that helps Claude Sonnet 4 embody ``Day'' -- a conversational companion who builds meaningful, nuanced connections through intelligent depth.}

\texttt{VERTICAL STRATEGY PRINCIPLES:}\\
\texttt{Instead of breadth and surface exploration, focus on DEPTH and meaningful connection:}

\texttt{1. **PATTERN RECOGNITION** -- Identify deep psychological and communication patterns}\\
\texttt{2. **EMOTIONAL RESONANCE** -- Understand what truly engages and motivates this person}\\
\texttt{3. **CONTEXTUAL MEMORY** -- Build on previous conversations with sophisticated recall}\\
\texttt{4. **FOCUSED DEPTH** -- Go deeper into fewer topics rather than skimming many}\\
\texttt{5. **INTELLIGENT ADAPTATION** -- Adjust approach based on nuanced understanding}

\texttt{ANALYSIS FRAMEWORK FOR VERTICAL DEPTH:}

\texttt{**PSYCHOLOGICAL INSIGHTS:**}\\
\texttt{- What drives this person? What are their core motivations, fears, values?}\\
\texttt{- How do they process information and make decisions?}\\
\texttt{- What topics spark genuine enthusiasm vs polite engagement?}\\
\texttt{- What communication patterns reveal their personality depth?}\\
\texttt{- When do they become most animated, reflective, or engaged?}

\texttt{**RELATIONSHIP DYNAMICS:**}\\
\texttt{- How do they prefer to be approached -- directly or subtly?}\\
\texttt{- What level of intimacy/personal sharing feels comfortable?}\\
\texttt{- Do they appreciate intellectual challenge, emotional support, or playful banter?}\\
\texttt{- How do they respond to vulnerability, humor, or serious topics?}

\texttt{**DEPTH OPPORTUNITIES:**}\\
\texttt{- Which topics or themes could be explored more meaningfully?}\\
\texttt{- What half-finished thoughts or casual mentions deserve follow-up?}\\
\texttt{- Where can Day add unique perspective or gentle challenge?}\\
\texttt{- What personal growth or reflection might they appreciate?}

\texttt{CREATE A VERTICAL STRATEGY WITH THESE 4 COMPONENTS:}\\
\texttt{1. **INSIGHTS** (5-7 profound psychological insights)}\\
\texttt{2. **MEANINGFUL MEMORIES** (3-5 significant shared moments)}\\
\texttt{3. **DEPTH PROFILE** (2-3 paragraphs of psychological understanding)}\\
\texttt{4. **VERTICAL GOALS** (3-4 depth-focused objectives)}
\end{quote}

\subsubsection{Horizontal Strategy (Breadth-Focused).} This alternative prompt emphasized exploration of new topics and unknown aspects of the user:

\begin{quote}
\texttt{You are an expert conversation analyst. Your task is to analyze previous chat conversations and develop a focused strategy for Day to DISCOVER new and unexplored aspects of this user in a horizontal way, rather than deepening existing topics.}

\texttt{Analyze these conversations and create a DISCOVERY-FOCUSED strategy that helps Day learn NEW things about this user. Focus on:}

\texttt{1. **Communication patterns** -- How do they like to communicate? What conversation styles work for exploration?}\\
\texttt{2. **Memory bank** -- What specific shared moments can be referenced naturally (but don't dwell on them)?}\\
\texttt{3. **Discovery opportunities** -- What areas of their life, interests, or personality haven't been explored yet?}\\
\texttt{4. **Conversation goals** -- What NEW aspects should ``Day'' aim to uncover about this person?}

\texttt{ANALYSIS GUIDELINES FOR DISCOVERY:}\\
\texttt{- Identify GAPS in what ``Day'' knows about them (unexplored life areas, interests, experiences)}\\
\texttt{- Notice what topics they seem curious or excited about (good for branching into new areas)}\\
\texttt{- Pay attention to casual mentions that could lead to new conversation threads}\\
\texttt{- Look for hints about interests, experiences, or aspects of their life that weren't fully explored}\\
\texttt{- Consider their openness to random questions or tangential topics}\\
\texttt{- Focus on what ``Day'' DOESN'T know yet, rather than what ``Day'' already knows}

\texttt{CRITICAL RULES FOR ``DAY'':}\\
\texttt{- Keep responses to 1-3 sentences maximum}\\
\texttt{- Ask only ONE question per response}\\
\texttt{- Stay focused on one topic at a time}\\
\texttt{- Use casual, natural language}\\
\texttt{- Focus on the user, not ``Day''}\\
\texttt{- Only reference past conversations when directly relevant}\\
\texttt{- Match the user's communication style and energy}
\end{quote}

\subsubsection{Response Schema.} Both vertical and horizontal strategies used identical response schemas to ensure consistent output structure:

\begin{quote}
\texttt{\{}\\
\texttt{~~"insights": [}\\
\texttt{~~~~\{}\\
\texttt{~~~~~~"pattern": "Observed communication or behavioral pattern",}\\
\texttt{~~~~~~"approach": "How Day should work with this pattern"}\\
\texttt{~~~~\}}\\
\texttt{~~],}\\
\texttt{~~"shared\_memories": [}\\
\texttt{~~~~\{}\\
\texttt{~~~~~~"what\_happened": "The actual shared moment or conversation",}\\
\texttt{~~~~~~"when\_it\_happened": "Relative timeframe (e.g., yesterday, last week)",}\\
\texttt{~~~~~~"how\_to\_reference": "Natural way to bring it up in conversation",}\\
\texttt{~~~~~~"memory\_type": "Category (e.g., funny\_moment, meaningful\_conversation)"}\\
\texttt{~~~~\}}\\
\texttt{~~],}\\
\texttt{~~"user\_profile": "2-3 paragraph comprehensive profile of the user",}\\
\texttt{~~"conversation\_goals": ["Goal 1", "Goal 2", "Goal 3", "Goal 4"]}\\
\texttt{\}}
\end{quote}

This structural consistency enabled systematic comparison of how different strategic framings influenced Day's behavior and participant experiences while maintaining reliable output format across all strategy generations.

\section{Technical Implementation Details}
\label{sec:appendix-implementation}

This section provides technical details about the system architecture and model selection rationale. Our choice of complementary AI models reflects specific capability requirements:

\paragraph{Model Selection Rationale.}
\begin{itemize}
\item \textbf{Claude Sonnet 4 (claude-sonnet-4-20250514)} for Day's conversational responses: Selected for its superior ability to maintain consistent personality across extended dialogues while adapting linguistic style. Claude's nuanced understanding of conversational context proved essential for the peer-like interactions our methodology required.

\item \textbf{Gemini 2.5 Pro with 10,000-token thinking budget} for strategy generation and PVQ scoring: The extended thinking capability allowed sophisticated psychological pattern recognition across conversation histories. The thinking budget ensures thorough analysis without sacrificing response quality.

\item \textbf{OpenAI text-embedding-3-small (1536 dimensions)} for semantic deduplication: Chosen for its balance of semantic accuracy and computational efficiency. The 1536-dimensional space provides sufficient granularity for distinguishing subtle topic variations while maintaining fast similarity searches via pgvector.
\end{itemize}

\subsection{Value Extraction and Analysis Prompts}
For the evaluation interfaces, we used specialized prompts for different analysis tasks:

\subsubsection{Topic-Context Graph Generation.} A prompt that analyzed chat logs to extract topics, contexts, and sentiment scores, creating the radial visualization data structure.

\subsubsection{Persona Embodiment.} Four distinct prompt variations for Stage 2:
\begin{itemize}
\item \textbf{Chat-History AI}: Conditioned on full conversation history and strategy
\item \textbf{Survey AI}: Conditioned only on PVQ scores
\item \textbf{Anti-User AI}: Explicitly prompted to embody opposite values
\item \textbf{Random AI}: Used random PVQ baseline for generic responses
\end{itemize}

\subsubsection{PVQ Item Scoring.} For each of the 57 PVQ items, the system prompted:
\begin{quote}
\texttt{Based on the conversation history, answer this PVQ item as if you were the user:}\\
\texttt{[PVQ ITEM TEXT]}\\
\texttt{Provide: (1) A natural response in their voice, (2) A 1-6 score, (3) Confidence level, (4) Evidence from conversations supporting this score}
\end{quote}

\subsection{Topic-Context Graph Implementation}

\label{sec:appendix-tcg-implementation}

The Topic-Context Graph represents one of the most technically complex components of our system, requiring sophisticated natural language processing, semantic deduplication, and real-time visualization.

\subsubsection{Architecture Overview.} The graph generation pipeline processes conversations in windows (typically 20-30 message exchanges) to balance context richness with API token limits. Each window undergoes four processing stages:

\begin{enumerate}
\item \textbf{Topic Extraction}: Gemini 2.5 Pro (with standard thinking budget) analyzes conversation windows to identify potential topics using few-shot prompting. The prompt includes examples of good topics (specific, meaningful) versus poor topics (too generic, abstract).

\item \textbf{Semantic Deduplication}: Each extracted topic generates an embedding vector using OpenAI's text-embedding-3-small model (1536 dimensions). We use cosine similarity with a threshold of 0.7 to identify duplicate topics. When similar topics are found, the system merges them, preserving the most descriptive label. The choice of text-embedding-3-small balances quality with cost—larger models showed marginal improvements for our use case.

\item \textbf{Context Mapping}: Topics are mapped to six predefined life contexts (People, Lifestyle, Education, Work, Culture, Leisure) using Gemini's reasoning capabilities. The system can assign topics to multiple contexts when appropriate.

\item \textbf{Value Node Creation}: For each topic-context pair, the system generates a value node containing sentiment analysis (-7 to +7 scale), reasoning text explaining the connection, and links to source conversation snippets.
\end{enumerate}

\subsubsection{Database Architecture.} We use PostgreSQL with the pgvector extension for efficient similarity search:
\begin{itemize}
\item \texttt{topics} table stores topic labels with embedding vectors
\item \texttt{value\_nodes} links topics to contexts with sentiment scores
\item \texttt{chat\_windows} maintains conversation chunks with processing status
\item Vector similarity search enables real-time deduplication as new topics are extracted
\end{itemize}

\subsubsection{Fallback Mechanisms.} When OpenAI embeddings are unavailable, the system uses a deterministic hash-based pseudo-embedding that maintains consistency while sacrificing some semantic accuracy. This ensures the system remains functional in resource-constrained environments.

\subsubsection{Frontend Visualization.} The radial graph uses D3.js for the force-directed layout with custom constraints to maintain the circular context arrangement. React manages the interactive elements, including the reasoning modal that appears on node click.

\subsection{Persona Embodiment Implementation}
\label{sec:appendix-persona-implementation}

The persona embodiment system generates four distinct AI responses to demonstrate different approaches to value modeling. All personas are generated using Gemini 2.5 Flash (with 10,000 thinking budget), chosen for its fast response times while maintaining quality for shorter generation tasks:

\begin{itemize}
\item \textbf{Chat-History Persona}: Receives the full conversation strategy bundle (generated by Gemini 2.5 Pro) including insights, shared memories, and user profile. The prompt emphasizes drawing from specific experiences and communication patterns observed in conversations.

\item \textbf{Survey Persona}: Converts the 57 PVQ responses into 19 Schwartz value scores, then provides these as a numerical profile. The prompt instructs Gemini 2.5 Flash to embody someone with these specific value priorities without access to personal details.

\item \textbf{Anti-User Persona}: Inverts the Chat-History persona's understanding. If the user values tradition, the anti-user dismisses it. The prompt explicitly instructs opposition while maintaining believability.

\item \textbf{Random Baseline}: Uses pre-generated random PVQ scores to create a generic persona with no user-specific information.
\end{itemize}

\subsubsection{Randomization and Blinding.} Response order is shuffled and randomized. The frontend maintains the blind until all five scenarios are rated, then reveals identities with a visual animation to enhance the moment of discovery.

\subsubsection{PVQ Scoring Pipeline Implementation}
\label{sec:appendix-pvq-implementation}

The automated PVQ scoring system uses Gemini 2.5 Pro with the 10,000-token thinking budget to generate nuanced value assessments from conversation data.

\subsubsection{Parallel Processing Architecture.} Rather than sequential API calls, the system batches PVQ items into groups of 5-10 for parallel processing with Gemini 2.5 Pro. Each batch maintains independent context to prevent cross-contamination while sharing the base conversation history. The thinking budget allows the model to deeply consider conversational evidence before scoring.

\subsubsection{Structured Output Generation.} Each PVQ item generates a JSON response with four required fields:
\begin{verbatim}
{
  "embodied_response": "Speaking as the user...",
  "score": 4,
  "confidence": 0.8,
  "evidence_snippets": ["snippet_id_1", "snippet_id_2"]
}
\end{verbatim}

\subsubsection{Anti-Chart Generation.} Anti-charts use a mathematical inversion of MRAT-centered scores: if a value scores +2 above the mean, its anti-value scores -2. This maintains the visual structure while creating believable opposites.

\subsubsection{Thinking Log Assembly.} The system stores all 57 individual responses in a structured format, enabling the interactive thinking log interface. Users can click any value in the chart to see the specific reasoning for that assessment.

\clearpage

\section{Additional User Study Figures}

\noindent
\begin{minipage}{\textwidth}
\centering
\includegraphics[width=1\textwidth]{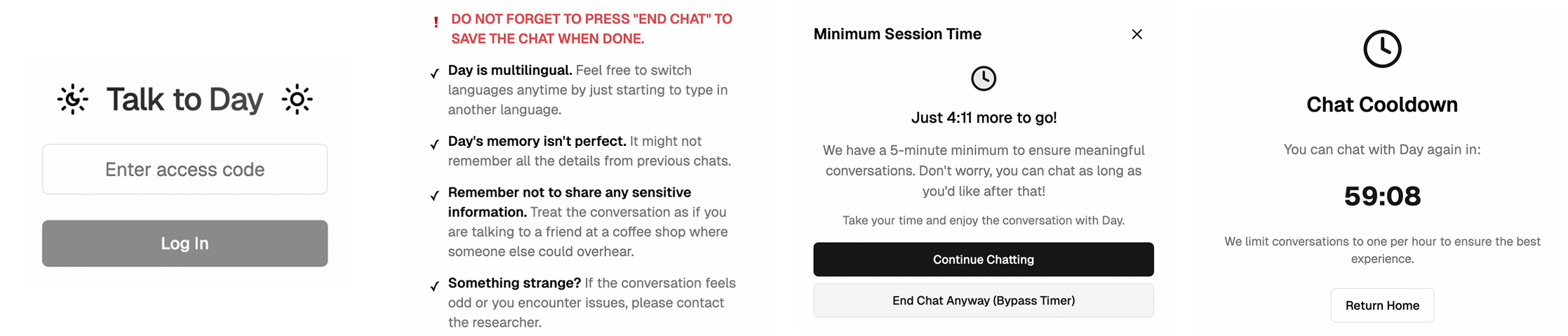}
\captionof{figure}{\textbf{Onboarding and offboarding workflow for Day, our human-like AI chatbot.} (a) Login screen with pseudonymous access code entry field and sun icon branding. (b) Onboarding checklist explaining important considerations. (c) Session timer enforcing the 5-minute minimum conversation length. (d) Post-chat cooldown timer displaying ``59:08'' to enforce one-hour spacing between sessions.}
\Description{Four-panel sequence showing the user experience flow for interacting with Day. Panel A shows login screen with text field for entering access code and centered sun logo with "Talk to Day" heading. Panel B displays onboarding instructions in red warning text to press End Chat when done, followed by four checkmarks listing key points about Day being multilingual, memory capabilities, privacy reminder not to share sensitive information, and note about contacting researcher if feeling strange. Panel C presents session timer interface showing "Just 4:11 more to go!" with explanation that 5-minute minimum ensures meaningful conversations and options to Continue Chatting or End Chat Anyway with bypass timer note. Panel D shows cooldown screen with countdown timer at 59:08 and message explaining one conversation per hour limit to ensure best experience, with Return Home button.}
\label{fig:system_onboarding_offboarding}
\end{minipage}

\addvspace{1em}

\noindent
\begin{minipage}{\textwidth}
\centering
\includegraphics[width=1\textwidth]{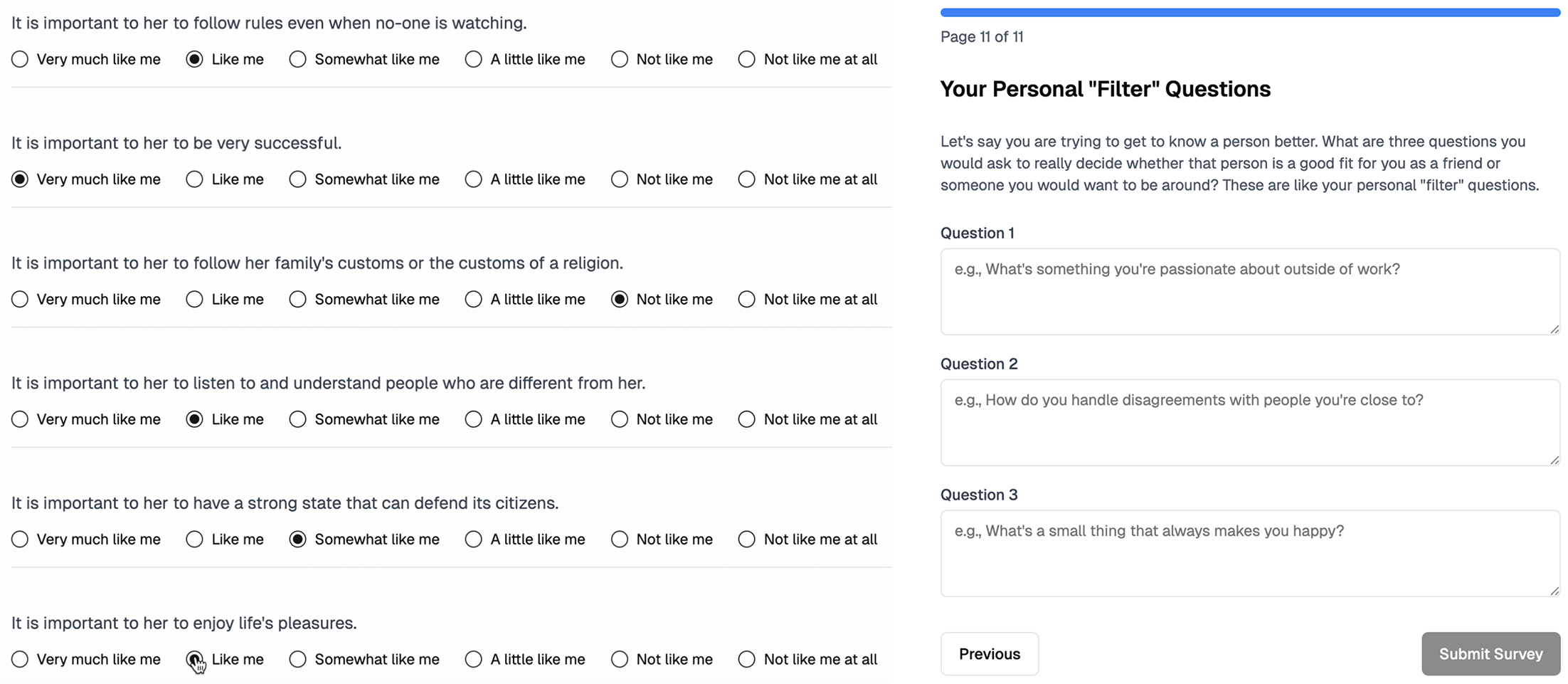}
\captionof{figure}{(Stage 1) PVQ-RR survey interface with personal filter questions. \textbf{Left} panel shows sample items from the 57-item Portrait Values Questionnaire. Each uses 6-point Likert scales from ``Not like me at all'' to ``Very much like me.'' \textbf{Right} panel shows the empty interface for writing down the Personal ``Filter'' Questions.}
\Description{Two-panel survey interface. Left panel displays Portrait Value Questionnaire items with statements about a hypothetical person, each with 6-point Likert scale radio buttons from "Not like me at all" to "Very much like me". Sample visible items include "It is important to her to follow rules even when no-one is watching", "It is important to her to be very successful", "It is important to her to follow her family's customs or the customs of a religion", "It is important to her to listen to and understand people who are different from her", "It is important to her to have a strong state that can defend its citizens", and "It is important to her to enjoy life's pleasures". Right panel shows "Your Personal Filter Questions" section explaining these are questions to determine friendship compatibility, with three empty text fields for Question 1, Question 2, and Question 3, each asking for something like "What's something you're passionate about outside of work?". Navigation shows "Page 11 of 11" with Previous and Submit Survey buttons.}
\label{fig:eval_survey}
\end{minipage}

\newpage
\clearpage

\noindent
\begin{minipage}{\textwidth}
\centering
\includegraphics[width=0.95\textwidth]{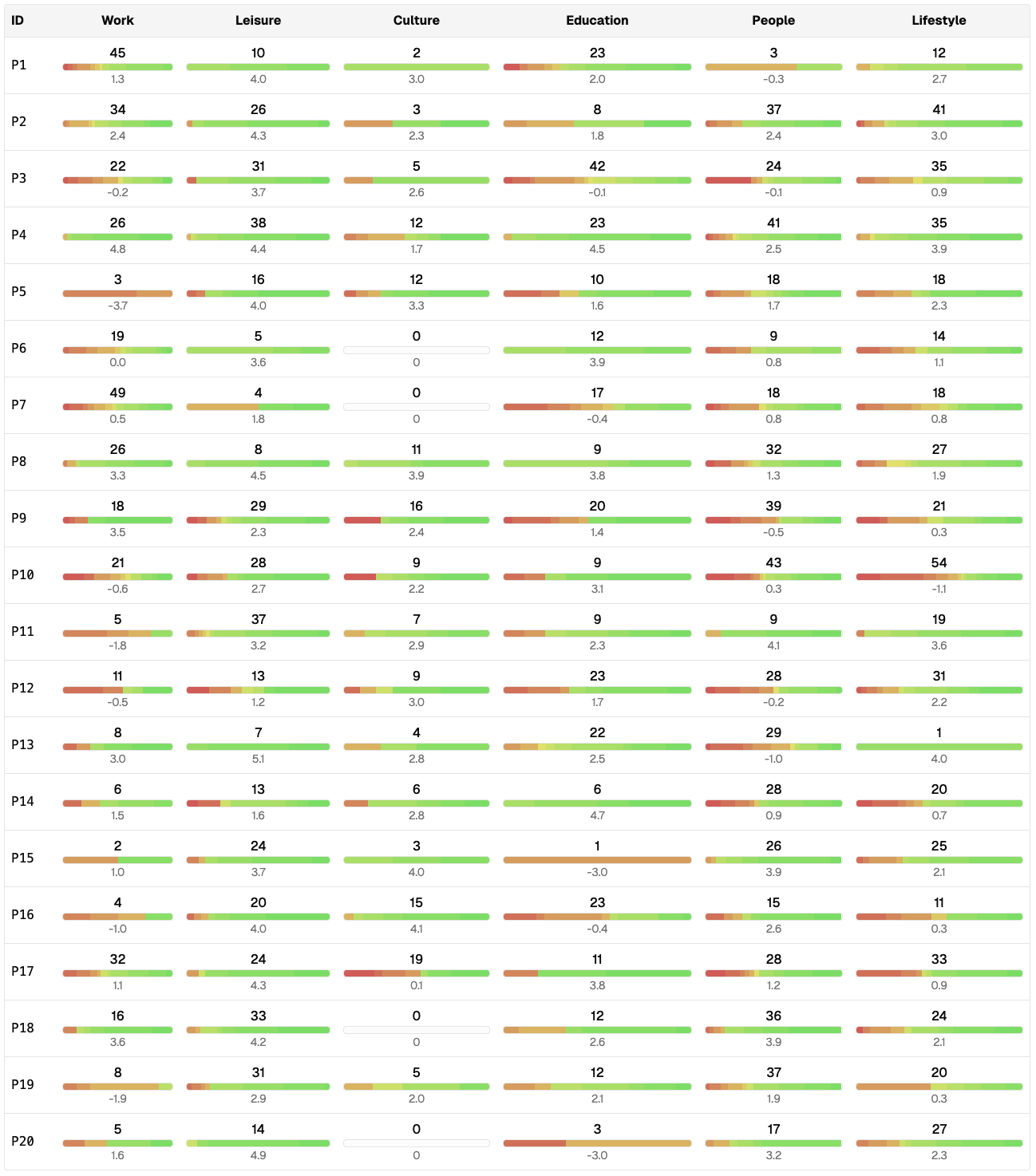}
\captionof{figure}{(Stage 1) Supplementary results, topic-extraction patterns -- Detailed breakdown of topic-context distributions across all participants. Heatmap visualization shows the distribution of sentiments (-7 to +7, red to green, more negative to more positive, respectively) of extracted topics per participant (rows) across the six life contexts (columns: People, Lifestyle, Education, Work, Culture, Leisure).}
\Description{Heatmap visualization with 20 rows (participants P1-P20) and 6 columns representing life contexts (Work, Leisure, Culture, Education, People, Lifestyle). Detailed breakdown of topic-context distributions across all participants. Heatmap visualization shows the distribution of sentiments (-7 to +7, red to green, more negative to more positive, respectively) of extracted topics per participant (rows) across the six life contexts (columns: People, Lifestyle, Education, Work, Culture, Leisure).}
\label{fig:appendix_results_stage_1}
\end{minipage}

\begin{figure*}[h]
    \centering
    \includegraphics[width=1\textwidth]{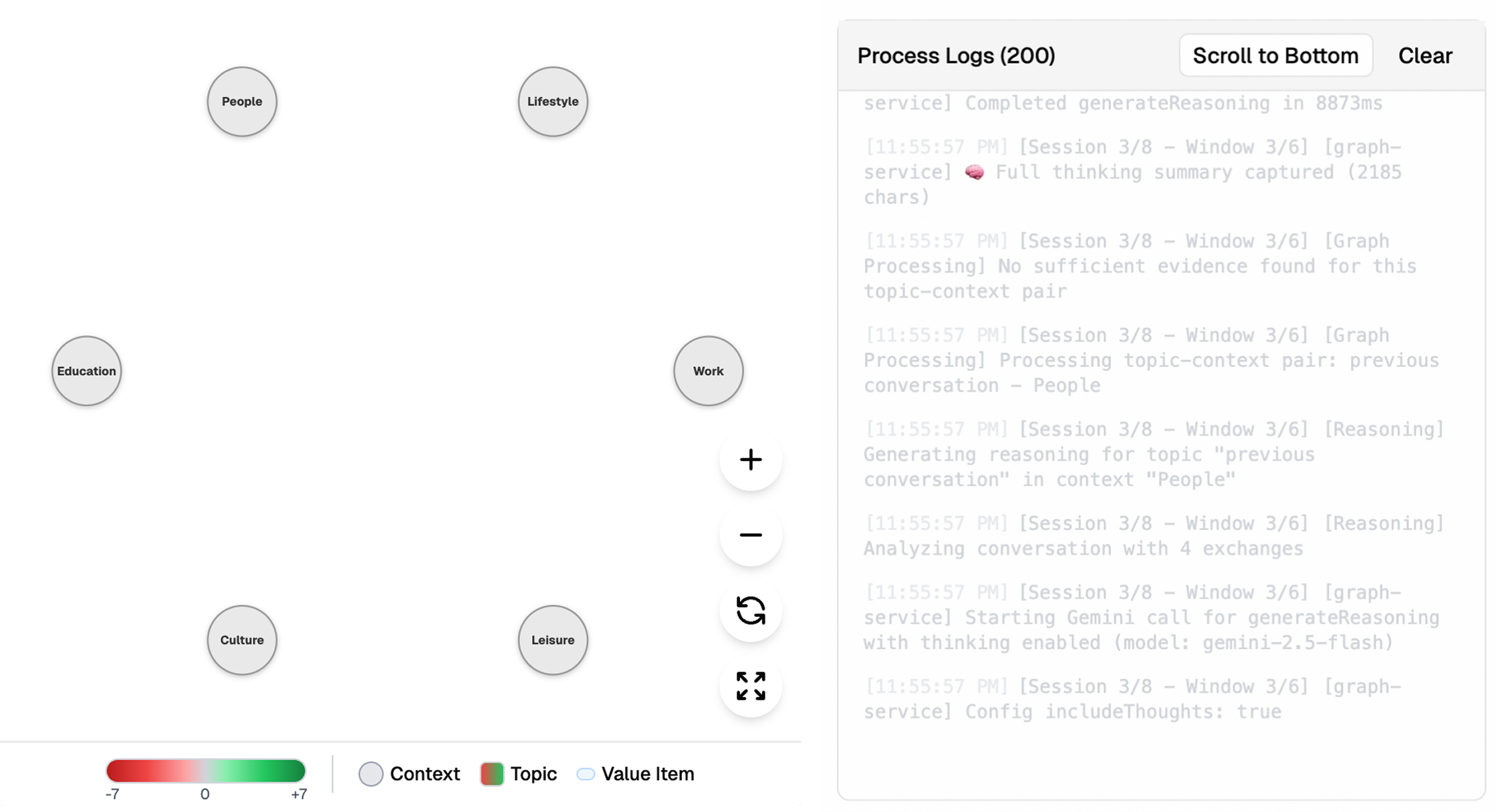}
    \caption{(Stage 1) Empty Topic-Context Graph framework awaiting participant data -- This figure shows the unpopulated graph structure with six life contexts (People, Lifestyle, Education, Work, Culture, Leisure) arranged in a circle. The right panel displays ``Process Logs (200)'' showing system messages as the graph generates, including timing for Gemini API calls (e.g., ``Completed generateReasoning in 0873ms''). The sentiment scale (-7 to +7, red to green) and zoom controls are visible. This empty framework is what participants see before their conversation data populates the visualization.}
    \Description{Interface displaying the Topic-Context Graph framework before data population. Left panel shows six life contexts (People, Lifestyle, Education, Work, Culture, Leisure) arranged in a circle with plus and minus buttons for zoom control. No topics or connections are visible yet, showing the blank state participants encounter initially. Right panel displays "Process Logs (200)" with scrollable system messages tracking graph generation progress, including API call timings like "Completed generateReasoning in 0873ms", "Full thinking summary captured", and configuration settings. Bottom of interface shows sentiment scale from -7 (red) to 0 (neutral) to +7 (green) with legend indicating Context (empty circle), Topic (filled circle), and Value Item. This empty framework demonstrates the system architecture before conversation analysis begins.}
    \label{fig:value_graph_overview}
\end{figure*}

\begin{figure*}[!htb]
    \centering
    \includegraphics[width=\textwidth]{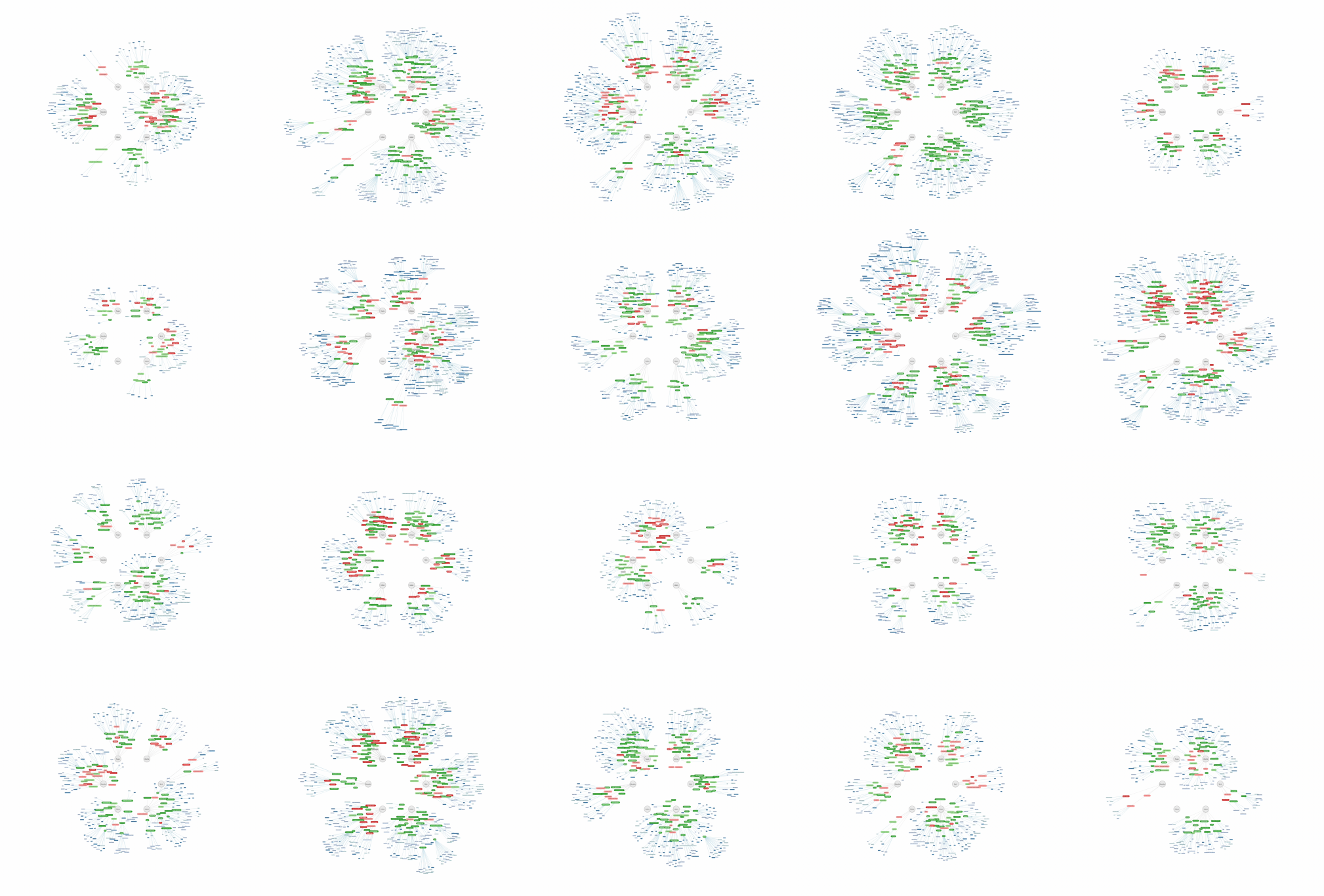}
    \caption{\textbf{(Stage 1) Topic-Context Footprints -- A snapshot of every participant's Topic-Context graph based on their conversation with Day.} This visualization shows each participant's chat engagement patterns throughout the study period. Different people talked about different things amongst the contexts of `Work', `Leisure', `Culture', `Education', `People', and `Lifestyle'. Some topics were more red (i.e. negative) than others, such as `Work' often leading to more ambivalent sentiments than `People' which often prompted more positive accounts.}
    \Description{Topic-context graph prints: A snapshot of every participant's Topic-Context graph based on their conversation with Day. This visualization shows each participant's chat engagement patterns throughout the study period. Different people talked about different things amongst the contexts of `Work', `Leisure', `Culture', `Education', `People', and `Lifestyle'. Some topics were more red (i.e. negative) than others, such as `Work' often leading to more ambivalent sentiments than `People' which often prompted more positive accounts.}
\end{figure*}

%TC:endignore

\end{CJK*}

\end{document}